\pdfoutput=1
\documentclass[a4paper, 11pt]{article}
\usepackage[pdftex]{graphicx,color} 
\usepackage{jheppub}
\usepackage{amsmath}
\usepackage{amssymb}
\usepackage{comment}
\usepackage{multirow}
\usepackage{mathtools}
\usepackage[boxsize=1em,centertableaux]{ytableau}
\usepackage{dsdshorthand}
\usepackage{braket}
\usepackage{tocloft}

\usepackage{tikz}
\usetikzlibrary{arrows,calc,shapes,decorations.pathmorphing,decorations.markings}
\tikzset{
>=stealth',
help lines/.style={dashed, thick},
axis/.style={<->},
important line/.style={thick},
connection/.style={thick, dotted},
  cross/.style={
    cross out,
    draw=black, 
    minimum size=7pt, 
    inner sep=0pt,
    outer sep=0pt
  },
  branchcut/.style={
    decoration={
      snake,
      amplitude=1pt,
      segment length=6pt,
    },
    decorate,
    thick
  },
->-/.style={decoration={
  markings,
  mark=at position #1 with {\arrow{>}}},postaction={decorate}},
  twopt/.style={
    circle,
    draw,
    fill=black,
    inner sep=1pt,
    minimum size=1pt
  },
  scalar/.style={
    thick,
    dashed,
    postaction={
      decorate,
      decoration={
        markings,
        mark=at position 0.5 with {\arrow{>}}
      }
    }
  },
  spinning/.style={
    thick,
    postaction={
      decorate,
      decoration={
        markings,
        mark=at position 0.5 with {\arrow{>}}
      }
    }
  },
  scalar no arrow/.style={
    thick,
    dashed,
  },
  spinning no arrow/.style={
    thick,
  },
  finite/.style={
    decoration={
      snake,
      amplitude=1pt,
      segment length=6pt,
    },
    decorate,
    thick
  },
  axis/.style={
    thick,
    postaction={
      decorate,
      decoration={
        markings,
        mark=at position 1 with {\arrow{>}}
      }
    }
  },
}

\tikzset{snake it/.style={decorate, decoration=snake}}

\newcommand{\bea}{\begin{equation}\begin{aligned}}
\newcommand{\eea}[1]{\label{#1}\end{aligned}\end{equation}}
\newcommand{\beq}{\begin{equation}}
\newcommand{\eeq}{\end{equation}}
\newcommand   \xb  {\bar{x}}
\newcommand   \zb  {\bar{z}}

\newcommand   \pb  {{\bar{p}}}

\DeclareMathOperator\Disc{Disc}
\newcommand{\dDisc}{\text{dDisc}}

\def\1{{\rm 1-loop}}

\newcommand{\diagramEnvelope}[1]{#1}

\newcommand   \bnu  {\boldsymbol{\nu}}
\newcommand   \bmu  {\boldsymbol{\mu}}

\newcommand   \balpha  {\boldsymbol{\alpha}}
\newcommand   \bbeta  {\boldsymbol{\beta}}
\newcommand   \bgamma  {\boldsymbol{\gamma}}
\newcommand   \bdelta  {\boldsymbol{\delta}}

\title{The perturbative CFT optical theorem and high-energy string scattering in AdS at one loop}
\author{Ant\'{o}nio Antunes$ ^{\dagger}$, Miguel S. Costa$ ^{\dagger}$, Tobias Hansen$ ^{\Diamond}$, Aaditya Salgarkar$ ^{\dagger}$, Sourav Sarkar$ ^{\dagger}$}
\affiliation{$ ^{\dagger}$Centro de F\'\i sica do Porto,
Departamento de F\'\i sica e Astronomia\\
Faculdade de Ci\^encias da Universidade do Porto\\
Rua do Campo Alegre 687,
4169--007 Porto, Portugal}
\affiliation{$ ^{\Diamond}$Mathematical Institute, University of Oxford, Andrew Wiles Building,\\
Radcliffe Observatory Quarter, Woodstock Road, Oxford, OX2 6GG, UK}
\affiliation{$ ^{\Diamond}$Department of Physics and Astronomy,
	Uppsala University,\\
	Box 516,
	SE-751 20 Uppsala,
	Sweden}
\emailAdd{alantunes@fc.up.pt, miguelc@fc.up.pt, tobias.hansen@maths.ox.ac.uk, salgarkaraaditya@fc.up.pt, ssarkar@fc.up.pt}

\abstract{We derive an optical theorem for perturbative CFTs which computes the double discontinuity of conformal correlators from the single discontinuities of lower order correlators, in analogy with the optical theorem for flat space scattering amplitudes. The theorem takes a purely multiplicative form in the CFT impact parameter representation used to describe high-energy scattering in the dual AdS theory. We use this result to study  four-point correlation functions that are dominated in the Regge limit by the exchange of the graviton Regge trajectory (Pomeron) in the dual theory. At one-loop the scattering is dominated by double Pomeron exchange and receives contributions from tidal excitations of the scattering states which are efficiently described by an AdS vertex function, in close analogy with the known Regge limit result for one-loop string scattering in flat space at finite string tension. We compare the flat space limit of the conformal correlator to the flat space results and thus derive constraints on the one-loop vertex function for type IIB strings in AdS and also on general spinning tree level type IIB amplitudes in AdS.}

\begin{document}
\maketitle


\section{Introduction}
\label{sec:intro}

In recent years it has been shown that powerful analytical results for scattering amplitudes in quantum field theory, namely the Froissart-Gribov formula and dispersion relations, have equally powerful CFT analogues in the Lorentzian inversion formula \cite{Caron_Huot_2017,Simmons_Duffin_2018,Kravchuk:2018htv,Lemos:2017vnx,Liendo:2019jpu} and the two-variable CFT dispersion relation \cite{Carmi:2019cub,Caron-Huot:2020adz}. Dispersion relations reconstruct a scattering amplitude from the discontinuity of the amplitude, while the Froissart-Gribov formula extracts the partial wave coefficients from the discontinuity and makes their analyticity in spin manifest. The utility of these methods as computational tools for scattering amplitudes stems from the fact that the discontinuity of an amplitude (or that of its integrand) in perturbation theory is determined in terms of lower-loop data by the optical theorem, which in turn is a direct consequence of unitarity. 

The CFT analogue of the discontinuities of amplitudes, which contain the dispersive data and are of central importance in the aforementioned analytical results, is the double discontinuity (dDisc) of CFT four-point functions. The Lorentzian inversion formula 
computes OPE data (anomalous dimensions and OPE coefficients) from the dDisc of four-point functions and establishes the analyticity in spin of OPE data. The CFT dispersion relation, much like its QFT inspiration, directly reconstructs the full correlator from the dDisc. There also exist simpler single-variable dispersion relations in terms of a single discontinuity (Disc) of the correlation function that determine only the OPE coefficients while the anomalous dimensions are required as inputs \cite{Bissi:2019kkx}. 

The unitarity based methods to compute amplitudes  inspire the development of  similar unitarity methods for CFT, in particular, 
for the dDisc of four-point functions one gains a loop or leg order for free.
It was first noticed in large spin expansions \cite{Alday:2016njk,Alday:2016jfr,Aharony:2016dwx} and later understood more generally in terms of the Lorentzian inversion formula
that OPE data at one-loop can be obtained from tree-level data \cite{Alday:2017vkk,Alday:2017zzv}. Generically, in perturbative CFT calculations the dDisc at a given order only depends on OPE data from lower order or lower-point correlators. More recently, in the context of the AdS/CFT correspondence \cite{Maldacena:1997re,Witten:1998qj,Gubser:1998bc},
these unitarity methods for CFT have been related to cutting rules for computing the dDisc of one-loop Witten diagrams \cite{Liu:2018jhs} from tree-level diagrams \cite{Ponomarev:2019ofr,Meltzer:2019nbs,Meltzer:2020qbr}. 
See also the earlier work of \cite{Fitzpatrick:2011dm}. 

However, so far we have been missing a direct adaptation of the optical theorem to CFT correlation functions. More concretely, we still lack 
the ability to express the dDisc of a perturbative correlator, at a given order in the perturbative parameter, in terms of lower order 
correlators, without the detour via the OPE data and without  making explicit reference to AdS Witten diagrams. 
In this paper we provide a direct CFT derivation of such unitarity relations. In particular we present an optical theorem for 1-loop four-point functions wherein the dDisc is fixed in terms of single discontinuities of lower-loop  correlators.

Let us briefly describe the logic that underlies the perturbative CFT optical theorem. Throughout this paper we will  consider the  correlator
\beq
A(y_i) = \< \cO_1 (y_1) \cO_2 (y_2) \cO_3 (y_3) \cO_4 (y_4)  \>\,.
\eeq
We begin by expanding the dDisc of this correlator in $t$-channel  conformal blocks. 
We may do this by expanding in conformal partial waves and then projecting out the contribution of the 
exchange  of the shadow operator $\tilde{\cO}$. The advantage of this procedure is that  when 
writing the partial waves as an integrated  product of three-points functions, the dDisc operation factorizes as a product of discontinuities,
\beq
\dDisc_t A(y_i) =
-\frac{1}{2}
\sum\limits_{\cO} \int dy dy' \,
\Disc_{23} \< \cO_2 \cO_3  \cO (y) \>  \<\tilde{\cO} (y) \tilde{\cO} (y') \>  \Disc_{14}  \< \cO_1 \cO_4  \cO (y') \>
\Big|_{\cO}\,,
\label{eq:general_contribution}
\eeq
where we use the shorthand notation $d^dy\equiv dy$. 
Notice that the sum runs over all operators in the theory. 
We give the precise definitions of the double and single discontinuities of the correlator in section \ref{sec:glue}.

Next, let us assume  that the correlator admits an expansion in a small parameter around mean field theory (MFT).
The example we have in mind is the 
$1/N^2$ expansion,
\beq
A = A_\text{MFT} + \frac{1}{N^2} A_\text{tree} + \frac{1}{N^4} A_\text{1-loop} + \cdots \,.
\label{eq:loop_expansion}
\eeq
We can then separate the sum over intermediate operators $\mathcal{O}$ into single-, double-, and higher-trace operators, and rewrite the multi-trace contributions as 
higher-point functions of single-trace operators.
The contribution of single-trace operators to the $t$-channel expansion of dDisc
in \eqref{eq:general_contribution} is left unchanged and is still given
in terms of discontinuities of three-point functions
\bea
\dDisc_t A(y_i)\Big|_\text{s.t.} =
-\frac{1}{2}
\sum\limits_{\cO \in s.t.} \int dy dy' \,
\Disc_{23} \< \cO_2 \cO_3  \cO (y) \> \<\tilde{\cO} (y) \tilde{\cO} (y') \>   \Disc_{14}  \< \cO_1 \cO_4  \cO (y') \>
\Big|_{\cO}\,.
\eea{eq:stcontribution}
Here no simplifications occur, however this contribution is already simple as
loop corrections come from corrections to the three-point functions of single-trace operators.

The essential simplification that we call the perturbative optical theorem arises for the
contributions of double-trace operators to \eqref{eq:general_contribution}, which are now expressed in terms of
discontinuities of four-point functions of single-trace operators
\bea
\dDisc_t A_{\rm 1-loop}(y_i)\Big|_\text{d.t.} \!
= -\frac{1}{2}
\sum\limits_{\substack{\mathcal{O}_5,\mathcal{O}_6\\ \in\  s.t.}}
 \!
 \int dy_5 dy_6 \, \Disc_{23}  A_{\rm tree}^{3652}(y_k) \; \bS_{5}\bS_{6}\Disc_{14} A_{\rm tree}^{1564}(y_k)  \Big|_{[\cO_5 \cO_6]}\,.
\eea{eq:cft_optical_theorem_intro}
Here and henceforth, we shall use the notation $A^{abcd}(y_k)=\< \cO_a \cO_b  \cO_c \cO_d \>$ 
to denote the correlator of a set of operators other than $\< \cO_1 \cO_2  \cO_3 \cO_4 \>$, which we denote simply as $A(y_i)$.
$\bS_{5}\bS_{6} A^{1564}$ is defined as the shadow transform of $A^{1564}$ with respect to 
the operators $\cO_{5}$ and $\cO_{6}$.
The operators $\cO_5$ and $\cO_6$ are summed over all single-trace operators for which the tree-level correlators exist. These may have spin, in which case the indices are contracted between the two tree-level correlators.
Importantly, in this case dDisc is of order $1/N^4$ and can be computed from the product of 
the discontinuities of tree-level four-point functions, each of order  $1/N^2$.
  
Together equations (\ref{eq:stcontribution}) and (\ref{eq:cft_optical_theorem_intro})
compute the full double discontinuity at one-loop in large $N$ CFTs, since the contributions from higher traces will start at higher loops. 
 Their analogue is of course the optical theorem for amplitudes which computes discontinuities of one-loop amplitudes in terms of two- and one-line cuts. Note that although we use the notation 
 $A_\text{tree}$ and $A_\text{1-loop}$, these refer to conformal correlation functions and in general are not Witten diagrams. The notation with the terms ``one-loop" and ``tree" for the correlators is used only because we always refer to a perturbative expansion. The result is valid for CFTs with an expansion in a small parameter around MFT. The fact that it naturally handles cuts of spinning particles gives an advantage over previous CFT unitarity methods that work in terms of OPE data.

\begin{figure}
\begin{equation*}
\dDisc_t \,
\diagramEnvelope{\begin{tikzpicture}[anchor=base,baseline]
	\node [coordinate] (BL) at (-1.4,-.7) {};
	\node [coordinate] (TL) at (-1.1, .9) {};
	\node [coordinate] (BR) at ( 1.4,-.7) {};
	\node [coordinate] (TR) at ( 1.1, .9) {};
	\node at (-1.4,-1.2) {$1$};
	\node at (-1.2, 1.1) [] {$2$};
	\node at ( 1.4,-1.2) [] {$4$};
	\node at ( 1.2, 1.1) [] {$3$};
    \draw [thick] (0,0) ellipse (1.7 and 1.2);
    \fill[cyan,fill opacity=0.4] (BR) to [out=160,in=20] (BL)
    to [out=40,in=300] (TL)
    to [out=330,in=210] (TR)
    to [out=230,in=120] cycle;
	\draw [thick] (BR) to [out=160,in=20] (BL);
	\draw [thick] (BL) to [out=40,in=300] (TL);
	\draw [thick] (TL) to [out=330,in=210] (TR);
	\draw [thick] (TR) to [out=230,in=120] (BR);
    \fill[white] (-.28,.07) to [out=20,in=160] (.28,.07)
    to [out=340,in=200] cycle;
	\draw [thick] (-.3,.05) to [out=20,in=160] (.3,.05);
	\draw [thick] (-.5,.1) to [out=340,in=200] (.5,.1);
\end{tikzpicture}} \,
\sim \sum\limits_{\cO_5, \cO_6} \int dy_{5,6}
\Disc_{23}
\diagramEnvelope{\begin{tikzpicture}[anchor=base,baseline]
	\node [coordinate] (BL) at (-1.1,-.5) {};
	\node [coordinate] (TL) at (-.9, .7) {};
	\node [coordinate] (BR) at ( 1.1,-.5) {};
	\node [coordinate] (TR) at ( .9, .7) {};
	\node at (-1.1,-1.1) {$5$};
	\node at (-1., 1.) [] {$2$};
	\node at ( 1.1,-1.1) [] {$6$};
	\node at ( 1., 1.) [] {$3$};
    \draw [thick] (0,0) ellipse (1.3 and 1.);
    \fill[cyan,fill opacity=0.4] (BR) to [out=160,in=20] (BL)
    to [out=40,in=300] (TL)
    to [out=330,in=210] (TR)
    to [out=230,in=120] cycle;
	\draw [thick] (BR) to [out=160,in=20] (BL);
	\draw [thick] (BL) to [out=40,in=300] (TL);
	\draw [thick] (TL) to [out=330,in=210] (TR);
	\draw [thick] (TR) to [out=230,in=120] (BR);
\end{tikzpicture}}
\Disc_{14}
\diagramEnvelope{\begin{tikzpicture}[anchor=base,baseline]
	\node [coordinate] (BL) at (-1.1,-.5) {};
	\node [coordinate] (TL) at (-.9, .7) {};
	\node [coordinate] (BR) at ( 1.1,-.5) {};
	\node [coordinate] (TR) at ( .9, .7) {};
	\node at (-1.1,-1.1) {$1$};
	\node at (-1., 1.) [] {$\tl 5$};
	\node at ( 1.1,-1.1) [] {$4$};
	\node at ( 1., 1.) [] {$\tl 6$};
    \draw [thick] (0,0) ellipse (1.3 and 1.);
    \fill[cyan,fill opacity=0.4] (BR) to [out=160,in=20] (BL)
    to [out=40,in=300] (TL)
    to [out=330,in=210] (TR)
    to [out=230,in=120] cycle;
	\draw [thick] (BR) to [out=160,in=20] (BL);
	\draw [thick] (BL) to [out=40,in=300] (TL);
	\draw [thick] (TL) to [out=330,in=210] (TR);
	\draw [thick] (TR) to [out=230,in=120] (BR);
\end{tikzpicture}}
\end{equation*}
\caption{In the Regge limit the dDisc of the genus one closed string amplitude in AdS is given by the perturbative CFT optical theorem in terms of genus zero amplitudes.}
\label{fig:optical_theorem_string_pictures}	
\end{figure}
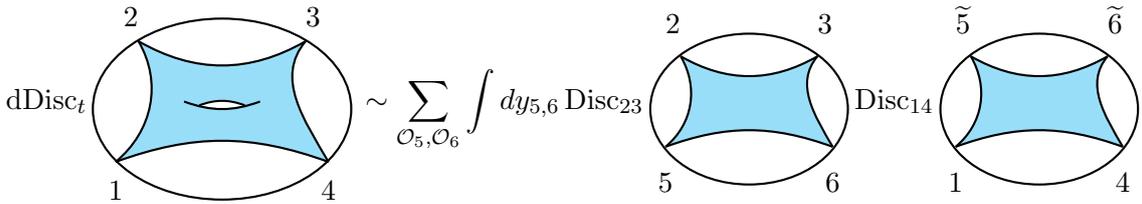
In the second part of the paper, we employ the perturbative CFT optical theorem in the context of the AdS/CFT correspondence \cite{Maldacena:1997re,Witten:1998qj,Gubser:1998bc}
to study high-energy scattering of strings in AdS, which is governed by the CFT Regge limit \cite{Cornalba:2006xk,Cornalba:2006xm}. This is illustrated in figure \ref{fig:optical_theorem_string_pictures}.
High-energy string scattering in flat space has been of interest for a long time, both in the fixed angle case \cite{Gross:1987kza,Gross:1987ar} and in the fixed momentum transfer Regge regime \cite{Amati:1987wq,Amati:1987uf,Amati:1988tn}. This second set of works studied the effects of the finite string size on the exponentiation of the phase shift (eikonalization) in the Regge limit. In particular, 
it was shown that the amplitudes indeed eikonalize provided we allow the phase shift to become an operator acting on the string Hilbert space, whose matrix elements account for the possibility of the external particles becoming intermediate excited string states, known as tidal excitations.
The phase shift $\de(s,b)$, which depends on the Mandelstam $s$ and on the impact parameter $b$, is 
obtained by Fourier transforming the amplitude with respect to momentum transfer 
in the directions transverse to the scattering plane. This gives a multiplicative optical theorem of the form
	\bea
		\Im \de_{\text{1-loop}}(s,b) 
		={}& \frac{1}{2} \sum\limits_{\substack{m_5,\rho_5,\e_5\\m_6,\rho_6,\e_6}}
		\de_\text{tree}^{3652}(s,-b)^* \,\de_\text{tree}^{1564}(s,b)\,,
	\eea{eq:impact_optical_theorem_flat}
where the sum is over all possible exchanged particles, characterized by their mass $m_i$ and Little group representation $\rho_i$,
and their polarization tensors $\epsilon_i$.
In \cite{Amati:1987uf} the one-loop amplitude for four-graviton scattering in type IIB string theory was presented in a particularly nice form, where the tidal excitations, which constitute a complicated sum in \eqref{eq:impact_optical_theorem_flat}, are packaged into a single explicit scalar function, the so-called vertex function.
 
To study the analogous process in AdS we derive an AdS/CFT analogue of \eqref{eq:impact_optical_theorem_flat} by transforming the correlators in the CFT optical theorem \eqref{eq:cft_optical_theorem_intro} to AdS impact parameter space \cite{Cornalba:2006xk,Cornalba:2006xm}. This gives the following multiplicative optical theorem for CFTs
\beq
-\Re \cB_{\text{1-loop}}(p,\bar{p})  \Big|_{\text{d.t.}} = \frac{1}{2}
\sum\limits_{\cO_5, \cO_6 \in s.t.}
\cB^{3652}_\text{tree}(-\pb,-p)^*  \, \cB^{1564}_\text{tree}(p,\pb)  \Big|_{[\cO_5 \cO_6]} \,.
\label{eq:gluing_stripped_intro}
\eeq
Here $\cB$ denotes the impact parameter transform of $A$. 
These transforms depend on two cross ratios $S$ and $L$, respectively interpreted as the square of the energy 
and as the impact parameter of the AdS scattering process,
that can be expressed in terms of two $d$-dimensional vectors $p$ and $\bar{p}$, as will be detailed below.
When ${\cO_5}$ or ${\cO_6}$ have spin,  $\cB$ has tensor structures that depend on
$p$ and $\bar{p}$.
Equations \eqref{eq:gluing_stripped_intro} and \eqref{eq:impact_optical_theorem_flat} are related through the flat space limit for the impact parameter representation, where the radius of AdS is sent to infinity and where $\cB(p,\bar{p})$ is mapped to $i \de(s,b)$.
In this way, each of the infinite number of tree-level correlators with spinning particles 5 and 6 that appear on the right hand side of \eqref{eq:gluing_stripped_intro} is partially fixed by the corresponding flat space phase shift. Moreover, we will be able to efficiently describe the summed result in terms of an AdS vertex function, 
which is in turn constrained by the one-loop flat space vertex function, as constructed for example for type IIB strings in  \cite{Amati:1987uf}.

For neutral scalar operators of dimension four in $d=4$, the four-point function considered here is dual to the scattering of four dilatons in the bulk of AdS$_5$. There are two expansion parameters that we need to consider, the loop order parameter $1/N^2$, and the t'Hooft coupling $\lambda$.
The large $\lambda$ limit is given by supergravity in AdS.
In this limit the tree-level four-point function is dominated by graviton exchange \cite{Cornalba:2006xk,Cornalba:2006xm} and beyond tree-level one can safely resum  the $1/N$ expansion by exponentiating the single graviton exchange \cite{Cornalba:2007zb,Brower:2007qh}.
For finite $\lambda$, string effects are included at tree-level via Pomeron exchange \cite{Brower:2006ea} and can
be described using conformal Regge theory \cite{Cornalba:2007fs,Costa:2012cb}. 
A very non-trivial question we address in this paper is the inclusion of string effects beyond tree-level.

To account for such effects in the Regge limit,  
the earlier works  \cite{Cornalba:2007fs,Cornalba:2008qf,Brower:2007xg}
conjectured the exponentiation of the tree-level Pomeron  phase shift, assuming stringy tidal excitations to be negligible \cite{Cornalba:2009ax}. 
More recently \cite{Meltzer:2019pyl}, the loop effects of Pomeron exchange were systematically taken into account from the CFT side in the AdS high-energy limit $S \gg \lambda \gg 1 $, with the crucial use of CFT unitarity to obtain higher-loop amplitudes from the lower-loop ones. This work also pointed out the suppression of tidal excitations in the supergravity limit $\lambda \gg 1$, in agreement with \cite{Cornalba:2007fs,Cornalba:2008qf,Brower:2007xg}. In the present work, we take finite $\lambda$ (or $\alpha'$) and include all tidal or stringy corrections. This is made possible because the perturbative CFT optical theorem is able to describe cuts involving spinning operators, so we can take into account
intermediate massive string excitations that are exchanged in the t-channel. 

This paper has the following structure. In section \ref{sec:glue}  we first motivate how \eqref{eq:general_contribution} for double-trace operators leads  to the perturbative CFT optical theorem \eqref{eq:cft_optical_theorem_intro} using the technique of ``conglomeration" \cite{Fitzpatrick:2011dm}, and then give a detailed derivation of \eqref{eq:cft_optical_theorem_intro} using tools from  harmonic analysis of the conformal group. Then in section \ref{sec:flat_space} 
we review some important ideas from flat space scattering, including impact parameter space, unitarity cuts and the vertex function, both to guide the AdS version and to serve as a target for the flat space limit. We subsequently move to the holographic case in section \ref{sec:ads}, where we transform the correlator to CFT impact parameter space to write a multiplicative optical theorem for phase shifts.
We use conformal Regge theory in the case of arbitrary spinning operators leading to the derivation of the AdS vertex function. In section \ref{sec:Appendix_tchannel} we recover the results for the one-loop correlator in the large $\lambda$ limit \cite{Meltzer:2019pyl} and also derive new t-channel constraints on CFT data at finite $\lambda$. We give the details of the flat space limit prescription in section \ref{sec:flat_space_limit}, and consider the specific four-dilaton amplitude of type IIB strings in section \ref{sec:IIB_AdS_flat}, constraining  several spinning tree-level correlators of the dual ${\cal N}=4$ SYM theory.
We conclude and briefly discuss some generalizations and applications of our work in section \ref{sec:conclusions}. Many technical details and additional considerations about spinning amplitudes are relegated to the appendices.


\section{Perturbative CFT optical theorem}
\label{sec:glue}

In this section we will give a derivation for the perturbative CFT optical theorem in \eqref{eq:cft_optical_theorem_intro} using results from harmonic analysis of the conformal group following \cite{Karateev:2018oml}, but first let us motivate \eqref{eq:cft_optical_theorem_intro} and \eqref{eq:stcontribution} using the conglomeration of operators \cite{Fitzpatrick:2011dm}. 

Unitarity in CFT can be formulated as  completeness of the set of states corresponding to local operators
\beq
1 = \sum_{\mathcal{O}} |\cO|\,.
\eeq
The right hand side is a sum over projectors associated to a primary operator $\cO$.
Such projectors can be formulated in terms of a conformally invariant pairing 
known as the shadow integral \cite{Ferrara:1972uq,Simmons_Duffin_2014}
\beq
|\cO| = \int dy\, | \cO(y)\> \< \bS[\cO](y) |  \Big|_{\cO}\,,
\label{eq:shadow_projector}
\eeq
which defines the projector to the conformal family with primary operator $\cO$,
automatically taking into account the contribution of descendants of $\cO$.
Here we used the shadow transform, defined by
\be
\label{eq:shadowtransform}
\bS[\cO](y) &= \frac{1}{N_{\cO}}\int  dx\, \<\tl \cO(y) \tl \cO^\dag(x)\>\cO(x)\,,
\ee
with an index contraction implied for spinning operators. We normalize the two-point functions to unity and 
\be
N_{\cO} = \pi^{d}(\Delta-1)_{|\rho|}(d-\Delta-1)_{|\rho|}\,
\frac{\Gamma\big(\Delta-\frac{d}{2}\big)\Gamma\big(\frac{d}{2}-\Delta\big)}{\Gamma(d-\Delta+|\rho|)\Gamma(\Delta+|\rho|)} \,. 
\label{eq:shadownormali}
\ee
Note that with this normalization of $\bS[\cO]$, $\bS^{2}$ is $1/N_{\cO}$  times the identity map.
$|\rho|$ is the number of indices of the operator $\cO$.
The shadow transform is a map
from the operator $\cO$ to $\tl\cO$, where $\tl \cO$ is in the representation labeled by $(\tl \De = d-\De, \rho)$.
$\cO^\dag$ is an operator with scaling dimension $\Delta$ but transforming in the dual $SO(d)$ representation $\rho^{*}$.

Inserting the projector \eqref{eq:shadow_projector} into a four-point function, one finds the contribution of the t-channel 
conformal partial wave $\Psi_\cO$ to the four-point function
\beq
\< \cO_2 \cO_3 |\cO| \cO_1 \cO_4 \> \propto \Psi^{3214}_\cO\,.
\eeq
The conformal partial wave is a linear combination of the conformal blocks for exchange of $\cO$ and its shadow $\tl \cO$.
This explains  the notation $|_{\cO}$ adopted in \eqref{eq:shadow_projector}, since we need to project onto the contribution from $\cO$ and discard that of $\tl \cO$.

In the large $N$ expansion of CFTs, there exists a complete basis of states spanned by the multi-trace operators. In a one-loop four-point function of single trace operators, with an expansion as shown in \eqref{eq:loop_expansion},
only single- and double-trace operators appear
\beq
A(y_i)
= \sum\limits_{\cO \in \cO_\text{s.t.}, \,\cO_\text{d.t.}}\< \cO_2 \cO_3 |\cO| \cO_1 \cO_4 \>\,.
\label{eq:4pt_projected}
\eeq
The right hand side involves three-point functions with single- and double-trace operators.
The double-trace operators are composite operators of the schematic form
\beq
[\cO_5 \cO_6]_{n,\ell} \sim \cO_5 \partial^{2n} \partial_{\mu_1} \ldots \partial_{\mu_\ell}
\cO_6\,,
\eeq
and have conformal dimensions
\beq
\Delta_5 +\Delta_6 +2n +\ell + O \big( 1/N^2\big)\,.
\eeq
Below we often omit the $n$ and $\ell$ labels when talking about a family of double-trace operators.
To obtain an optical theorem resembling the one in flat space, we would like to project onto states created by products of single-trace operators $| \cO_5(y_5) \cO_6(y_6) \>$, rather than the often infinite sum over $n$ and $\ell$ of the double-trace operators $| [\cO_5 \cO_6]_{n,\ell} (y) \>$.
This can be achieved by relating these two states using the technique of conglomeration \cite{Fitzpatrick:2011dm}, 
which amounts to using the formula
\beq
\label{eq:intro_conglo}
| [\cO_5 \cO_6]_{n,\ell}(y) \> = \int dy_5 dy_6\, | \cO_5 (y_5) \cO_6 (y_6) \> \<  \bS[\cO_5](y_5) \bS[\cO_6](y_6) [\cO_5 \cO_6]_{n,\ell} (y) \> \,.
\eeq
This shows that we can define a projector
onto double-trace operators in terms of a double shadow integral
\beq
\label{eq:conglomeration_projector}
| \cO_5 \cO_6 | = \int dy_5 dy_6 \, | \cO_5 (y_5) \cO_6 (y_6) \> \< \bS[\cO_5](y_5) \bS[\cO_6](y_6) | \Big|_{[\cO_5 \cO_6]}\,,
\eeq
and thus \eqref{eq:intro_conglo} is just the projection
\beq
| \cO_5 \cO_6 | [\cO_5 \cO_6]_{n,\ell} \> = | [\cO_5 \cO_6]_{n,\ell} \>\,.
\eeq
The notation $|_{[\cO_5 \cO_6]}$ means that we project onto the contributions from the double-traces of the physical operators and discard contributions coming from the shadows, which, as we will discuss below, can be generated when using this bi-local projector.
Using this projector, together with (\ref{eq:shadow_projector}) for the single-traces,
we can write the one-loop four-point function in \eqref{eq:4pt_projected} as
\beq
A(y_i)
= \sum\limits_{\cO \in \cO_\text{s.t.}}\< \cO_2 \cO_3 |\cO| \cO_1 \cO_4 \>
+\sum\limits_{\cO_5, \cO_6 \in \cO_\text{s.t.}}\< \cO_2 \cO_3 |\cO_5 \cO_6| \cO_1 \cO_4 \>\,.
\label{eq:4pt_projected_conglomerated}
\eeq
The important step in \eqref{eq:4pt_projected_conglomerated} is that we replaced the sum over double-trace operators with a double sum over the corresponding single-trace operators. This is already close to the single- and double-line cuts that appear in the flat-space optical theorem at one-loop.

The main difference of \eqref{eq:4pt_projected_conglomerated} with the flat space optical theorem is that in flat space one needs to sum only over cuts of internal lines, while if we express \eqref{eq:4pt_projected_conglomerated} in terms of Witten diagrams it would also contain contributions from external line cuts.
Another way to see this is that even the disconnected correlator for $\cO_1=\cO_2$ and $\cO_4=\cO_3$ has contributions of the form
\beq
\< \cO_2 \cO_3 |\cO_2 \cO_3| \cO_2 \cO_3 \>\,,
\eeq
while internal double line cuts in a diagram can only appear starting at one-loop.
This problem is resolved by acting on \eqref{eq:4pt_projected_conglomerated} with the double discontinuity. This procedure shifts the contributions of external double-traces to a higher order in $\frac{1}{N}$. In the context of \eqref{eq:cft_optical_theorem_intro} that we propose for conformal correlation functions (and not for Witten diagrams specifically), taking the double discontinuity suppresses the contributions of the external double-trace operators $[\cO_{2}\cO_{3}]$ and $[\cO_{1}\cO_{4}]$. We will expand on this further in section \ref{sec:Nexpansion}. 

We will make the definitions of the double discontinuity and the single discontinuities more precise in sec.\ \ref{sec:Nexpansion} but for now, let us mention that the double discontinuity can be written in the following factorized form
\beq
\dDisc_t A(y_i) = - \frac{1}{2} \Disc_{14} \Disc_{23} A(y_i)\,.
\label{eq:dDisc_y}
\eeq
The discontinuities on the right hand side are defined in terms of analytic continuations of the distances $y_{14}^2$ and $y_{23}^2$ to the negative real axis,
\beq
\label{eq:discdefnorg}
\Disc_{jk} A(y_i)=
A(y_i)|_{y_{jk}^{2} \to y_{jk}^{2} e^{\pi i}}
-  A(y_i)|_{y_{jk}^{2} \to y_{jk}^{2} e^{- \pi i}}\,.
\eeq
Note that each term in this discontinuity is defined through a Wick rotation of the two coordinates $y_j$ and $y_k$ while we hold the other points Euclidean (or spacelike separated).

The result \eqref{eq:stcontribution} for the exchange of single-trace operators comes from the first term on the right hand side of \eqref{eq:4pt_projected_conglomerated} with the double discontinuity taken on both sides. This are simply the single-trace terms in the conformal block expansion of the correlator. For the more interesting result \eqref{eq:cft_optical_theorem_intro}, let us use the explicit form of the projector \eqref{eq:conglomeration_projector} in the second term on the right hand side of \eqref{eq:4pt_projected_conglomerated}. This gives
\be
\label{eq:optical_inter}
\sum\limits_{\cO_5, \cO_6 \in \cO_\text{s.t.}} \, \int dy_5 dy_6 \, \< \cO_3 \cO_2 | \cO_5 (y_5) \cO_6 (y_6) \> \< \bS[\cO_5](y_5) \bS[\cO_6](y_6) | \cO_1 \cO_4 \> | \Big|_{[\cO_5 \cO_6]}  \,.
\ee  
We can now take the double discontinuity on the left hand side using \eqref{eq:dDisc_y}, while on the right hand side we can take $\Disc_{23}$ on the first correlator and $\Disc_{14}$ on the second. This gives the result  \eqref{eq:cft_optical_theorem_intro}.
In the next subsections we provide a detailed proof of this perturbative CFT optical theorem using results from  harmonic analysis of the conformal group   \cite{Karateev:2018oml}.  

\subsection{Conformal blocks and partial waves}
\label{sec:blocks_waves}

A conformal correlator can be expanded in $s$-channel conformal blocks as follows,
\beq
A(y_i)=T^{1234}(y_i) {\cal A}^{1234} (z,\bar{z})\,, \quad
{\cal A}^{1234}(z,\bar{z})=\sum\limits_{\cO}c_{12\cO}c_{34\cO}\,g^{1234}_{\cO}(z,\bar{z})\,,
\label{eq:OPE_s}
\eeq
with the kinematical prefactor 
\begin{align}
T^{1234}(y_i)=\frac{1}{y_{12}^{\Delta_1+\Delta_2}y_{34}^{\Delta_3+\Delta_4}}\left(\frac{y^2_{14}}{y^2_{24}}\right)^{\frac{\Delta_{21}}{2}}\left(\frac{y^2_{14}}{y^2_{13}}\right)^{\frac{\Delta_{34}}{2}}\,,
\label{eq:T}
\end{align}
where  $\Delta_{ij}=\Delta_{i}-\Delta_{j}$ and the cross-ratios are defined as
\begin{align}
z\bar{z}=\frac{y_{12}^{2}y_{34}^{2}}{y_{13}^{2}y_{24}^{2}}, \qquad (1-z)(1-\bar{z})=\frac{y_{14}^{2}y_{23}^{2}}{y_{13}^{2}y_{24}^{2}}\,.
\label{eq:crossratios}
\end{align}
The $t$-channel OPE  is obtained by exchanging the labels 1 and 3, thus
\beq
A(y_i)=T^{3214}(y_i) {\cal A}^{3214} (z,\bar{z})\,, \quad
 {\cal A}^{3214}(z,\bar{z})=\sum\limits_{\cO}c_{32\cO}c_{14\cO}\,g^{3214}_{\cO}(1-z,1-\bar{z})\,,
\label{eq:OPE_t}
\eeq
Note that although $A^{jklm}(y_i)$ is invariant under permutations of the $jklm$ labels, the ordering of the labels is meaningful in ${\cal A}^{jklm}(z,\zb)$ because of the pre-factor $T^{jklm}(y_i)$. For the conformal blocks we will also use the notation
\beq
G^{1234}_{\cO}(y_k) = T^{1234}(y_k)\, g^{1234}_{\cO}(z,\bar{z})\,,
\label{eq:G}
\eeq
and similarly for $t$-channel blocks.

In order to perform harmonic analysis of the conformal group, one expands the four-point function not in conformal blocks but in conformal partial waves of principal series representations $\De = \frac{d}{2} + i \nu$, $\nu \in \mathbb{R}^+$ \cite{Dobrev:1977qv}. A conformal correlator can be expanded in terms of $s$-channel conformal partial waves as follows
\be
\label{eq:partialwaveexpansion}
A(y_i) &= \sum_{\rho} \int_{\frac d 2}^{\frac d 2 + i\oo} \frac{d\De}{2\pi i}  \,I^{1234}_{ab}(\De,\rho) \Psi^{1234(ab)}_\cO(y_i) + \textrm{discrete},
\ee
where the operator $\cO$ is labeled by the scaling dimension $\Delta$ and a finite dimensional irreducible representation $\rho$ of $SO(d)$, which we take to be bosonic.
$I_{ab}$ is the spectral function carrying the OPE data, and it can be extracted from the correlator using the Euclidean inversion formula. We will assume that there are no discrete contributions.
The conformal partial waves are defined as a pairing of three-point structures
\be
\label{eq:partialwavedefinition}
\Psi^{1234(ab)}_\cO(y_i) &= \int d y \,\<\cO_1 \cO_2 \cO(y)\>^{(a)} \< \cO_3 \cO_4 \tl \cO^\dag(y)\>^{(b)}\,,
\ee
where $a$ and $b$ label different tensor structures in case the external operators have spin. 
The conformal partial wave $ \Psi^{1234(ab)}_\cO$ is related to the conformal block $G^{1234(ab)}_{\cO}$ and to the block for the exchange of the shadow by
\be
\label{eq:expressionforpartialwave}
\Psi^{1234(ab)}_\cO &= S(\cO_3\cO_4[\tl\cO^\dag])^b{}_c \,G^{1234(ac)}_{\cO} + S(\cO_1\cO_2[\cO])^a{}_c \,G^{1234(cb)}_{\tl \cO}\,.
\ee
The matrices $S(\cO_i \cO_j [\cO_k])^a{}_b$ are part of the action of the shadow transform \eqref{eq:shadowtransform} on three-point functions,
\be
\label{eq:shadowcoeffdef}
\<\cO_1 \cO_2 \bS[\cO_3]\>^{(a)} &=  \frac{S(\cO_1\cO_2[\cO_3])^a{}_b}{N_{\cO_3}} \, \<\cO_1 \cO_2 \tl \cO_3\>^{(b)}  \,,
\ee 
with $N_{\cO_3}$ as defined in \eqref{eq:shadownormali}. Acting with the shadow transform on an operator within a three-point structure also rotates into a different basis of tensor structures. The shadow coefficients/matrices $S$ act as a map between the two bases.
Note that the inverse of $S(\cO_1\cO_2[\cO_3])^a{}_b$ is $(1/N_{\cO_3})S(\cO_1\cO_2[\tl \cO_3])^a{}_b$.

The usual conformal block expansion \eqref{eq:OPE_s} can be obtained from \eqref{eq:partialwaveexpansion} by inserting \eqref{eq:expressionforpartialwave} and using the identity \cite{Simmons_Duffin_2018}
\be
\label{eq:Iide}
I_{ab}(\Delta,\rho)\, S(\cO_3\cO_4[\tl\cO^\dag])^b{}_c = I_{bc}\big(\tl\Delta,\rho\big)\, S(\cO_1\cO_2[\tl \cO])^b{}_a  \, ,
\ee
to replace the contribution of the shadow block with an extension of the integration region to $\frac{d}{2}-i\infty$,
\be
A(y_i) = \sum_{\rho} \int_{\frac d 2 - i\oo}^{\frac d 2 + i\oo} \frac{d\De}{2\pi i}\, C^{1234}_{ab}(\De,\rho) \,G^{1234(ab)}_\cO(y_i) \,,
\label{eq:confblockexpansionprincipal}
\ee
where
\be
C^{1234}_{ab}(\De,\rho) = I^{1234}_{ac}(\De,\rho) \,S(\cO_3\cO_4[\tl\cO^\dag])^c{}_b \, .
\label{eq:C1234}
\ee
The conformal block decays for large real $\De > 0$, so the contour can be closed to the right and the integral is the sum of residues
\bea
-\underset{\De \to \De^*}\Res C^{1234}_{ab}(\De,\rho^*) 
&=
\sum_{I} c_{12\cO^*,a}^{I} c_{34\cO^*,b}^{I} \,.
\eea{eq:residue}
The sum over $I$ in \eqref{eq:residue} is over degenerate operators with the quantum numbers $\left(\Delta^{*},\rho^{*}\right)$. 
Degeneracies among multi-trace operators are natural in expansions around mean field theory.\footnote{A simple example built with spin 1 operators are the families $\mathcal{O}_5^\mu \square^n \mathcal{O}_{6,\mu}$ and $\mathcal{O}_5^\mu \partial_\mu\partial_\nu\square^{n-1} \mathcal{O}_6^\nu$, which we wrote schematically. Both these sets of operators have quantum numbers $\Delta= \Delta_5 +\Delta_6+2n$ and $\rho= \bullet$.}

In section \ref{sec:HAproof} we will use the partial wave expansion of the shadow transformed four-point function. To obtain it let us now apply the shadow transform in \eqref{eq:shadowtransform} to $\cO_1$ and $\cO_2$ on both sides of the partial wave expansion \eqref{eq:partialwaveexpansion}.
Using  \eqref{eq:partialwavedefinition} this gives
\bea
\<\mathbf{S}[\cO_1] \mathbf{S}[\cO_2] \cO_3 \cO_4\>& = \sum_{\rho} \int_{\frac d 2}^{\frac d 2 + i\oo} \frac{d\De}{2\pi i} \,I^{1234}_{ab}(\De,\rho)\\
&\int d y \, \<\mathbf{S}[\cO_1] \mathbf{S}[\cO_2] \cO(y)\>^{(a)} \< \cO_3 \cO_4 \tl \cO^\dag(y)\>^{(b)}  \,.
\eea{eq:appshadow}
From \eqref{eq:shadowcoeffdef}, we thus obtain the partial wave expansion of the shadow transformed correlator
\be
\<\mathbf{S}[\cO_1] \mathbf{S}[\cO_2] \cO_3 \cO_4\>  = \sum_{\rho} \int_{\frac d 2}^{\frac d 2 + i\oo} \frac{d\De}{2\pi i}\, I^{\mathbf{S}[1]\mathbf{S}[2]34}_{ab}(\De,\rho)\, \Psi^{\tl 1\tl 2 34(ab)}_\cO(y_i)  \, ,
\label{eq:shadowtrans_exp}
\ee
where
\bea
I^{\mathbf{S}[1]\mathbf{S}[2]34}_{ab} & = I^{1234}_{mb}(\De,\rho)\,   \frac{S(\cO_1 [\cO_2] \cO)^{m}{}_{n}}{N_{\cO_2}} \frac{S([\cO_1] \tl\cO_2 \cO)^n{}_a}{N_{\cO_1}} \,,   
\\
\Psi^{\tl 1\tl 2 34(ab)}_\cO(y_i) & = \int dy\, \<\tl\cO_1 \tl\cO_2 \cO(y)\>^{(a)} \< \cO_3 \cO_4 \tl \cO^\dag(y)\>^{(b)}   \,.
\eea{eq:shadowtrans_exp2}
There are examples of the $S$ coefficients computed in \cite{Karateev:2018oml} which tell us that they have the appropriate zeroes to kill the double-trace poles in $I^{1234}$ and replace them with the poles for the double-traces of the shadows, as would be appropriate for $I^{\mathbf{S}[1]\mathbf{S}[2] 34}$.

\subsection{A derivation using harmonic analysis}
\label{sec:HAproof}

We are ready to begin the derivation of the perturbative CFT optical theorem \eqref{eq:cft_optical_theorem_intro}. $\bS_{5}\bS_{6} A^{1564}_{\text{tree}}(y_i)$ in \eqref{eq:cft_optical_theorem_intro} is the coefficient of 
$1/N^2$ in the correlator $\< \mathbf{S}[\cO_{6}]^\dag \mathbf{S}[\cO_5]^\dag  \cO_1 \cO_4\>$, and $A^{3652}_{\text{tree}}(y_i)$ is the coefficient of $1/N^2$ in $\< \cO_{3} \cO_2  \cO_5 \cO_6\>$.
Consider the following conformally invariant pairing of two four-point functions
\begin{align}
&\int dy_5 dy_6 \,
\<\cO_3 \cO_2 \cO_5 \cO_6\> \< \mathbf{S}[\cO_{6}]^\dag \mathbf{S}[\cO_5]^\dag  \cO_1 \cO_4\> \label{eq:gluing1}=
\\
& \sum_{\rho,\rho'} \int_{\frac d 2}^{\frac d 2 + i\oo} \frac{d\De}{2\pi i} \frac{d\De'}{2\pi i} \,  I^{3256}_{ab}(\De,\rho)\, I^{\mathbf{S}[6] \mathbf{S}[5] 14}_{cd}(\De',\rho')
\int dy_5 dy_6 \, \Psi^{3256(ab)}_\cO(y_i)\, \Psi^{\tl 6 \tl 5 14(cd)}_{\cO'}(y_i)\,.\nonumber
\end{align}
To compute the $y_5$ and $y_6$ integrals, we use \eqref{eq:partialwavedefinition} and 
the following result for the pairing of the three-point structures by two legs,
which is known as the bubble integral,
\be
\label{eq:bubbleintegral}
\int dy_1 dy_2 \< \cO_1 \cO_2 \cO(y) \>^{(a)} \<\tl \cO^\dag_1 \tl \cO^\dag_2 \tl \cO^{'\dag}(y') \>^{(b)}
=
\frac{\p{\< \cO_1 \cO_2 \cO \>^{(a)},\<\tl \cO^\dag_1 \tl \cO^\dag_2 \tl \cO^{\dag} \>^{(b)}}}{\mu(\De,\rho)}  \mathbf{1}_{yy'} \de_{\cO\cO'}\,, 
\ee
with $\de_{\cO\cO'} \equiv 2\pi \de(s-s') \de_{\rho \rho'}$.
Here $\mu(\De,\rho)$ is the Plancherel measure and the brackets denote 
a conformally invariant pairing of 3-point functions, given by
\be
\label{eq:structurepairing}
\p{\<\cO_1 \cO_2 \cO_3\>,\<\tl \cO_1^\dag \tl \cO_2^\dag \tl \cO_3^\dag\>} &= \int \frac{d y_1 d y_2 d y_3}{\vol\SO(d+1,1)}\,\<\cO_1 \cO_2 \cO_3\>\<\tl \cO_1^\dag \tl \cO_2^\dag \tl \cO_3^\dag\>\,.
\ee
Using \eqref{eq:partialwavedefinition} and the bubble integral in \eqref{eq:bubbleintegral} we find
\be
\label{eq:cpw_bubble}
\int dy_5 dy_6  \Psi^{3256(ab)}_\cO(y_i) \Psi^{\tl 6 \tl 5 14(cd)}_{\cO'}(y_i) =  
\frac{\p{\< \cO_5 \cO_6 \tl \cO^{\dag} \>^{(b)},\<\tl \cO^\dag_6 \tl \cO^\dag_5 \cO \>^{(c)}}}{\mu(\De,\rho)}  \, \de_{\cO\cO'} \Psi^{3214(ad)}_\cO(y_i)\,.
\ee
We can now plug \eqref{eq:cpw_bubble} into \eqref{eq:gluing1} which gives  
\begin{align}
&\int dy_5 dy_6 
\<\cO_3 \cO_2 \cO_5 \cO_6\> \< \mathbf{S}[\cO_{6}]^\dag \mathbf{S}[\cO_5]^\dag  \cO_1 \cO_4\> =
\label{eq:gluing2}
\\
& \sum_{\rho} \int_{\frac d 2}^{\frac d 2 + i\oo} \frac{d\De}{2\pi i} \, I^{3256}_{ab}(\De,\rho) I^{\mathbf{S}[6] \mathbf{S}[5] 14}_{cd}(\De,\rho)
\frac{\p{\< \cO_5 \cO_6 \tl \cO^{\dag} \>^{(b)},\<\tl \cO^\dag_6 \tl \cO^\dag_5 \cO \>^{(c)}}}{\mu(\De,\rho)} \,  \Psi^{3214(ad)}_\cO(y_i)\,.
\nonumber
\end{align}

In the next steps we will show that the factor $\big( {\< \cO_5 \cO_6 \tl \cO^{\dag} \>,\<\tl \cO^\dag_6 \tl \cO^\dag_5 \cO \>}\big)$
in (\ref{eq:gluing2})
, along with the various shadow coefficients,
will cancel the contribution of the OPE coefficients $c^{\text{MFT}}_{56[56]}$ in the spectral functions $I^{3256}$ and $I^{\mathbf{S}[6] \mathbf{S}[5] 14}$. In the simple case where at least one of the spectral functions in \eqref{eq:gluing2} belong to scalar MFT correlators (which requires pairwise equal operators) this is particularly easy to see, since  \cite{Karateev:2018oml}
\be
\label{eq:MFTspec}
I^{\text{MFT}}(\De,\rho) = \frac{\mu(\De,\rho)}{\p{\< \cO_1 \cO_2 \tl \cO^{\dag} \>,\<\tl \cO^\dag_1 \tl \cO^\dag_2 \cO \>}} \, S([\tl \cO_1] \tl \cO_2 \cO) \, S(\cO_1 [\tl \cO_2 ] \cO)   \,,
\ee
so that the pairing of three-point functions can be canceled directly with one of the spectral functions.
The general case is less obvious because the cancellation happens on the level of OPE coefficients, not spectral functions.
Here we use \eqref{eq:shadowtrans_exp} in \eqref{eq:gluing2}, and extend the range of the principal series integral as in \eqref{eq:confblockexpansionprincipal} by repeated use of \eqref{eq:Iide}. This gives 
\begin{align}
&\int dy_5 dy_6 
\<\cO_3 \cO_2 \cO_5 \cO_6\> \< \mathbf{S}[\cO_{6}]^\dag \mathbf{S}[\cO_5]^\dag  \cO_1 \cO_4\>
={} \sum_{\rho} \int_{\frac d 2 -i\oo}^{\frac d 2 + i\oo} \frac{d\De}{2\pi i} \, I^{3256}_{ab} I^{6514}_{md}  S(\cO_1\cO_4[\tl\cO^\dag])^d{}_l     
 \nonumber
\\
&
\qquad\quad
   \,\frac{S(\cO_6 [\cO_5] \cO)^{m}{}_{n}}{N_{\cO_5}} \frac{S([\cO_6] \tl\cO_5 \cO)^n{}_c}{N_{\cO_6}} \,
 \frac{\p{\< \cO_5 \cO_6 \tl \cO^{\dag} \>^{(b)},\<\tl \cO^\dag_6 \tl \cO^\dag_5 \cO \>^{(c)}}}{\mu(\De,\rho)}   
 \,G^{3214(al)}_\cO(y_i)\,.
\label{eq:gluing3}
 \end{align}
Using  \eqref{eq:C1234} we can express \eqref{eq:gluing3} as
\bea
{}\int dy_5 dy_6 
\<\cO_3 \cO_2 \cO_5 \cO_6\> \< \mathbf{S}[\cO_{6}]^\dag \mathbf{S}[\cO_5]^\dag  \cO_1 \cO_4\>
={} \sum_{\rho} \int_{\frac d 2 - i\oo}^{\frac d 2 + i\oo} \frac{d\De}{2\pi i}  \,C^{3256}_{ak} C^{6514}_{md} \, Q^{km}_{65\cO} \,   G^{3214(ad)}_\cO\,,
\eea{eq:gluing4}
where,
\be
\label{eq:hiddenMFT}
Q^{km}_{65\cO} = \frac{S(\cO_6 [\cO_5] \cO)^{m}{}_{n}}{N_{\cO_5}} \frac{S([\cO_6] \tl\cO_5 \cO)^n{}_c}{N_{\cO_6}}  \frac{\p{\< \cO_5 \cO_6 \tl \cO^{\dag} \>^{(b)},\<\tl \cO^\dag_6 \tl \cO^\dag_5 \cO \>^{(c)}}}{\mu(\De,\rho)} \,\frac{S(\cO_5\cO_6[\cO^\dag])^k{}_b}{N_{\cO}} \,    \, .
\ee
Next we analyze the pole structure of the spectral function in \eqref{eq:gluing4} and close the integration contour to obtain the block expansion.
First let us consider the simple poles at the dimensions of the double-trace operators $\cO_{[56]}$ in each of $C^{3256}$ and $C^{6514}$. We will show that $Q_{65\cO}(\Delta,\rho)$ has a zero at each 
of these dimensions, canceling one of the two poles from $C^{3256}$ and $C^{6514}$.
This ensures that in the MFT limit the spectral function in \eqref{eq:gluing4} has a simple pole for each double-trace dimension.
This can be seen explicitly in specific examples for the $S$ coefficients computed in \cite{Karateev:2018oml}, but in general let us note the following identity, which can be derived by applying Euclidean inversion on the expansion \eqref{eq:partialwaveexpansion} for the MFT correlator \cite{Karateev:2018oml}
\be
\frac{I_{ab}^{6565,\mathrm{MFT}}(\De,\rho)}{\mu(\De,\rho)}
\p{\<\cO_6^\dag \cO_5^\dag \tl \cO^\dag \>^{(b)},\<\tl \cO_6 \tl \cO_5 \cO\>^{(c)}} 
&=
S([\tl \cO_6]\tl \cO_5 \cO)^{c}{}_{l} \, S(\cO_6 [\tl \cO_5]\cO)^l{}_a \,.
\label{eq:mftcoeffs}
\ee
Since all operators are bosonic, \eqref{eq:mftcoeffs} can be expressed as 
\bea
(-1)^{2J}  \,\frac{I_{ab}^{6556,\mathrm{MFT}}(\De,\rho)}{\mu(\De,\rho)} \, \frac{S(\cO_6 [\cO_5] \cO)^{m}{}_{n}}{N_{\cO_5}} \frac{S([\cO_6] \tl\cO_5 \cO)^n{}_c}{N_{\cO_6}}  \p{\<\cO_5 \cO_6 \tl \cO^\dag \>^{(b)},\<\tl \cO_6^\dag \tl \cO_5^\dag \cO\>^{(c)}} \,  = \delta^{m}_{a}     \, .
\eea{eq:mftid1}
Using \eqref{eq:confblockexpansionprincipal} and \eqref{eq:hiddenMFT}, we rephrase \eqref{eq:mftid1} as 
\be
\label{eq:mftid2}
 C_{ak}^{6556,\mathrm{MFT}}(\Delta,\rho)\; Q^{km}_{65\cO}(\Delta,\rho) =  \delta^{m}_{a} \, .
\ee
Let $(\Delta,\rho)$ be $(\Delta^{*},\rho^{*})$ for the double-trace operators $\cO_{[56]^{*}}^{I}$,
where $I$ labels degenerate operators, as discussed previously.
The coefficient $C^{6556,\mathrm{MFT}}_{ak}$ has a simple pole at this location and therefore \eqref{eq:mftid2} implies that $Q^{km}_{65\cO}(\Delta,\rho)$ is its inverse matrix and has a corresponding zero at this value. Evaluated at $\De = \De^*$, \eqref{eq:mftid2} takes the form
\be
\label{eq:mftid3}
\left(\sum_{I} c_{65[56]^{*},a}^{MFT,I}c_{56[56]^{*},k}^{MFT,I}\right) \, q^{km} = \delta^{m}_{a} \,,
\ee
where $c_{65[56]^{*},a}^{MFT,I}c_{56[56]^{*},k}^{MFT,I}$ is the contribution to the residue of $C^{6556,\mathrm{MFT}}_{ak}$ corresponding to $\cO_{[56]^{*}}^{I}$ and $q^{km}$ is the coefficient of the first order zero of $Q^{km}_{65\cO}$ at $\De^*$.

Note that the matrix of OPE coefficients $c_{65[56]^{*},a}^{MFT,I}c_{56[56]^{*},k}^{MFT,I}$ for a specific double-trace operator is singular. In general, \eqref{eq:mftid2} and \eqref{eq:mftid3} imply that there are sufficiently many degenerate double-trace families so the matrix obtained by summing over all of them is not singular. In the case where there is a unique tensor structure, such as when $\mathcal{O}_5$ and $\mathcal{O}_6$ are scalars, the $1\times1$ matrix is of course non-degenerate, so degenerate double-trace operators need not exist.  Contracting both sides of \eqref{eq:mftid3} with $c_{65[56]^{*},m}^{MFT,J}$, we obtain
\be
\label{eq:mftid4}
 c_{56[56]^{*},k}^{MFT,I} \, q^{km} \, c_{65[56]^{*},m}^{MFT,J} = \delta^{IJ} \, .
\ee
Finally,
using \eqref{eq:residue} and \eqref{eq:mftid4} we obtain the contribution of the $(\De^{*},\rho^{*})$ pole to the spectral integral in \eqref{eq:gluing4}
\be
\label{eq:contribonepole}
-\underset{\De \to \De^*}\Res  C^{3256}_{ak} C^{6514}_{md} \, Q^{km}_{65\cO} \,   G^{3214(ad)}_\cO \big|_{\rho \to \rho^*}
= \sum_{I} c_{32[56]^{*},a}^{I} c_{14[56]^{*},d}^{I} \, G^{3214(ad)}_{[56]^{*}}(y_i)   \,.
\ee
Given that this is precisely the contribution of the double-trace operators $[\cO_5 \cO_6]$ to the correlator $A^{3214}(y_i)$, this shows that the conformally invariant pairing
we started with in \eqref{eq:gluing1} computes precisely this contribution, to leading order in $1/N^2$ because we used MFT expressions along the way. Thus
\beq
\Big(1 + O\big(1/N^2\big) \Big)\, A(y_k) \big|_{[\cO_5 \cO_6]} = 
 \int dy_5 dy_6 \, A^{3652}(y_k) \, \bS_5 \bS_6 A^{1564}(y_k)  \Big|_{[\cO_5 \cO_6]}
\,.
\label{eq:gluing_nodisc}
\eeq
In the context of the the projector defined in the previous section in \eqref{eq:conglomeration_projector}, this result can be phrased as
\be
\sum_{n,\ell,I} \Big| [\cO_5 \cO_6]_{n,\ell}^{I} \Big|  
= | \cO_5 \cO_6 | + O\big(1/N^2\big)   \,.
\label{eq:proj_gluing}
\ee
The labels $n,\ell$ sum over the double-trace operators with different dimensions and spins, while $I$ sums over degenerate operators.
The projection $|_{[\cO_5 \cO_6]}$ appears on the two sides of \eqref{eq:gluing_nodisc} for different reasons. On the left hand side it selects one family of double-trace operators among all the operators appearing in the OPE, while on the right hand side it serves to discard poles from shadow operators that we would pick up when we close the contour in \eqref{eq:gluing4}. For example, it is evident from the first equation in \eqref{eq:shadowtrans_exp2} that $Q(\De,\rho)$ has poles at the double-traces $\cO_{[\tl 5\tl 6]}^{I}$ composed of $\tl\cO_{5}$ and $\tl\cO_{6}$ and we pick up these contributions too. Let us take for simplicity the case with  $\cO_{5}$ and $\cO_{6}$ scalars and $\cO$ with integer spin $\ell$ in 4 dimensions. 
The corresponding three-point function has only one tensor structure and the expressions for $S(\cO_6 [\cO_5] \cO)$ and $S([\cO_6] \tl\cO_5 \cO)$ 
are known \cite{Karateev:2018oml}
\be
\label{eq:footnoteScoeff}
S(\cO_6 [\cO_5] \cO) \sim \frac{\Gamma\left(\frac{\De_{6}+\tl\De_{5}-\De + \ell }{2}\right)}{\Gamma\left(\frac{\De_{6}+\De_{5}-\De + \ell }{2}\right)}\,, \qquad \qquad 
S([\cO_6] \tl\cO_5 \cO) \sim \frac{\Gamma\left(\frac{\tl\De_{6}+\tl\De_{5}-\De + \ell }{2}\right)}{\Gamma\left(\frac{\De_{6}+\tl\De_{5}-\De + \ell }{2}\right)}    \, .
\ee
Therefore, the product has poles at the double-traces $[\tl\cO_5\tl\cO_6]$ (and zeroes at the double-traces $[\cO_5\cO_6]$).
To determine such contributions in the same way as above, we should express $I^{3256}$ in terms of $I^{32\mathbf{S}[5]\mathbf{S}[6]}$ by inverting \eqref{eq:shadowtrans_exp} at \eqref{eq:gluing2} in the derivation above. We can follow the remaining steps  and use an identity for the MFT spectral function similar to \eqref{eq:mftcoeffs} (see \cite{Karateev:2018oml}). This gives the contribution from the double-traces of shadows $\cO_{[\tl 5\tl 6]}^{I}$ to be of the same form as in \eqref{eq:contribonepole}. Note that in the case of scalar MFT correlators, these poles in $Q(\De,\rho)$ are canceled by zeros in the MFT spectral function \eqref{eq:MFTspec} and hence we do not have these contributions from the double-traces of shadows.

\subsection{Discontinuities in the large $N$ expansion}
\label{sec:Nexpansion}

Equation \eqref{eq:gluing_nodisc} by itself is not very useful because of the $O(\frac{1}{N^2})$ error term. External double traces contribute already at $O(N^0)$ so that their contributions at $O(\frac{1}{N^2})$ are already not attainable by  \eqref{eq:gluing_nodisc}.
This problem is solved by taking the double discontinuity of  \eqref{eq:gluing_nodisc}, which will ensure that both sides of the equation are valid to $O(\frac{1}{N^4})$ for all double traces $[\cO_5 \cO_6]$, both external and internal.

The discontinuities are given by commutators in Lorentzian signature, hence we analytically continue the correlators to Lorentzian signature and take the difference of different operator orderings. Euclidean correlators can be continued to Wightman functions using the following prescription \cite{Hartman:2015lfa}
\be
\< \cO_1(t_1 , \vec{x}_1) \cO_2(t_2 , \vec{x}_2) \cdots \cO_n(t_n , \vec{x}_n) \> = \lim_{\epsilon_i \rightarrow 0}  \< \cO_1(t_1 - i\epsilon_1 , \vec{x}_1) \cdots \cO_n(t_n - i\epsilon_n , \vec{x}_n) \>  \,,   \label{eq:wightman}
\ee
with $\tau_i = i t_i$ where $\tau$ is Euclidean and $t$ Lorentzian time. The limits are taken assuming $\epsilon_1 > \epsilon_2 > \cdots > \epsilon_n $. 

Let us assume without loss of generality that $\cO_4$ is in the future of $\cO_1$,  that $\cO_2$ is in the future of $\cO_3$ and that all other pairs of operators are spacelike from each other. Now we apply the epsilon prescription to $\< \cO_1 \cO_2 \cO_3 \cO_4 \>$ with $\epsilon_4 >\epsilon_1$ and $\epsilon_2 >\epsilon_3$. The relative ordering of epsilons is unimportant for the spacelike separated pairs. This gives  the Lorentzian correlator $A^{\circlearrowleft} = \< \cO_2 \cO_3 \cO_4 \cO_1 \>$, which is equal to the time ordered correlator for the assumed kinematics. Similarly, we obtain $A^{\circlearrowright}= \< \cO_3 \cO_2 \cO_1 \cO_4 \>$ from the ordering $\epsilon_4 <\epsilon_1$, $\epsilon_2 <\epsilon_3$.  The Euclidean configurations $A_{\text{Euc}}$ correspond to the mixed orderings $\epsilon_4 >\epsilon_1$, $\epsilon_2 <\epsilon_3$ and $\epsilon_4 <\epsilon_1$, $\epsilon_2 >\epsilon_3$. We can then relate  the $\dDisc_t$  to these four configurations by
\be
\dDisc_t A(y_i) = A_{\text{Euc}}(y_i) - \frac{1}{2}\left(A^{\circlearrowleft}(y_i) + A^{\circlearrowright}(y_i)\right) = -\frac{1}{2}\< \left[\cO_2 ,\cO_3 \right] \left[\cO_4 ,\cO_1 \right]\>   \,.       \label{eq:dDisc_commutator} 
\ee
Using \eqref{eq:OPE_s} this gives  the conventional definition of the double discontinuity \cite{Caron_Huot_2017}
\bea
\dDisc_{t} A(y_i) &= T^{1234}(y_i)   \bigg[ \cos \big(\pi (a+b)\big) {\cal A}^{1234}(z,\bar{z}) -
\\
&-
\frac{1}{2}\left(
e^{i \pi(a+b)}  {\cal A}^{1234}(z,\bar{z}^\circlearrowleft)
+  e^{-i \pi(a+b)}  {\cal A}^{1234}(z,\bar{z}^\circlearrowright)
\right) \bigg] ,
\eea{eq:dDisc_conventional}
where $a=\De_{21}/2$ and $b=\De_{34}/2$. $\bar{z}^\circlearrowleft$ and $\bar{z}^\circlearrowright$ denote that $\zb$ is analytically continued by a full circle counter-clockwise and clockwise around $\zb=1$, respectively.\footnote{The relation between \eqref{eq:dDisc_commutator} and \eqref{eq:dDisc_conventional} can be obtained by assigning the phases $y_{ij}^2 \to y_{ij}^2 e^{\pm i \pi}$ to the timelike distances $y_{14}$ and $y_{23}$.}

The gluing of correlators on the right hand side in \eqref{eq:gluing_nodisc}, with the shadow integrals now written explicitly, is a sum of terms of the form
\be
\frac{1}{N_{\cO_5}N_{\cO_6}}\int dy_5 \, dy_6 \, dy_7 \, dy_8 \, \<\cO_2 \cO_3 \cO_6 \cO_5\> \, \<\tl \cO_5 \tl{\cO}_{7}^{\dag}\> \, \<\tl \cO_6 \tl{\cO}_{8}^{\dag}\> \, \<\cO_7 \cO_8 \cO_1 \cO_4 \>       \,. \label{eq:gluing_expan}
\ee 
Note that $\cO_5 = \cO_7$ and $\cO_6 = \cO_8$  but we have used the different labels to denote the insertion points. We can apply the same 
$\epsilon$-prescriptions on \eqref{eq:gluing_expan} while we hold $y_5 ,\, y_6 ,\, y_7 ,\, y_8$ to be Euclidean. Taking the same combinations as in \eqref{eq:dDisc_commutator} we arrive at
\be
\frac{1}{N_{\cO_5}N_{\cO_6}}\int dy_5 \, dy_6 \, dy_7 \, dy_8 \, \<\left[\cO_2 ,\cO_3 \right] \cO_6 \cO_5\> \, \<\tl \cO_5 \tl{\cO}_{7}^{\dag}\> \, \<\tl \cO_6 \tl{\cO}_{8}^{\dag}\> \, \<\cO_7 \cO_8 \left[\cO_4 ,\cO_1 \right] \>       \,. \label{eq:gluing_comm}
\ee
The commutators in \eqref{eq:gluing_comm} give discontinuities in the correlator as defined in \eqref{eq:discdefnorg}. 

We will now show that taking the $\dDisc$ of \eqref{eq:gluing1} ensures that the external double-traces $[\cO_{1}\cO_{4}]$ and $[\cO_{2}\cO_{3}]$ which usually appear at $O(N^0)$ are suppressed in $1/N$ so that they appear at the same order as other double trace operators. To this end, let us briefly discuss the $1/N$ expansion of correlators and associated CFT data.
The leading contribution is $A_\text{MFT}$, which is simply the disconnected correlator if the external operators are pairwise equal and is absent otherwise.
Because of this, the only operators that appear at $O(N^0)$ are the ones appearing in the disconnected correlator,
\bea
c_{ij [\cO_i \cO_j]_{n,\ell}}
&= c^\text{MFT}_{ij [\cO_i \cO_j]_{n,\ell}}
+ \frac{1}{N^2} \, c^{(1)}_{ij [\cO_i \cO_j]_{n,\ell}} + \cdots\,,\\
\De_{[\cO_i \cO_j]_{n,\ell}} &= \De_i + \De_j + 2n + \ell + \frac{1}{N^2} \,\gamma_{[\cO_i \cO_j]_{n,\ell}}+ \cdots \,.
\eea{eq:dt_expansion}
Other double-trace operators can only appear at higher orders in the OPE, therefore
\beq
\label{eq:otheroper}
c_{ij [\cO_k \cO_l]_{n,\ell}}
= \frac{1}{N^2} \,c^{(1)}_{ij [\cO_k \cO_l]_{n,\ell}} + \cdots\,, \qquad i,j \neq k,l\,.
\eeq
The analytic continuation of a $t$-channel conformal block to the Regge sheet is given by the following simple expression
\beq
g^{3214}_{\cO}\big(1-z,(1-\bar{z})e^{i \b}\big) = e^{i \b \frac{\tau_{\cO}}{2}} g^{3214}_{\cO}(1-z,1-\bar{z})\,.
\label{eq:block_monodromy}
\eeq
As a result, the action of the single and double discontinuities on the $t$-channel block expansion in \eqref{eq:OPE_t} is given by
\begin{align}
{} \Disc_{14}  A^{1ij4}(y_k)
&=  \sum\limits_{\cO} 2 i
\sin\left(\tfrac{\pi}{2} (\tau_\cO-\De_1-\De_4) \right)
c_{ij\cO}c_{14\cO} \, G^{ij14}_{\cO}(y_k)\,,\nonumber\\
{} \Disc_{23} A^{3ji2}(y_k)
&=  \sum\limits_{\cO} 2 i
\sin\left(\tfrac{\pi}{2} (\tau_\cO-\De_2-\De_3) \right)
c_{32\cO}c_{ij\cO}\,G^{32ij}_{\cO}(y_k)\,,\label{eq:dDisc_tchan}\\
 \dDisc_{t}  A(y_k)
&=  \sum\limits_{\cO} 2 
\sin\left(\tfrac{\pi}{2} (\tau_\cO-\De_1-\De_4) \right)
\sin\left(\tfrac{\pi}{2} (\tau_\cO-\De_2-\De_3) \right)
c_{32\cO}c_{14\cO}\,G^{3214}_{\cO}(y_k)\,.\nonumber
\end{align}
The sines in the expansions are responsible for suppressing the contribution of external double-traces. 
Therefore, using \eqref{eq:dDisc_tchan}, \eqref{eq:dt_expansion} and \eqref{eq:otheroper}, the leading contribution to the discontinuity of a correlator is $O(1/N^2)$
\begin{align}
{}&\Disc_{14}  A (y_i) = \frac{1}{N^2} \Disc_{14} A_\text{tree}(y_i) + O(1/N^4) =
\label{eq:leading_Disc} \\
&  \sum\limits_{\cO = [\cO_1 \cO_4]}  i \pi \frac{\gamma_\cO}{N^2} \,c^\text{MFT}_{14\cO}
c_{32\cO}G^{3214}_{\cO}
+\sum\limits_{\cO \neq [\cO_1 \cO_4]} 2 i
\sin\big(\tfrac{\pi}{2} (\tau_\cO-\De_1-\De_4) \big)
\frac{c^{(1)}_{14\cO}}{N^2} \,c_{32\cO}G^{3214}_{\cO}\,,
\nonumber
\end{align}
and similarly  the leading contribution to the double discontinuity is $O(1/N^4)$
\begin{align}
\label{eq:dDisc_order}
&\dDisc_{t} A(y_i) = \frac{1}{N^4} \,\dDisc_{t}  A_\text{1-loop}(y_i) + O(1/N^6)\,.
\end{align}
In particular, when acting with \eqref{eq:dDisc_y} on the left hand side of \eqref{eq:gluing1} we have
\bea
\Disc_{23} A^{3652}  
&=  \hspace{-9pt}\sum\limits_{\cO = [\cO_2 \cO_3]}\hspace{-9pt}  i \pi \frac{\gamma_\cO}{N^2}\, c^\text{MFT}_{32\cO}
c_{56\cO}\, G^{3256}_{\cO}
\\
&+ \hspace{-9pt} \sum\limits_{\cO \neq [\cO_2 \cO_3]} \hspace{-9pt}
2 i \sin\big(\tfrac{\pi}{2} (\tau_\cO-\De_2-\De_3) \big)\,
\frac{c^{(1)}_{32\cO}}{N^2} \,c_{56\cO}\,G^{3256}_{\cO}\,,
\\
\Disc_{14} A^{1\tl 5 \tl 6 4} 
&= \hspace{-9pt}\sum\limits_{\cO = [\cO_1 \cO_4]}\hspace{-9pt}  i \pi \frac{\gamma_\cO}{N^2} \,c_{\tl 6 \tl 5\cO} c^\text{MFT}_{14\cO}
G^{\tl 6 \tl 514}_{\cO}
\\
&+\hspace{-9pt}\sum\limits_{\cO \neq [\cO_1 \cO_4]}\hspace{-9pt} 2 i
\sin\big(\tfrac{\pi}{2} (\tau_\cO-\De_1-\De_4) \big)\,
c_{\tl 6 \tl 5\cO} \frac{c^{(1)}_{14\cO}}{N^2} \,G^{\tl 6 \tl 514}_{\cO}\,.
\eea{eq:leading_Disc_56} 
Since every term is these expansions already has an explicit factor of $1/N^2$, the only operators that can contribute at this order are the ones with $c_{56\cO} = O(N^0)$ or $c_{\tl 6\tl 5\cO} = O(N^0)$, which are the double-traces $\cO = [\cO_5 \cO_6]$ and $\cO = [\tl \cO_5 \tl \cO_6]$.
Applying the discontinuities to both sides of \eqref{eq:gluing_nodisc}
leaves us with one of our main results, the perturbative optical theorem for the contributions of double-trace operators to the 1-loop 
$\dDisc$ of the correlator, as stated in the introduction 
\bea
\dDisc_t \, A_{\rm 1-loop}(y_i)\Big|_\text{d.t.}  \!= -\frac{1}{2}
\sum\limits_{\substack{\mathcal{O}_5,\mathcal{O}_6\\ \in\  s.t.}}
 \!
 \int dy_5 dy_6 \, \Disc_{23}  A_{\rm tree}^{3652}(y_k) \, \bS_5 \bS_6 \Disc_{14} A_{\rm tree}^{1564}(y_k)  \Big|_{[\cO_5 \cO_6]}\,.
\eea{eq:cft_optical_theorem}
The integrals in this formula are over Euclidean space. 
It would be very interesting to derive a fully Lorentzian generalization of this formula.
In \cite{Kravchuk:2018htv} it was shown that there is a Lorentzian version of the shadow integral which computes the conformal block without the need to project out shadow operators. A Lorentzian version of \eqref{eq:cft_optical_theorem} might have this feature as well.
In section \ref{sec:ads} we will propose a Lorentzian fomula that is valid in the Regge limit.

To obtain the full double discontinuity, this generally has to be supplemented by the contributions of single trace operators, which already had the appropriate form in terms of three-point functions of single trace operators from the start, as shown in \eqref{eq:stcontribution}. The two types of contributions are analogous to double and single line cuts of scattering amplitudes in the S-matrix optical theorem.


\section{Review of flat space amplitudes}
\label{sec:flat_space}

In this section we review the Regge limit in $D$-dimensional flat space. Then we review 
the optical theorem in impact parameter space and explain how the notion of a one-loop vertex function arises. 
Not only does this serve as a hopefully more familiar introduction before discussing the same concepts in AdS, but it also provides the results we need later when we take the flat space limit of our AdS results and match them to known flat space expressions. To mimic the $1/N$ expansion in the CFT, it will be convenient to define an expansion in $G_N$ for the flat space scattering amplitude
\beq
A(s,t) =
\frac{2 G_N}{\pi} 
A_{\text{tree}}(s,t) +
 \left( \frac{2 G_N}{\pi}\right)^2 
A_{\text{1-loop}}(s,t)+ \dots \,,
\label{eq:amplitude_loop_expansion}
\eeq
and we use an identical expansion for the phase shifts $\delta(s,b)$ defined below.
\subsection{Regge limit and Regge theory}
\label{sec:regge_limit_flat}
We start by introducing the impact parameter representation, following \cite{Camanho:2014apa}.
Let us consider a tree-level scattering process with incoming momenta $k_1$ and $k_3$ that have large momenta along different lightcone directions.
For simplicity we assume for now that all external particles are massless scalars. This process is dominated by $t$-channel exchange diagrams of the type
	\beq
	\diagramEnvelope
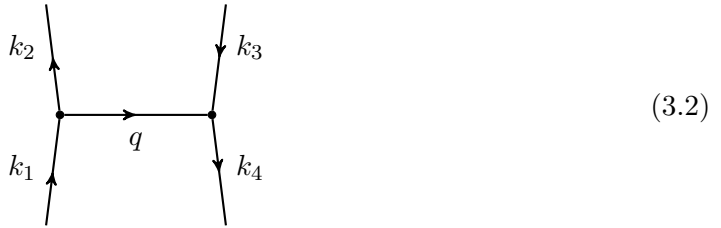
{\begin{tikzpicture}[anchor=base,baseline]
		\node (vertL) at (-1,0) [twopt] {};
		\node (vertR) at ( 1,0) [twopt] {};
		\node (opO1) at (-1.2,-1.6) [] {};
		\node (opO2) at (-1.2, 1.6) [] {};
		\node (opO3) at ( 1.2, 1.6) [] {};
		\node (opO4) at ( 1.2,-1.6) [] {};
		\node at (-1.5,-0.8) {$k_1$};
		\node at (-1.5,0.8) {$k_2$};
		\node at (1.5,0.8) {$k_3$};
		\node at (1.5,-0.8) {$k_4$};
		\node at (0,-0.4) {$q$};
		\draw [spinning] (opO1) -- (vertL);
		\draw [spinning] (vertL)-- (opO2);
		\draw [spinning] (vertL)-- (vertR);
		\draw [spinning] (opO3) -- (vertR);
		\draw [spinning] (vertR)-- (opO4);
	\end{tikzpicture}}
	\eeq
and the amplitude can be expressed in terms of the Mandelstam variables
	\beq
		s=-(k_1+k_3)^2\,, \qquad t= -(k_1-k_2)^2\,.
	\eeq
The amplitude now depends only on $s$ and the momentum exchange $q$ in the transverse directions, because we are considering the following configuration of null momenta, written in light-cone coordinates $p=(p^u,p^v,p_\perp)$
	\bea
		k_1^\mu &= \left( k^u, \frac{q^2}{4 k^u},\frac{q}{2} \right)\,, \qquad
		&& k_3^\mu = \left( \frac{q^2}{4k^v},k^v,-\frac{q}{2} \right)\,, \\
		k_2^\mu &= \left( k^u, \frac{q^2}{4 k^u},-\frac{q}{2} \right)\,, \qquad
		&& k_4^\mu = \left( \frac{q^2}{4 k^v},k^v,\frac{q}{2} \right)\,.
	\eea{eq:external_momenta} 
Notice that we reserve the letter $q$ for $(D-2)$-dimensional vectors in the transverse momentum space.
In the Regge limit $k^u \sim k^v \to \infty$ the  Mandelstams are given by 
	\beq
		s\approx k^u k^v \,, \qquad t=-q^2\,.
		\label{eq:impact_mandelstams}
	\eeq
The tensor structures in such amplitudes are fixed in terms of the on-shell three-point amplitudes.
For the case with two external scalars
and an intermediate particle (labeled $I$) with spin $J$
there is only one possible tensor structure for the three-point amplitudes given by
	\beq
		  \tilde{A}^{12I} = a_J (\e_I \cdot k_1)^{J}\,, \qquad \qquad
		  \tilde{A}^{34I} = a_J (\e_I \cdot k_3)^{J}
		\,,
	\eeq
where we encode traceless and transverse polarization tensors in terms of
vectors satisfying $\e_i^2 =  \e_i \cdot k_i = 0$.
We can then   write the four-point amplitude as
	\beq
		 A_{(m,J)} (s,t)=  
		\frac{\sum_{\e_I} \tilde{A}^{12I} \tilde{A}^{34I}}{t-m^2}\approx
		  \frac{a_J^2\, s^J}{t-m^2} \,,
		\label{eq:A_spinning_flat}
	\eeq
where  we used that for large $s$ the sum over polarizations is dominated by $\e_{Iu} k_1^u \sim k^u$ and $\e_{Iv} k_3^v \sim k^v$.
The $s^J$ behavior is naively problematic at high energies, especially if the spectrum contains particles of large spin, as is the case in string theory. However, boundedness of the amplitude in the Regge limit means there is a delicate balance between the infinitely many contributions in the sum over spin.\footnote{The couplings $a_J$  are dimensionful,  $[a_J]= 3- D/2 -J$, and accommodate for higher derivatives in the couplings to higher spin fields. In string theory the dimensionful scale is $\alpha'$ and the dimensionless couplings are all proportional to the string coupling $g_s$.} The precise framework to describe this phenomenon is Regge theory \cite{Regge:1959mz}, which was reviewed for flat space in \cite{Costa:2012cb,Caron-Huot:2020nem}.

In the Regge limit one has to consider the particle with the maximum spin $j(m^2)$ for each mass. The function $j(m^2)$ is called the leading Regge trajectory and the contributions from these particles get resummed into an effective particle with continuous spin $j(t)$.
In this work we will focus on the leading trajectory with   vacuum  quantum numbers known as the Pomeron.
At tree-level the amplitude for Pomeron exchange factorizes into three-point amplitudes involving a Pomeron and the universal Pomeron propagator $\b(t)$. For example,  in the case of 4-dilaton scattering in type IIB strings we have
\beq
 A_\text{tree} (s,t)  =   \frac{8 }{\alpha'} A^{12P} \beta(t) A^{34P} \left(\frac{\alpha's}{4}\right)^{j(t)}\,,
\label{eq:A_tree_regge}
\eeq
with
	\beq
		\beta(t) =   2 \pi^2 \frac{\G(- \frac{\a'}{4} t)}{\G(1+\frac{\a'}{4} t)}\,
		e^{- \frac{i \pi \a'}{4} t} \,.
		\label{eq:pomeron_propagator}
	\eeq
$A^{ijP}$ are the three-point amplitudes between the external scalars and the Pomeron with the 
 $s$-dependence factored out and normalized such that in the case of 4-dilaton scattering   $A^{ijP}=1$. 
 This is convenient since later on, when we consider more general string states with spin, 
the string three-point amplitudes defined this way will contain just tensor structures.
Diagrammatically we can write (\ref{eq:A_tree_regge}) as
\beq
	\diagramEnvelope{\begin{tikzpicture}[anchor=base,baseline]
		\node (vertL) at (-0.8,0.1) [twopt] {};
		\node (vertR) at ( 0.8,0.1) [twopt] {};
		\node (opO1) at (-1.2,-1.2) [] {$1$};
		\node (opO2) at (-1.2, 1.2) [] {$2$};
		\node (opO3) at ( 1.2, 1.2) [] {$3$};
		\node (opO4) at ( 1.2,-1.2) [] {$4$};
		\node at (0,-0.4) {$P$};
		\draw [spinning no arrow] (opO1) -- (vertL);
		\draw [spinning no arrow] (vertL)-- (opO2);
		\draw [finite] (vertL)-- (vertR);
		\draw [spinning no arrow] (opO3) -- (vertR);
		\draw [spinning no arrow] (vertR)-- (opO4);
	\end{tikzpicture}}
=
	\diagramEnvelope{\begin{tikzpicture}[anchor=base,baseline]
		\node (vertL) at (-0.8,0.1) [twopt] {};
		\node (vertR) at ( 0.5,0.1) [] {};
		\node (opO1) at (-1.2,-1.2) [] {$1$};
		\node (opO2) at (-1.2, 1.2) [] {$2$};
		\node at (0,-0.4) {$P$};
		\draw [spinning no arrow] (opO1) -- (vertL);
		\draw [spinning no arrow] (vertL)-- (opO2);
		\draw [finite] (vertL)-- (vertR);
	\end{tikzpicture}}
\times
	\diagramEnvelope{\begin{tikzpicture}[anchor=base,baseline]
		\node (vertL) at (-0.5,0.1) [] {};
		\node (vertR) at ( 0.8,0.1) [twopt] {};
		\node (opO3) at ( 1.2, 1.2) [] {$3$};
		\node (opO4) at ( 1.2,-1.2) [] {$4$};
		\node at (0,-0.4) {$P$};
		\draw [finite] (vertL)-- (vertR);
		\draw [spinning no arrow] (opO3) -- (vertR);
		\draw [spinning no arrow] (vertR)-- (opO4);
	\end{tikzpicture}}
\times \frac{2 G_N}{\pi}  \frac{8 }{\alpha'}
\,\b(t) \left(\frac{\alpha's}{4}\right)^{j(t)}
\,.
\label{eq:Pomeron_factorization}
\eeq
Amplitudes involving a Pomeron can be computed in string theory using the Pomeron vertex operator \cite{Ademollo:1989ag,Ademollo:1990sd,Brower:2006ea}. The factorization into three-point functions and a Pomeron propagator holds for general external string states \cite{Brower:2006ea,DAppollonio:2013mgj}.

\subsection{Optical theorem and impact parameter space}
\label{sec:impact_optical_theorem_flat}

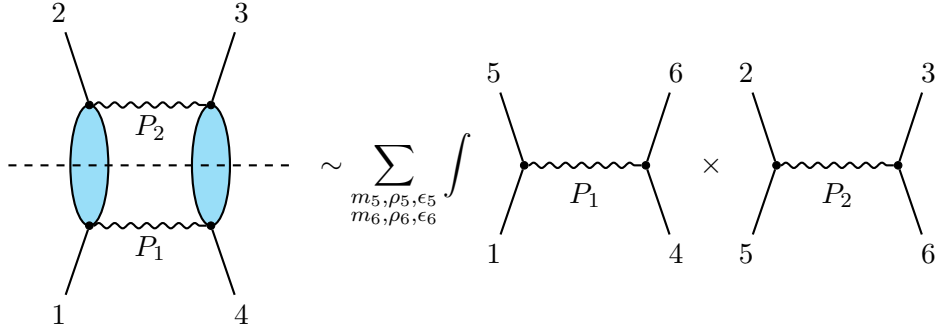
\begin{figure}
\begin{equation*}
	\diagramEnvelope{\begin{tikzpicture}[anchor=base,baseline]
        \draw [thick,fill=cyan,fill opacity=0.4] (-0.8,0.1) ellipse (0.25 and 0.8);
        \draw [thick,fill=cyan,fill opacity=0.4] (0.8,0.1) ellipse (0.25 and 0.8);
		\node (vert1) at (-.8,-.7) [twopt] {};
		\node (vert2) at (-.8, .9) [twopt] {};
		\node (vert3) at ( .8, .9) [twopt] {};
		\node (vert4) at ( .8,-.7) [twopt] {};
        \node (cut1) at (-2,.1) [] {};
        \node (cut2) at (2,.1) [] {};
		\node (opO1) at (-1.2,-2) [] {$1$};
		\node (opO2) at (-1.2, 2) [] {$2$};
		\node (opO3) at ( 1.2, 2) [] {$3$};
		\node (opO4) at ( 1.2,-2) [] {$4$};
		\node at (0,-1.1) {$P_1$};
		\node at (0,0.5) {$P_2$};
		\draw [spinning no arrow] (opO1) -- (vert1);
		\draw [spinning no arrow] (vert2) -- (opO2);
		\draw [spinning no arrow] (opO3) -- (vert3);
		\draw [spinning no arrow] (vert4) -- (opO4);
		\draw [finite] (vert1) -- (vert4);
		\draw [finite] (vert2) -- (vert3);
		\draw [scalar no arrow] (cut1) -- (cut2);
	\end{tikzpicture}}
\sim
\sum\limits_{\substack{m_5,\rho_5,\e_5\\m_6,\rho_6,\e_6}} \int
	\diagramEnvelope{\begin{tikzpicture}[anchor=base,baseline]
		\node (vertL) at (-0.8,0.1) [twopt] {};
		\node (vertR) at ( 0.8,0.1) [twopt] {};
		\node (opO1) at (-1.2,-1.2) [] {$1$};
		\node (opO2) at (-1.2, 1.2) [] {$5$};
		\node (opO3) at ( 1.2, 1.2) [] {$6$};
		\node (opO4) at ( 1.2,-1.2) [] {$4$};
		\node at (0,-0.4) {$P_1$};
		\draw [spinning no arrow] (opO1) -- (vertL);
		\draw [spinning no arrow] (vertL)-- (opO2);
		\draw [finite] (vertL)-- (vertR);
		\draw [spinning no arrow] (opO3) -- (vertR);
		\draw [spinning no arrow] (vertR)-- (opO4);
	\end{tikzpicture}}
\times
	\diagramEnvelope{\begin{tikzpicture}[anchor=base,baseline]
		\node (vertL) at (-0.8,0.1) [twopt] {};
		\node (vertR) at ( 0.8,0.1) [twopt] {};
		\node (opO1) at (-1.2,-1.2) [] {$5$};
		\node (opO2) at (-1.2, 1.2) [] {$2$};
		\node (opO3) at ( 1.2, 1.2) [] {$3$};
		\node (opO4) at ( 1.2,-1.2) [] {$6$};
		\node at (0,-0.4) {$P_2$};
		\draw [spinning no arrow] (opO1) -- (vertL);
		\draw [spinning no arrow] (vertL)-- (opO2);
		\draw [finite] (vertL)-- (vertR);
		\draw [spinning no arrow] (opO3) -- (vertR);
		\draw [spinning no arrow] (vertR)-- (opO4);
	\end{tikzpicture}}
\end{equation*}
\caption{Optical theorem in the Regge limit in terms of Feynman diagrams. The tree-level correlators are dominated by $s$-channel Pomeron exchange. The ellipses on the l.h.s.\ indicate that all string excitations are taken into account.}
\label{fig:optical_theorem_flat}	
\end{figure}

Next we consider the expression for the two-line cut of the one-loop amplitude in the impact parameter representation, which will be given in terms of the tree-level pieces we have discussed so far.
The two-line cut receives a contribution from two-Pomeron exchange, which is the leading term in the Regge limit of the one-loop amplitude.
Consider the following configuration of momenta
	\beq
	\diagramEnvelope{\begin{tikzpicture}[anchor=base,baseline]
		\node (vert1) at (-1,-1) [twopt] {};
		\node (vert2) at (-1, 1) [twopt] {};
		\node (vert3) at ( 1, 1) [twopt] {};
		\node (vert4) at ( 1,-1) [twopt] {};
		\node (opO1) at (-1.2,-2.6) [] {};
		\node (opO2) at (-1.2, 2.6) [] {};
		\node (opO3) at ( 1.2, 2.6) [] {};
		\node (opO4) at ( 1.2,-2.6) [] {};
		\node at (-1.5,-1.8) {$k_1$};
		\node at (-1.5,1.8) {$k_2$};
		\node at (1.5,1.8) {$k_3$};
		\node at (1.5,-1.8) {$k_4$};
		\node at (0,-1.4) {$l_1$};
		\node at (0,0.6) {$l_2$};
		\node at (-1.3,0) {$k_5$};
		\node at (1.3,0) {$k_6$};
		\draw [spinning] (opO1) -- (vert1);
		\draw [spinning] (vert2) -- (opO2);
		\draw [spinning] (opO3) -- (vert3);
		\draw [spinning] (vert4) -- (opO4);
		\draw [spinning] (vert1) -- (vert2);
		\draw [spinning] (vert1) -- (vert4);
		\draw [spinning] (vert2) -- (vert3);
		\draw [spinning] (vert3) -- (vert4);
	\end{tikzpicture}}\,.
\label{fig:momenta_1loop}
	\eeq
The external momenta are again in the configuration \eqref{eq:external_momenta} with Mandelstams \eqref{eq:impact_mandelstams}.
The optical theorem tells us to cut the internal lines of the diagram, putting the corresponding legs on-shell. This implies the following equation for the discontinuity of the amplitude
	\beq
	2 \Im A_{\text{1-loop}}(s,q) = \! \sum\limits_{\substack{m_5,\rho_5,\e_5\\m_6,\rho_6,\e_6}}
	\int \! \frac{d l_1}{(2 \pi)^{D}} \, 2\pi i \de(k_5^2 + m_5^2) \, 2\pi i \de(k_6^2 + m_6^2) 
	A_\text{tree}^{3652} (s,l_2)^*
	A_\text{tree}^{1564} (s,l_1)\,,
	\label{eq:optical_theorem_start}
	\eeq
where one sums over all possible particles 5 and 6 with masses $m$, in Little group representations $\rho$ and with 
polarization tensors $\e$. The sums over polarizations can be evaluated using completeness relations.
In order to remove the delta functions we express $k_5$ and $k_6$ in terms of $l_1$ and the external momenta \eqref{eq:external_momenta}.
Then we write the loop momentum as $l_1^\mu = (l^u,l^v,q_1)$
and use the delta-functions to fix the forward components of the loop momentum $l_u$ and $l_v$ to
\bea
l^u = \frac{m_6^2 +q_1^2 +q \cdot q_1}{ k^v}\,,
		    \qquad\qquad
l^v = -\frac{m_5^2 +q_1^2 -q \cdot q_1}{ k^u}\,,
\eea{eq:onshell_solutions_flat}
leaving only the transverse integration over $q_1$.
We arrive at the equation
	\beq
		\Im A_{\text{1-loop}}(s,q) = \sum\limits_{\substack{m_5,\rho_5,\e_5\\m_6,\rho_6,\e_6}}
		\int \frac{dq_1 dq_2}{(2 \pi)^{D-2}}  \, \frac{\de(q-q_1-q_2)}{4 s}
		A_\text{tree}^{3652} (s,q_2)^*
		A_\text{tree}^{1564} (s,q_1)\,,
		\label{eq:optical_theorem_flat}
	\eeq
where we introduced the transverse momentum $q_2=q-q_1$  to write the expression 
in a more symmetrical way. 	
Using that the tree-level amplitudes are given 
in the Regge limit by Pomeron exchange, we can write (\ref{eq:optical_theorem_flat}) diagrammatically as in figure \ref{fig:optical_theorem_flat}.
The optical theorem can be simplified even further by transforming it to impact parameter space. To this end the amplitude is expressed in terms of the impact parameter $b$, which is a vector in the transverse impact  parameter space $\mathbb{R}^{D-2}$, using the following transformation
	\beq
		\de (s,b) = \frac{1}{2s} \int \frac{d q}{(2\pi)^{D-2}} \,e^{i  q\cdot b}  A(s,t)\,.
		\label{eq:impact_flat}
	\eeq
We can use this definition together with \eqref{eq:optical_theorem_flat} to compute
	\bea
		\Im  \de_{\text{1-loop}}(s,b)  
		={}& \frac{1}{2} \sum\limits_{\substack{m_5,\rho_5,\e_5\\m_6,\rho_6,\e_6}}
		\de_\text{tree}^{3652}(s,-b)^* \,\de_\text{tree}^{1564}(s,b)\,.
	\eea{eq:impact_optical_theorem}
We conclude that the impact parameter representation absorbs the remaining phase space integrals in the optical theorem, resulting in a purely multiplicative formula. In fact, in the case where the particles on the left and right of the diagram do not change (i.e.\ 1,5,2 and 3,6,4 are identical particles), such a statement holds to all-loops, leading to exponentiation of the tree-level phase shift, which is the basis for the famous eikonal approximation.

\subsection{Vertex function}
\label{sec:vertex_function_flat}

Another notion we will use is that of the vertex function, which arises when combining
the optical theorem \eqref{eq:optical_theorem_flat} with the factorization of the tree-level amplitudes \eqref{eq:A_tree_regge} into three-point amplitudes. By combining the two results one sees that the sums over particles and their polarizations factorize into separate sums for particles 5 and 6, which we call the vertex function $V$
\beq
V(q_1,q_2) \equiv \sum\limits_{m_5,\rho_5,\e_5} A^{15P_1}(q_1) A^{25P_2}(q_2)\,.
\label{eq:V_flat_def}
\eeq
Moreover, such a sum over representations and polarizations for each mass is given by tree-level unitarity as the residue of the four-point amplitudes with two external Pomerons
\beq
\underset{k_5^2=-m_5^2}\Res A^{12 P_1 P_2}(k_5,q_1,q_2) = \sum\limits_{\rho_5,\e_5} A^{15P_1}(q_1) A^{25P_2}(q_2)\,.
\label{eq:residue_generic}
\eeq
In terms of diagrams this reads
\beq
V(q_1,q_2) \equiv \sum\limits_{m_5,\rho_5,\e_5}
	\diagramEnvelope{\begin{tikzpicture}[anchor=base,baseline]
		\node (vertU) at (0,0.9) [] {$5$};
		\node (vertD) at (0,-0.3) [twopt] {};
		\node (opO1) at (-1.2,-0.8) [] {$1$};
		\node (opOP1) at ( 1.2,-0.8) [] {$P_1$};
		\draw [spinning no arrow] (opO1) -- (vertD);
		\draw [spinning no arrow] (vertU)-- (vertD);
		\draw [finite] (vertD)-- (opOP1);
	\end{tikzpicture}}
\times
	\diagramEnvelope{\begin{tikzpicture}[anchor=base,baseline]
		\node (vertU) at (0,0.4) [twopt] {};
		\node (vertD) at (0,-0.8) [] {$5$};
		\node (opO2) at (-1.2, 0.9) [] {$2$};
		\node (opOP2) at ( 1.2, 0.9) [] {$P_2$};
		\draw [spinning no arrow] (vertU)-- (opO2);
		\draw [spinning no arrow] (vertU)-- (vertD);
		\draw [finite] (opOP2) -- (vertU);
	\end{tikzpicture}}
=\sum\limits_{m_5} \underset{k_5^2=-m_5^2}\Res
	\diagramEnvelope{\begin{tikzpicture}[anchor=base,baseline]
		\node (vertU) at (0,0.7) [twopt] {};
		\node (vertD) at (0,-0.7) [twopt] {};
		\node (opO1) at (-1.2,-1.2) [] {$1$};
		\node (opO2) at (-1.2, 1.2) [] {$2$};
		\node (opOP2) at ( 1.2, 1.2) [] {$P_2$};
		\node (opOP1) at ( 1.2,-1.2) [] {$P_1$};
		\node at (-.3,0) {$5$};
		\draw [spinning no arrow] (opO1) -- (vertD);
		\draw [spinning no arrow] (vertU)-- (opO2);
		\draw [spinning no arrow] (vertU)-- (vertD);
		\draw [finite] (opOP2) -- (vertU);
		\draw [finite] (vertD)-- (opOP1);
	\end{tikzpicture}}\,.
\label{eq:V_diagrams}
\eeq
The vertex function combines all information about the exchanges of possibly spinning particles 5 and 6 into a single scalar function. In terms of the vertex function the optical theorem \eqref{eq:optical_theorem_flat} in the Regge limit becomes
\begin{align}
\Im A_{\text{1-loop}}(s,q) =  -\frac{1}{4s} 
&\int  \frac{dq_1  dq_2}{(2\pi)^{D-2}} \, \de(q-q_1-q_2)
\label{eq:optical_theorem_V_flat}
\\
&\left( \frac{8}{\alpha'}\right)^2 \b(t_1)^* \b(t_2) V(q_1,q_2)^2 \left(\frac{\alpha's}{4}\right)^{j(t_1) + j(t_2)}\,,
\nonumber
\end{align}
where $t_i = -q_i^2$.

\subsection{Spinning three-point amplitudes}
\label{sec:3pt_amplitudes}

Since it will be important later to compare tensor structures in AdS and flat space, we will provide here some more details on the tensor structures of the three-point amplitudes that appear in the unitarity cut of the four-point amplitude $A^{12 P_1 P_2}$ discussed above.
The external momentum $k_1$ and the exchanged momentum $l_1$, with light-cone components given in the Regge limit by
(\ref{eq:onshell_solutions_flat}), fix the momentum $k_5=k_1-l_1$ as shown in the figure below. We may, however, change frame such that 
$k_5$ has no transverse momentum \cite{DAppollonio:2013mgj}. Such change of frame does not alter the fact that the light-cone components of $l_1$ are 
subleading. The same applies to $l_2$. Thus in the Regge limit we can safely write
\beq
	\diagramEnvelope{\begin{tikzpicture}[anchor=base,baseline]
		\node (vertU) at (0,0.7) [twopt] {};
		\node (vertD) at (0,-0.7) [twopt] {};
		\node (opO1) at (-1.2,-1.2) [] {$k_1$};
		\node (opO2) at (-1.2, 1.2) [] {$k_2$};
		\node (opOP2) at ( 1.2, 1.2) [] {$l_2$};
		\node (opOP1) at ( 1.2,-1.2) [] {$l_1$};
		\node at (-.3,0) {$k_5$};
		\draw [spinning] (opO1) -- (vertD);
		\draw [spinning] (vertU)-- (opO2);
		\draw [spinning] (vertD)-- (vertU);
		\draw [spinning] (vertU) -- (opOP2);
		\draw [spinning] (vertD)-- (opOP1);
	\end{tikzpicture}}
\qquad \quad
\begin{aligned}
k_5 &\approx  \left( k_5^u ,\frac{m_5^2}{k_5^u},0 \right)\,, &\\
l_2 &\approx  (0,0, q_2)\,, \qquad& k_2 = k_5 - l_2\,, \\
l_1 &\approx  (0,0, q_1)\,, \qquad& k_1 = k_5 + l_1 \,.
\end{aligned}
\eeq
We focus on the three-point amplitude $A^{15P_1}(q_1)$, which is related to the four-point amplitude via the tree-level unitarity \eqref{eq:residue_generic}.
In this relation we have a sum over a basis of possible polarizations $\e_5$,
which can be evaluated using completeness relations, e.g.\ for massive bosons \cite{Boels:2014dka}
	\beq
		\sum\limits_{\e_5} \e_{5}^{\mu_1 \ldots \mu_{|\rho|}}
		\e_{5}^{\nu_1 \ldots \nu_{|\rho|}} 
		= P^{\mu_1}_{5 \g_1} \ldots P^{\mu_{\rho}}_{5 \g_{\rho}}
		\pi_\rho^{\g_1 \ldots \g_{|\rho|}; \,\s_1 \ldots \s_{|\rho|}}
		P^{\nu_1}_{5 \s_1} \ldots P^{\nu_{\rho}}_{5 \s_{\rho}} \,,
		\label{eq:completeness_relation}
	\eeq
where $\pi_\rho$ is the projector to the irreducible $SO(D-1)$ representation $\rho$ and
	\beq
		P_{5 \nu}^{\mu} = \de^{\mu}_{\nu} - \frac{k_{5}^{\mu} k_{5 \nu}}{k_5^2}\,,
	\eeq
is a projector to the space transverse to $k_5$.
We will always absorb the projectors $P_{5 \nu}^{\mu}$ into the three-point amplitudes, i.e.\ consider amplitudes in a transverse configuration. That means that the indices corresponding to particle 5 have to be constructed from the projected momenta of the other particles, which are identical
\beq
P_{5 \nu}^{\mu} l_{1\mu} = P_{5 \nu}^{\mu} k_{1\mu}\,.
\label{eq:P5l}
\eeq
Apart from that, massive particles can also have a longitudinal polarization $v$ which satisfies
\beq
v \cdot k_5 = 0 \,, \qquad v^2 = 1\,,
\eeq
and is given in this frame explicitly by
\beq
v_\mu =\frac{1}{m_5} \left(  k_5^u, - \frac{m_5^2}{k_5^u},  0 \right)\,.
\eeq
For the case that particle 1 is a scalar, we can then construct $A^{15P}$ in terms of the following manifestly transverse tensor structures
	\beq
		A^{15P}_{m_5,\rho_5,\bmu} = \sum\limits_{k=0}^{|\rho_5|} 
		a_{m_5,\rho_5}^{k}(t_1) \,
i^{|\rho_5|} \sqrt{\a'}^{|\rho_5|-k}
v_{\mu_1} \ldots v_{\mu_k} q_{1\mu_{k+1}} \ldots q_{1\mu_{|\rho_5|}}\,,
		\label{eq:3pt_spinning}
	\eeq
where we introduced boldface indices $\bmu$ as multi-indices that stand for the $|\rho|$ indices for the irrep $\rho$.	
By abuse of language we defined the vector $q_1\equiv(0,0,q_1)$, since $q_1$ is transverse.
If particle 1 carries spin as well, as will be the case for the gravitons considered later on, we construct the polarization tensors out of the vector $\xi_1=(\xi_1^u, \xi_1^v, \e_1)$. In this case, again defining
$\epsilon_1\equiv (0,0,\epsilon_1)$, the amplitudes take the following form in the Regge limit
	\bea
		A^{15P}_{m_5,\rho_5,\bmu} =  \sum\limits_{n=0}^{\ell_1}  \sum\limits_{k=0}^{|\rho_5|-n}
		&a_{m_5,\rho_5}^{k,n}(t_1) \,
i^{|\rho_5|} 
\sqrt{\a'}^{|\rho_5|+\ell_1-2n-k}
(\e_1\cdot q_1)^{\ell_1 - n}
\\
& \e_{1\mu_1} \ldots \e_{1\mu_n}  v_{\mu_{n+1}} \ldots v_{\mu_{n+k}} q_{1\mu_{n+k+1}} \ldots q_{1\mu_{|\rho_5|}}\,,
	\eea{eq:3pt_spinning_more}
as can be checked by comparing with the explicit amplitudes computed in \cite{DAppollonio:2013mgj}. These choices for the tensor structures are particularly convenient since $q_1 \cdot v = \e_1 \cdot v = 0$. Contact with the momentum frame used in the previous subsections is made by identifying the Lorentz invariant $A^{12 P_1 P_2}$.


\section{AdS impact parameter space}
\label{sec:ads}

Our goal in this section is to compute the Regge limit of a scalar four-point function in a perturbative large $N$ CFT at one-loop and finite
$\Delta_{\text{gap}}$. At finite $\Delta_{\text{gap}}$ we have to consider the $t$-channel exchange of all possible double-trace operators and also single-trace operators, which are respectively dual to tidal excitations of the external scattering states and  to long-string creation in the string theory context. It was shown in \cite{Meltzer:2019pyl} that the exchange of single-trace operators dual to the long-string creation effects is subleading in the Regge limit. Therefore we only need to consider the exchange of double-trace operators.
This is where the new perturbative CFT optical theorem \eqref{eq:cft_optical_theorem_intro} takes a central role, as it allows us to compute the contributions of double-trace operators to the correlator starting from the corresponding tree-level correlators. The contribution of the leading Regge trajectory to the scalar tree-level correlators is known to leading order in the Regge limit \cite{Cornalba:2007fs,Costa:2012cb}. 

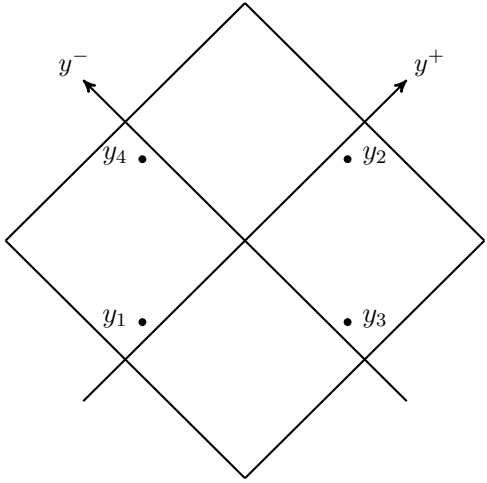
\begin{figure}
	\begin{center}
	\begin{tikzpicture}[anchor=base,baseline,scale=0.9, transform shape]
		\node (vertLT) at (-2.5, 2.5) [] {};
		\node (vertRT) at ( 2.5, 2.5) [] {};
		\node (vertLB) at (-2.5,-2.5) [] {};
		\node (vertRB) at ( 2.5,-2.5) [] {};
		\node (opO1) at (-1.2,-1.6) [] {};
		\node (opO2) at (-1.2, 1.6) [] {};
		\node (opO3) at ( 1.2,-1.6) [] {};
		\node (opO4) at ( 1.2, 1.6) [] {};
		\node at (-1.9,-1.2) {\large $y_1$};
		\node at (-1.9,1.2) {\large $y_4$};
		\node at (1.9,-1.2) {\large $y_3$};
		\node at (1.9,1.2) {\large $y_2$};
		\node at (-1.5,-1.2) [twopt] {};
		\node at (-1.5,1.2) [twopt] {};
		\node at (1.5,-1.2) [twopt] {};
		\node at (1.5,1.2) [twopt] {};
		\node at (2.7,2.5) {$y^+$};
		\node at (-2.5,2.5) {$y^-$};
		\draw [axis] (vertLB)-- (vertRT);
		\draw [axis] (vertRB)-- (vertLT);
		\draw [thick, black] (0,3.5) -- (-3.5,0);
		\draw [thick, black] (0,-3.5) -- (-3.5,0);
		\draw [thick, black] (3.5,0) -- (0,3.5);
		\draw [thick, black] (3.5,0) -- (0,-3.5);
	\end{tikzpicture}
	\end{center}
	\caption{Kinematics in the central Poincar\'{e} patch with coordinates $y_i$. Time is on the vertical axis, transverse directions are suppressed.}
	\label{fig:regge_figure}	
	\end{figure}

In this section we will therefore study \eqref{eq:cft_optical_theorem_intro} in the Regge limit, and this time we expand the tree-level correlators in the $s$-channel. In the Regge limit the four external points are 
in Lorentzian kinematics as depicted in figure \ref{fig:regge_figure}.
In this configuration all distances between points are spacelike except for $y_{14}^2, y_{23}^2 < 0$. 
The Regge limit is reached by sending the four-points to infinity along the light cones
\beq
y_{1}^{+} \rightarrow-\infty, \quad y_{2}^{+} \rightarrow+\infty, \quad y_{3}^{-} \rightarrow-\infty, \quad y_{4}^{-} \rightarrow+\infty\,.
\eeq
The Regge limit can be directly applied to the left hand side of \eqref{eq:cft_optical_theorem_intro}. The terms on the Regge sheets $A^{\circlearrowleft}(y_i)$ and $A^{\circlearrowright}(y_i)$ are dominant over the Euclidean terms in this limit. However, we cannot apply the Regge limit directly to the right hand side of \eqref{eq:cft_optical_theorem_intro} as the shadow integrals range over Euclidean configurations. Hence we will apply Wick rotations on the points $y_{5},\,y_{6},\,y_{7},\,y_{8}$ to obtain a gluing of the discontinuities of Lorentzian correlators. We will assume that in the Regge limit the dominant contribution to the gluing formula comes from the domain where the individual tree-level correlators are in the Regge limit themselves. We do not provide a proof of this assumption but we justify it in section \ref{sec:reggeandimpact}.

When each four-point function in \eqref{eq:gluing_expan} is in the Regge limit, the points $y_{5},\,y_{6},\,y_{7},\,y_{8}$ are placed in the same positions as $y_{1},\,y_{4},\,y_{2},\,y_{3}$ in fig.\ \ref{fig:regge_figure}, respectively. Thus $y_7$ is in the future of $y_8$ and this pair is spacelike from $y_1 ,\, y_4 ,\, y_5 ,\, y_6$. Similarly, $y_6$ is in the future of $y_5$ and is spacelike from $y_2 ,\, y_3 ,\, y_7 ,\, y_8$. 
For the chosen kinematics
we put the pair $y_5 ,\, y_6$ in anti-time order  using the epsilon prescription of \eqref{eq:wightman} with $\epsilon_5 > \epsilon_6$, and the pair $y_7 ,\, y_8$ in time order using $\epsilon_7 > \epsilon_8$. Applying this on \eqref{eq:cft_optical_theorem} gives the following formula for the Regge limit of the double discontinuity
\begin{align}
\dDisc_t A_{\text{1-loop}}(y_k) \Big|_{\text{d.t.}} = -\frac{1}{2}
\sum\limits_{\cO_5, \cO_6}  \frac{1}{N_{\cO_5}N_{\cO_6}}\int & dy_5 dy_6 dy_7 dy_8\, \< [\cO_2 ,\cO_3 ] \cO_5 \cO_6 \>_{\text{tree}} \< \tl \cO_5 \tl{\cO}_{7}^{\dag}\>  \nonumber\\
&\<\tl \cO_6 \tl{\cO}_{8}^{\dag}\>
		 \< \cO_7 \cO_8 [\cO_4 , \cO_1]\>_{\text{tree}} \Big|_{\left[\cO_5\cO_6\right]} \,.
\label{eq:regge_gluing_form}
\end{align}
The relative ordering between $\epsilon_5 ,\epsilon_7$ and $\epsilon_6 ,\epsilon_8$ is irrelevant as the pairs, appearing in the two-point functions on the right hand side of \eqref{eq:regge_gluing_form}, are spacelike separated in the Regge configuration.

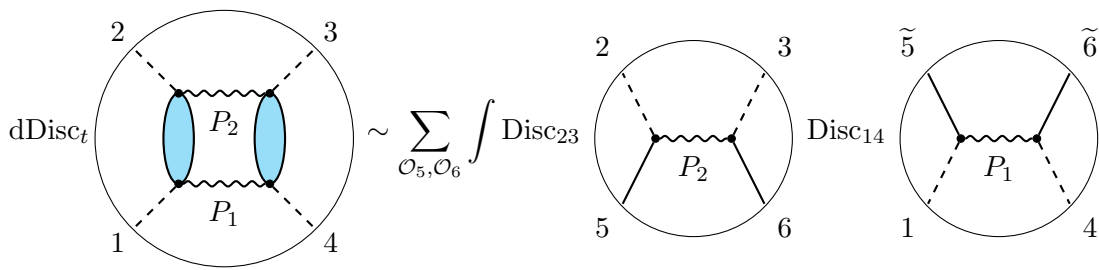
\begin{figure}
	\begin{center}
\begin{equation*}
\dDisc_t \,
\diagramEnvelope{\begin{tikzpicture}[anchor=base,baseline]
    \draw [thick,fill=cyan,fill opacity=0.4] (-0.6,0) ellipse (0.2 and 0.6);
    \draw [thick,fill=cyan,fill opacity=0.4] (0.6,0) ellipse (0.2 and 0.6);
	\node (vertLU) at (-0.6,0.6) [twopt] {};
	\node (vertRU) at ( 0.6,0.6) [twopt] {};
	\node (vertLD) at (-0.6,-0.6) [twopt] {};
	\node (vertRD) at ( 0.6,-0.6) [twopt] {};
	\node (opO1) at (-1.3,-1.3) [] {};
	\node (opO2) at (-1.3, 1.3) [] {};
	\node (opO3) at ( 1.3,-1.3) [] {};
	\node (opO4) at ( 1.3, 1.3) [] {};
	\node at (-1.4,-1.5) {$1$};
	\node at (-1.4, 1.3) [] {$2$};
	\node at ( 1.4,-1.5) [] {$4$};
	\node at ( 1.4, 1.3) [] {$3$};
	\node at (0,0.5) [below] {$P_2$};	
	\node at (0,-0.7) [below] {$P_1$};
	\draw [scalar no arrow] (vertLD)-- (opO1);
	\draw [scalar no arrow] (vertLU)-- (opO2);
	\draw [finite] (vertLU)-- (vertRU);
	\draw [finite] (vertLD)-- (vertRD);
	\draw [scalar no arrow] (vertRD)-- (opO3);
	\draw [scalar no arrow] (vertRU)-- (opO4);
    \draw (0,0) circle (1.7);
\end{tikzpicture}} \,
\sim \sum\limits_{\cO_5, \cO_6} \int 
\Disc_{23}
\diagramEnvelope{\begin{tikzpicture}[anchor=base,baseline]
	\node (vertL) at (-0.5,0) [twopt] {};
	\node (vertR) at ( 0.5,0) [twopt] {};
	\node (opO1) at (-1.,-1.) [] {};
	\node (opO2) at (-1., 1.) [] {};
	\node (opO3) at ( 1.,-1.) [] {};
	\node (opO4) at ( 1., 1.) [] {};
	\node at (-1.2,-1.3) {$5$};
	\node at (-1.2, 1.1) [] {$2$};
	\node at ( 1.2,-1.3) [] {$6$};
	\node at ( 1.2, 1.1) [] {$3$};
	\node at (0,-0.1) [below] {$P_2$};	
	\draw [spinning no arrow] (vertL)-- (opO1);
	\draw [scalar no arrow] (vertL)-- (opO2);
	\draw [finite] (vertL)-- (vertR);
	\draw [spinning no arrow] (vertR)-- (opO3);
	\draw [scalar no arrow] (vertR)-- (opO4);
    \draw (0,0) circle (1.3);
\end{tikzpicture}}
\Disc_{14}
\diagramEnvelope{\begin{tikzpicture}[anchor=base,baseline]
	\node (vertL) at (-0.5,0) [twopt] {};
	\node (vertR) at ( 0.5,0) [twopt] {};
	\node (opO1) at (-1.,-1.) [] {};
	\node (opO2) at (-1., 1.) [] {};
	\node (opO3) at ( 1.,-1.) [] {};
	\node (opO4) at ( 1., 1.) [] {};
	\node at (-1.2,-1.3) {$1$};
	\node at (-1.2, 1.1) [] {$\tl 5$};
	\node at ( 1.2,-1.3) [] {$4$};
	\node at ( 1.2, 1.1) [] {$\tl 6$};
	\node at (0,-0.1) [below] {$P_1$};	
	\draw [scalar no arrow] (vertL)-- (opO1);
	\draw [spinning no arrow] (vertL)-- (opO2);
	\draw [finite] (vertL)-- (vertR);
	\draw [scalar no arrow] (vertR)-- (opO3);
	\draw [spinning no arrow] (vertR)-- (opO4);
    \draw (0,0) circle (1.3);
\end{tikzpicture}}
\end{equation*}
\end{center}
\caption{Optical theorem in the Regge limit in terms of Witten diagrams. The tree-level correlators are dominated by $s$-channel Pomeron exchange. The external operators are scalars, while $\cO_5$ and $\cO_6$ are summed over all states that couple to the 
external scalars and the Pomeron (tidal excitations). The ellipses on the l.h.s.\ indicate that all string excitations are taken into account.}
\label{fig:optical_theorem_strings}	
\end{figure}

In this section we define the discontinuities as the commutators inserted into the fully Lorentzian correlators
\bea
\Disc_{14} A^{1874}(y_i) &\coloneqq \< \cO_7 \cO_8 [\cO_4 , \cO_1]\> = 
A^{1874}{}^\circlearrowleft(y_i) - A^{1874}_{\text{Euc}}(y_i)\,, \\
\Disc_{23}  A^{3652}(y_i) &\coloneqq \< [\cO_2 ,\cO_3 ] \cO_5 \cO_6 \> = 
A^{3652}_{\text{Euc}}(y_i) - A^{3652}{}^\circlearrowright(y_i)  \,.
\eea{eq:new_singledisc}
This definition differs slightly from the one in \eqref{eq:discdefnorg}. Stripping out the appropriate pre-factors from \eqref{eq:new_singledisc}, one can check that these single discontinuities can be equivalently defined as
\bea
\Disc_{14}  \mathcal{A}^{1234}(z,\bar{z}) &\coloneqq
e^{i \pi(a+b)} \mathcal{A}^{1234}(z,\bar{z}^\circlearrowleft) - e^{-i \pi(a+b)} \mathcal{A}^{1234}(z,\bar{z})\,, \\
\Disc_{23}  \mathcal{A}^{1234}(z,\bar{z}) &\coloneqq
\mathcal{A}^{1234}(z,\bar{z}) -  \mathcal{A}^{1234}(z,\bar{z}^\circlearrowright)\,, \\
\Disc_{23}  \mathcal{A}^{3412}(z,\bar{z}) & =
e^{-i \pi(a+b)} \mathcal{A}^{3412}(z,\bar{z}) - e^{i \pi(a+b)} \mathcal{A}^{3412}(z,\bar{z}^\circlearrowright)\,.
\eea{eq:Disc_conventional}
Starting from the discontinuity defined in \eqref{eq:discdefnorg}, these expressions result from continuing another half circle in $\zb$, so that the different terms are either evaluated at the original position or continued a full circle around 1. The extra phase comes from the additional Wick rotations.
The final  result matches the definition of the discontinuity in \cite{Caron-Huot:2020nem}.
Note that $\zb$ is continued an extra half circle in opposite directions for the first two lines in \eqref{eq:Disc_conventional}, so that with these definitions the relation to the $\dDisc$ in \eqref{eq:dDisc_y} remains valid.\footnote{For $t$-channel blocks, the new definitions for the discontinuities in \eqref{eq:Disc_conventional} are related to the old definition in \eqref{eq:discdefnorg} by a phase, for example, for $\Disc_{14}$ the relative phase is $e^{i\pi\tau_{\cO}/2}$.}
Therefore the optical theorem in the Regge limit can still be expressed as
\bea
\dDisc_t  A_{\rm 1-loop}(y_i)\Big|_\text{d.t.}  \!= -\frac{1}{2}
\sum\limits_{\substack{\mathcal{O}_5,\mathcal{O}_6\\ \in\  s.t.}}
 \!
 \int dy_5 dy_6 \, \Disc_{23}  A_{\rm tree}^{3652}(y_k) \, \bS_5 \bS_6 \Disc_{14} A_{\rm tree}^{1564}(y_k)  \Big|_{[\cO_5 \cO_6]}\,,
\eea{eq:cft_optical_theorem_s} 
with the discontinuities as defined in the first and third lines of \eqref{eq:Disc_conventional}, and the gluing and shadow integrals now ranging over Minkowski space. This formula is also depicted in figure \ref{fig:optical_theorem_strings} in terms of Witten diagrams.

We should also note that for real $z, \zb$, $\Disc_{23}$ in the third line of \eqref{eq:Disc_conventional} is related to $\Disc_{14}$ in the first line by
\beq
\Disc_{23}  \mathcal{A}^{3412}(z,\bar{z}) = - \left. \left( \Disc_{14} \mathcal{A}^{1234}(z,\bar{z}) \right)^* \right|_{(a,b)\to (-b,-a)}\,, \quad 0<z, \zb < 1\,.
\label{eq:disc_relation}
\eeq
Applied to the correlators appearing in \eqref{eq:cft_optical_theorem_s} this reads
\beq
\Disc_{23}  \mathcal{A}^{3652}(z,\zb) = - \left. \left( \Disc_{14} \mathcal{A}^{1564} (z,\zb)\right)^* \right|_{1564\to 3652}\,.
\label{eq:disc_relation56}
\eeq
To benefit from this useful relation, we will always strip out a pre-factor such that we obtain the correlator $\mathcal{A}^{3652}(z,\zb)$ on the right hand side of \eqref{eq:cft_optical_theorem_s}. Otherwise we would have to use the second line of \eqref{eq:Disc_conventional} for $\Disc_{23}$.
Finally, in the Regge limit the analytically continued correlators are dominant over the Euclidean contributions so that we have
\bea
\Disc_{14} \mathcal{A}^{1234}(z,\bar{z}) &\approx e^{i \pi(a+b)} \mathcal{A}^{1234}(z,\bar{z}^\circlearrowleft)\,,\\
\Disc_{23}  \mathcal{A}^{3412}(z,\bar{z}) &\approx - e^{i \pi(a+b)} \mathcal{A}^{3412}(z,\bar{z}^\circlearrowright)\,,\\
\dDisc_t \, \mathcal{A}^{1234}(z,\bar{z}) &\approx - \frac{1}{2} \left(
e^{i \pi(a+b)} \mathcal{A}^{1234}(z,\bar{z}^\circlearrowleft)
+ e^{-i \pi(a+b)} \mathcal{A}^{1234}(z,\bar{z}^\circlearrowright)
\right)\,,
\eea{eq:discs_Regge}
where the $\approx$ sign means we took the Regge limit.
In order to account for the tidal excitations, the operators $\cO_5$ and $\cO_6$ can carry spin, in which case their indices are contracted with the ones of $\tl \cO_5$ and $\tl \cO_6$ and sums over tensor structures are implied. In subsections \ref{sec:reggeandimpact},
 \ref{sec:impact} and 
\ref{sec:sdisc} below we will mostly suppress the aspect of spinning correlators. We will come back to this issue in subsection \ref{sec:vertex_function}.

\subsection{Regge limit}
\label{sec:reggeandimpact}

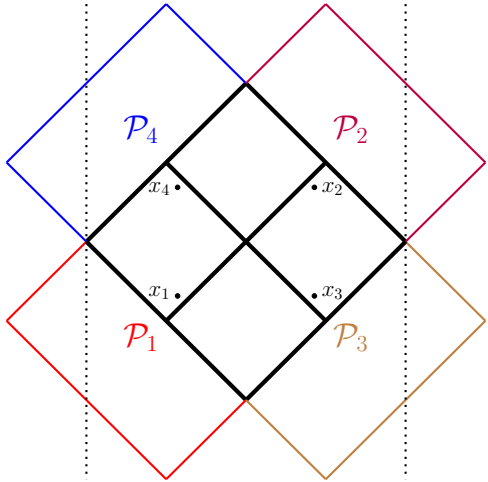
\begin{figure}
	\begin{center}
	\begin{tikzpicture}[anchor=base,baseline,scale=0.6, transform shape]
		\node (opO1) at (-1.2,-1.6) [] {};
		\node (opO2) at (-1.2, 1.6) [] {};
		\node (opO3) at ( 1.2,-1.6) [] {};
		\node (opO4) at ( 1.2, 1.6) [] {};
		\node at (-1.9,-1.2) {\Large $x_1$};
		\node at (-1.9,1.1) {\Large $x_4$};
		\node at (1.9,-1.2) {\Large $x_3$};
		\node at (1.9,1.1) {\Large $x_2$};
		\node at (-1.5,-1.2) [twopt] {};
		\node at (-1.5,1.2) [twopt] {};
		\node at (1.5,-1.2) [twopt] {};
		\node at (1.5,1.2) [twopt] {};
		\draw [ultra thick, black] (-1.75,-1.75)-- (1.75,1.75);
		\draw [ultra thick, black] (1.75,-1.75)-- (-1.75,1.75);
		\draw [ultra thick, black] (0,3.5) -- (-3.5,0);
		\draw [ultra thick, black] (0,-3.5) -- (-3.5,0);
		\draw [ultra thick, black] (3.5,0) -- (0,3.5);
		\draw [ultra thick, black] (3.5,0) -- (0,-3.5);
		\draw [thick, blue] (0,3.5) -- (-1.75,5.25);
		\draw [thick, blue] (-5.25,1.75) -- (-1.75,5.25);
		\draw [thick, blue] (-5.25,1.75) -- (-3.5,0);
		\node at (-2.3,2.3) [blue] {\huge $\mathcal{P}_4$};
		\draw [thick, red] (-3.5,0) -- (-5.25,-1.75);
		\draw [thick, red] (-5.25,-1.75) -- (-1.75,-5.25);
		\draw [thick, red] (-1.75,-5.25) -- (0,-3.5);
		\node at (-2.3,-2.3) [red] {\huge $\mathcal{P}_1$};
		\draw [thick, brown] (0,-3.5) -- (1.75,-5.25);
		\draw [thick, brown] (1.75,-5.25) -- (5.25,-1.75);
		\draw [thick, brown] (5.25,-1.75) -- (3.5,0);
		\node at (2.3,-2.3) [brown] {\huge $\mathcal{P}_3$};
		\draw [thick, purple] (3.5,0) -- (5.25,1.75);
		\draw [thick, purple] (5.25,1.75) -- (1.75,5.25);
		\draw [thick, purple] (1.75,5.25) -- (0,3.5);
		\node at (2.3,2.3) [purple] {\huge $\mathcal{P}_2$};
		\draw [thick, black, dotted] (-3.5, 5.25) -- (-3.5,-5.25);
		\draw [thick, black, dotted] (3.5, 5.25) -- (3.5,-5.25);
	\end{tikzpicture}
	\end{center}
	\caption{The external operators at coordinates $x_i$ in their respective Poincar\'e patches $\mathcal{P}_i$. The black dotted lines are identified when the Poincar\'e patches are wrapped on the boundary of the global AdS cylinder.}
	\label{fig:Poincare_patches}	
	\end{figure}

To obtain the impact parameter representation, we first change the coordinate system placing each point on a different Poincar\'{e} patch as shown in figure \ref{fig:Poincare_patches}. We use the following coordinate transformations
\beq
\begin{aligned}
x_{i} &=\left(x_{i}^{+}, x_{i}^{-}, x_{i \perp}\right)=-\frac{1}{y_{i}^{+}}\left(1, y_{i}^{2}, y_{i \perp}\right), & & i=1,2,5,7\,, \\
x_{i} &=\left(x_{i}^{+}, x_{i}^{-}, x_{i \perp}\right)=-\frac{1}{y_{i}^{-}}\left(1, y_{i}^{2}, y_{i \perp}\right), & & i=3,4,6,8\,.
\end{aligned}
\label{eq:xofy}
\eeq
In the new $x_i$ coordinates, the Regge limit corresponds to placing the four external points at the origin of their respective Poincar\'{e} patches,
\beq
x_1, x_2, x_3, x_4 \to 0\,.
\label{eq:regge_limit_x}
\eeq
However, $x_5$ to $x_8$ are integrated over in the CFT optical theorem.

Conformal correlators transform covariantly under the transformation \eqref{eq:xofy}.
In the scalar case we have
\beq
A\left(y_{i}\right)=(-y_{1}^{+})^{-\De_1} (y_{2}^{+})^{-\De_2} (-y_{3}^{-})^{-\De_3} (y_{4}^{-})^{-\Delta_4}  A\left(x_{i}\right)\,.
\label{eq:Ax_to_Ay}
\eeq
In the spinning case, one must additionally account for the Jacobian matrix $\partial y^a / \partial x^m$.\footnote{For external spinning operators, the conformal transformations have a non-trivial rotation matrix $\partial y^a / \partial x^m$. Conformal covariance of the correlators gives, in the representative example of two vectors and two scalars \cite{Cornalba:2009ax}, $A^{a b}\left(y_{i}\right)=\left(-y_{1}^{+} y_{2}^{+}\right)^{-1-\Delta_V}\left(-y_{3}^{-} y_{4}^{-}\right)^{-\Delta_S} \frac{\partial y_{1}^{a}}{\partial x_{1}^{m}} \frac{\partial y_{2}^{b}}{\partial x_{2}^{n}} A^{m n}\left(x_{i}\right)$. These matrices ensure that the inversion tensors are correctly mapped from $y_i$ to $x_i$ variables, preserving their form.}

Next we use conformal symmetry to express the correlator in terms of two vectors.
This is similar to expressing the correlator in terms of two scalar cross-ratios, with the difference that here we fix two, instead of the customary three, positions using translations and special conformal transformations to express the correlator in terms of the remaining two position vectors.
 We can follow \cite{Cornalba:2009ax} and use a translation to send $x_1$ to 0
which, due to the different transformations in \eqref{eq:xofy} (see also \cite{Kulaxizi_2018}), will act as a special conformal transformation on the Poincar\'{e} patches for $x_3$ and $x_4$,
\beq
x_1 \to 0\,, \quad x_2 \to x_2 - x_1\,, \quad
x_{3,4} \to \frac{x_{3,4} - x_{3,4}^2 x_1}{1-2 x_{3,4} \cdot x_1 + x_{3,4}^2 x_1^2}\,.
\eeq
Next we implement a translation on the $x_3$ and $x_4$ patches (acting as special conformal transformation on $x_{1,2}$) to also map $x_4$ to 0 in its own patch and find that the correlator as a function of the Poincar\'e patch coordinates $A(x_i)$, as defined in \eqref{eq:Ax_to_Ay}, can always be expressed as
\beq
A(x_1, x_2, x_3, x_4) \approx A(0, -x, \xb/\xb^2, 0) \equiv A(x,\xb)\,,
\label{eq:Axxbar}
\eeq
with
\beq
x\approx x_1 - x_2 \,, \qquad \xb \approx x_3 - x_4\,,
\eeq
in the Regge limit \eqref{eq:regge_limit_x}.

It is further convenient to implement the coordinate change using embedding space coordinates $P^M \in \mathbb{R}^{2,d}$
\beq
P^M = \big(P^+,P^-,P^m\big)\,, \qquad P\cdot P = -P^+ P^- + \eta_{m n} P^m P^n\,.
\eeq
These are related to the coordinates $y^m \in \mathbb{R}^{1,d-1}$ of physical Minkowski space by \cite{Cornalba:2009ax}
\beq
P^M = \big(y^+, y^-, 1, y^2,y_\perp\big) \quad \Rightarrow \quad
P_{ij} \equiv -2 P_i \cdot P_j = (y_i-y_j)^2\,,
\eeq
and to the coordinates $x_i$ by
\bea
P_1^M &= -y_1^+ \left(-1,-x_1^2,x_1^m\right)\,, & & 
P_2^M =  y_2^+ \left(-1,-x_2^2,x_2^m\right)\,,\\
P_3^M &= -y_3^- \left(-x_3^2,-1, x_3^m\right), & &
P_4^M =  y_4^- \left(-x_4^2,-1, x_4^m\right)\,.
\eea{eq:xiPatches}
One can easily show that the cross-ratios \eqref{eq:crossratios} are given in terms of $x$ and $\xb$ as
\beq
z\zb = x^2 \xb^2\,, \qquad (1-z)(1-\zb) = 1 + x^2 \xb^2 + 2 x \cdot \xb\,,
\label{eq:xxb_derivation}
\eeq
and the kinematic prefactor \eqref{eq:T} becomes
\beq
T^{1234} = \frac{(-y_{1}^{+})^{-\De_1} (y_{2}^{+})^{-\De_2} (-y_{3}^{-})^{-\De_3} (y_{4}^{-})^{-\Delta_4}}{x^{\De_1 + \De_2} \xb^{\De_3 + \De_4}}\,.
\eeq
When combining \eqref{eq:OPE_s} and \eqref{eq:Ax_to_Ay}, the numerator of the last expression cancels the Jacobian prefactor in \eqref{eq:Ax_to_Ay} to give,
\beq
A\left(x_{i}\right) = \frac{A^{1234}(z, \zb)}{x^{\De_1 + \De_2} \xb^{\De_3 + \De_4}}\,.
\label{eq:strip_A}
\eeq
 If we now study the correlator $A^{3652}(x_i)$, a priori we have to take into account that only $x_2$ and $x_3$ are affected by the Regge limit. However, we will assume that the integration will be dominated by the region where the integration points are also boosted. Using the embedding space coordinates
\bea
P_5^M =  -y_5^+ \left(-1,-x_5^2,x_5^m\right)\,, \qquad
P_6^M &= y_6^- \left(-x_6^2,-1, x_6^m\right)\,,
\eea{eq:x56Patches}
we find
\beq
x' \approx x_5 - x_2 \,, \qquad \xb' \approx x_3 - x_6\,.
\eeq
Where the primed variables are meant to emphasize that the points $1,4$ are replaced by $5,6$ in this correlator as compared to \eqref{eq:Axxbar}.
Performing these steps for all the correlators in \eqref{eq:cft_optical_theorem_s} we find,
\bea
\dDisc_t A_{\text{1-loop}}(x_{12},x_{34}) = -\frac{1}{2}
\sum\limits_{\cO_5, \cO_6}  \int dx_5 dx_6 \, \Disc_{23} A^{3652}_{\text{tree}}(x_{36},x_{52})   \\
									\bS_5 \bS_6	\Disc_{14} A^{1564}_{\text{tree}}(x_{15},x_{64}) \Big|_{\left[\cO_5\cO_6\right]}\,,
\eea{eq:cft_optical_theorem_x}
where we stop explicitly mentioning that we are dealing with only the contribution of double-trace operators as single-trace contributions are subleading in the limit considered.
Let us stress that in order to write the correlators on the right hand side in terms of two differences, we assumed that each of the individual tree-level correlators are in the Regge limit themselves. 
The easiest way to justify this is in Fourier space using the impact parameter transform defined below.
Each tree-level position space correlator is dominated by a power $\sigma^{1-j(\nu)}$ in the Regge limit, which maps to a power of the AdS center of mass energy $S^{j(\nu)-1}$ in impact parameter space. Since the optical theorem is multiplicative in impact parameter space, subleading Regge trajectories or kinematical corrections from the conformal block at finite boost get mapped to smaller powers of $S$, and therefore do not contribute to the leading behavior.
The eikonal approximation in AdS \cite{Cornalba:2007zb} gives additional intuition for this, since it means that even in AdS, the particles remain essentially undeflected, scattering forward each time they exchange a Pomeron. Furthermore, we will show in section \ref{sec:Appendix_tchannel} that this configuration reproduces the behavior at one-loop derived in \cite{Meltzer:2019pyl}.

\subsection{Impact parameter space}
\label{sec:impact}

Let us now consider the two-point functions $\<\tl \cO_5 \tl \cO_7\>$ and $\<\tl \cO_8 \tl \cO_6\>$ for the shadow transforms in \eqref{eq:cft_optical_theorem_x}.
In the Regge configuration, $x_5$ is the patch of $x_1$, $x_6$ is the patch of $x_4$, $x_7$ is the patch of $x_2$ and $x_8$ is the patch of $x_3$.
As explained in \cite{Cornalba:2007zb,Kravchuk:2018htv}, the two-point function between two coordinates on adjacent Poincar\'e patches has an additional phase factor $e^{i \pi \Delta}$ and this phase can be accounted for by switching from the $i\e$ prescription of a Feynman propagator to that of a Wightman propagator (see \cite{Cornalba:2007zb}).
The normalization of the shadow transform in \eqref{eq:shadowtransform} and \eqref{eq:shadownormali} is obtained from the Fourier transform of a two-point function as the shadow transform acts multiplicatively in Fourier space (see section 3.2 of \cite{Karateev:2018oml}). The normalization in \eqref{eq:shadownormali} is obtained from the Fourier transform of a Euclidean two-point function, which matches the one of a Lorentzian two-point function with Feynman $i\e$ prescription. The Wightman propagator in momentum space however has support only on the future lightcone and the coefficient of the Fourier transform is different (see Appendix B of \cite{Cornalba:2007zb} and 
section 2.1 of \cite{Gillioz:2018mto})
\bea
\int dx \frac{e^{-2iq\cdot x}}{\left[-(x^{0}-i\epsilon)^{2} + \vec{x}^{2}\right]^{\De}} =\cM_{\cO} 
 \,\Theta(q^{0}) \,\Theta(-q^{2})\left(-q^{2}\right)^{\De-\frac{d}{2}}\,,
\eea{eq:WightmanFT}
with 
\begin{equation}
 \cM_{\cO}  = \frac{2\pi^{\frac{d}{2}+1}}{\Gamma(\De)\Gamma\big(\De - \frac{d}{2} + 1\big)} \,.
\end{equation}
Consequently we change the normalization $N_\cO$ of the shadow transform in \eqref{eq:shadowtransform} to (for scalar operators)
\be
\mathcal{N}_\cO = \cM_\cO \cM_{\tl \cO}  \,. 
\label{eq:new_normali}
\ee

Next we define, following \cite{Cornalba:2007fs}, the impact parameter representation as the Fourier transform of the discontinuity of the correlator in the two remaining vectors%
\footnote{The AdS impact parameter representation was previously defined in \cite{Cornalba:2007fs,Cornalba:2006xm,Cornalba:2008qf} as the Fourier transform of the correlator on the second sheet $A(z,\zb^\circlearrowleft)$. Since this contribution dominates in the Regge limit, it is indistinguishable from the discontinuity of the correlator in this limit.
However, in the $t$-channel the two notions are clearly different and it was necessary to take the discontinuity to derive \eqref{eq:cft_optical_theorem_intro}.
In the next section we will see that the discontinuity is the better choice also in the $s$-channel.}
\beq
\Disc_{14} A^{1jk4}(x_{1j},x_{k4}) = \int dp \, d\bar{p} \, e^{-2i p \cdot x_{1j}-2i \bar{p} \cdot x_{k4}} B^{1jk4}(p,\bar{p})\,,
\label{eq:BtoA}
\eeq
where the function $B(p,\bar{p})$ has support only on the future Milne wedge of $p$ and $\bar{p}$.
Using \eqref{eq:disc_relation56} on \eqref{eq:BtoA} we get the Fourier transform of $\Disc_{23}$.
\be
\label{eq:BtoA23}
\Disc_{23} A^{3kj2}(x_{3k},x_{j2}) = -\int dp \, d\bar{p} \, e^{-2i p \cdot x_{3k}-2i \bar{p} \cdot x_{j2}} B^{3kj2}(-p,-\bar{p})^{*}\,.
\ee
The causal relations and thus the $i\epsilon$ prescription in \eqref{eq:BtoA23} are opposite to those in \eqref{eq:BtoA} and the complex conjugation prescribed in \eqref{eq:disc_relation56} compensates for that.
Inserting \eqref{eq:BtoA} into \eqref{eq:cft_optical_theorem_x} and using that $A^{1\tilde{5}\tilde{6}4}_{\text{tree}}$ is a double shadow transform of $A^{1784}_{\text{tree}}$ we obtain, upon using \eqref{eq:disc_relation56} for the $\Disc_{23}$,
\bea
{}& \dDisc_t A_{\text{1-loop}}(x_{12},x_{34})  = \frac{1}{2}
\sum\limits_{\cO_5, \cO_6}  \int dx_{5}\, dx_{6}\, dx_{7}\, dx_{8}\; 	\int dp \, d\pb \,  dp' \, d\bar{p}' 	  \\ 
	& \qquad 
	\times  e^{-2i (p' \cdot x_{36}+ \bar{p}' \cdot x_{52}+ p \cdot x_{17}+ \bar{p} \cdot x_{84})} B^{3652}_\text{tree} (-p',-\pb')^* B^{1564}_\text{tree}(p,\pb)  \\
&\qquad	
\times \frac{T^{(\rho_5)}(x_{75})}{\mathcal{N}_{\cO_{5}}\left[-(x^{0}_{75}-i\epsilon)^{2} + \vec{x}_{75}^{2}\right]^{d-\De_5}}  \frac{T^{(\rho_6)}(x_{68})}{\mathcal{N}_{\cO_{6}}\left[-(x^{0}_{68}-i\epsilon)^{2} + \vec{x}_{68}^{2}\right]^{d-\De_6}} \Bigg|_{\left[\cO_5\cO_6\right]} \,.
\eea{eq:ftgluing1}
$T^{(\rho)}(x_{ij})$ is the tensor structure for the two-point function of an operator with $SO(d)$ quantum number $\rho$. For example it is the familiar inversion tensor $\eta^{\mu\nu}-2\frac{x^{\mu}x^{\nu}}{x^{2}}$ for spin 1 operators. Note that we have $B^{1564}_\text{tree}$ instead of $B^{1784}_\text{tree}$, as the superscripts now only indicate the dimensions of the corresponding operators and $\Delta_5 = \De_7$, $\De_6 = \De_8$.

We can now express the two-point functions from the shadow transforms in Fourier space by inverting \eqref{eq:WightmanFT}
\be
\label{eq:shadow2ptFT}
\frac{T^{(\rho)}(x)}{\left[-(x^{0}-i\epsilon)^{2} + \vec{x}^{2}\right]^{d-\De}}  ={} \frac{\mathcal{M}_{\tl\cO}}{\pi^{d}}\int\limits_{M} dq\, e^{-2iq\cdot x}\, \widehat{T}^{(\rho)}(q) \, (-q^{2})^{\frac{d}{2}-\De} \,.
\ee
The Fourier space integral is over the future Milne wedge $M$ as the Fourier transform has support only on this domain.
$\widehat{T}^{(\rho)}(q)$ is the tensor structure of the two-point function in Fourier space,
which has been discussed for example in \cite{Gillioz:2018mto}.
It is a tensor composed of $q^\mu$ and $\eta^{\mu \nu}$ that can be factorized into a product of new tensors $t^{(\rho)}(q)$ as follows
\beq
\widehat{T}^{(\rho)}(q)^{\mu_1 \ldots \mu_{|\rho|}}_{\nu_1 \ldots \nu_{|\rho|}}
= t^{(\rho)}(q)^{\mu_1 \ldots \mu_{|\rho|}}_{\sigma_1 \ldots \sigma_{|\rho|}}
t^{(\rho)}(q)^{\sigma_1 \ldots \sigma_{|\rho|}}_{\nu_1 \ldots \nu_{|\rho|}}\,.
\label{eq:T_factorization}
\eeq
Using \eqref{eq:shadow2ptFT} in \eqref{eq:ftgluing1} we end up with four position integrals over $x_{5},x_{6},x_{7},x_{8}$ and six  integrals over $q,\bar{q}$ (from the four-point functions) and $p,\bar{p},p',\bar{p}'$. The four position integrals give  four Dirac delta functions with which we can eliminate the $q,\bar{q},p',\bar{p}'$ integrals to obtain
\bea
\dDisc_t A_{\text{1-loop}}(x_{12},x_{34})		= \ &
\frac{\pi^{2d}}{2}
\sum\limits_{\cO_5, \cO_6}  \frac{1}{\cM_{\cO_5}\cM_{\cO_6}}   \int dp \, d\pb\,
		e^{-2i (p \cdot x_{12}+ \pb \cdot x_{34})}      \\
	&	\frac{B^{3652}_\text{tree}(-\pb,-p)^* \;\,  \widehat{T}^{(\rho_5)}(p) \,  \widehat{T}^{(\rho_6)}(\bar{p}) \;\, B^{1564}_\text{tree}(p,\pb)}{(-p^{2})^{\De_5-\frac{d}{2}}(-\bar{p}^{2})^{\De_6-\frac{d}{2}}} \Bigg|_{\left[\cO_5\cO_6\right]} \,,
\eea{eq:circ_impact1}
with an implicit index contraction between $B^{1564}_\text{tree}$ and $B^{3652}_\text{tree}$ and the tensor structures $\widehat{T}^{(\rho)}$.
At this point we use \eqref{eq:T_factorization} and absorb the $ t^{(\rho)}(q)$ tensors into the definition of the phase shifts $B_\text{tree}$. This means that one needs to take it into account if one wants to relate tensor structures of CFT correlators and phase shifts, but we will not need to do such a basis change explicitly in this work.
Using \eqref{eq:discs_Regge} we see that the double discontinuity corresponds to the quantity
$- \Re B(p,\bar{p})$
in impact parameter space.
Thus we find the following gluing formula for the impact parameter representation, which is purely multiplicative,
\beq
- \Re B_{\text{1-loop}}(p,\bar{p}) = \frac{\pi^{2d}}{2}
\sum\limits_{\cO_5, \cO_6}  \frac{1}{\cM_{\cO_5}\cM_{\cO_6}} \, 
\frac{B^{3652}_\text{tree}(-\pb,-p)^* \,  B^{1564}_\text{tree}(p,\pb)}{(-p^{2})^{\De_5-\frac{d}{2}}(-\bar{p}^{2})^{\De_6-\frac{d}{2}}}  \Big|_{\left[\cO_5\cO_6\right]} \,.
\label{eq:gluing_B}
\eeq
Let us consider the case when $\cO_5 = \cO_1$ and $\cO_6 = \cO_3$. In this case, it is useful to strip out a scale factor similar to that in \eqref{eq:strip_A} from the impact parameter representation
\beq
B^{jjkk}(p, \bar{p})=\frac{\cM_{\cO_j} \cM_{\cO_k} \; \mathcal{B}^{jjkk}(p,\pb)}{\left(-p^{2}\right)^{\frac{d}{2}-\Delta_{j}}\left(-\bar{p}^{2}\right)^{\frac{d}{2}-\Delta_{k}}}     \, .
\label{eq:strip_pairwise}      
\eeq
Using \eqref{eq:WightmanFT} one sees that with this choice of normalization the impact parameter representation of the MFT correlator is $\mathcal{B}^{jjkk}_{\text{MFT}}=1$,
which is necessary for the eikonalization of the  phase shift in AdS gravity.
Inspired by this fact we choose the normalization
\beq
B^{ijkl}(p, \bar{p})=\frac{\sqrt{\cM_{\cO_i} \cM_{\cO_j} \cM_{\cO_k} \cM_{\cO_l}} \; \mathcal{B}^{ijkl}(p,\pb)}{\left(-p^{2}\right)^{\frac{d-\Delta_{i}-\Delta_{j}}{2}}\left(-\bar{p}^{2}\right)^{\frac{d-\Delta_{k}-\Delta_{l}}{2}}}     \, ,
\label{eq:strip}      
\eeq
for the general case.
This gives  the following compact form for the optical theorem in impact parameter space
\bea
- \Re \cB_{\text{1-loop}}(p,\bar{p}) =  \frac{1}{2}
\sum\limits_{\cO_5, \cO_6}    \; \cB^{3652}_\text{tree}(-\pb,-p)^* \; \cB^{1564}_\text{tree}(p,\pb)  \; \Bigg|_{\left[\cO_5\cO_6\right]}  \,.
\eea{eq:gluing_stripped}
In \cite{Meltzer:2019pyl}, the Regge limit of a one-loop four-point function of scalars was studied in the large $\lambda$ regime with $S\gg \lambda\gg 1$. The contribution of tidal excitations to the correlator is suppressed in this regime. It corresponds to just one term in the sum on the right hand side of \eqref{eq:gluing_stripped} i.e.\ with $\cO_5 = \cO_1$ and $\cO_6 = \cO_3$. We show in section \ref{sec:Dav_match} that this term from our formula \eqref{eq:gluing_stripped} reproduces the result from \cite{Meltzer:2019pyl} in the large $\lambda$ or equivalently the large $\De_{\text{gap}}^{2}$ limit. We do not need to discard any shadow double-trace contributions for this match. This motivates us to assume that the only non-zero contributions to the gluing of tree-level correlators in the Regge limit is from the physical double-traces $[\cO_5 \cO_6]$, and we will drop the explicit projections henceforth. This is compatible with the intuition that there is no need to project out shadow operators in a Lorentzian CFT optical theorem.

\subsection{$s$-channel discontinuities in the Regge limit}
\label{sec:sdisc}

Next we have to analyze the discontinuities on the right hand side of  the optical theorem \eqref{eq:cft_optical_theorem_s}. The discontinuity of the scalar $s$-channel block was recently computed in general without taking the Regge limit in \cite{Caron-Huot:2020nem} and we will review it here. The generalization to external spinning operators is done in section \ref{sec:vertex_function}, after taking the Regge limit. Let us take the conformal partial wave expansion \eqref{eq:partialwaveexpansion} in the $s$ channel and use the symmetry of the integrand to extend the integration region at the cost of a factor $1/2$
\be
A^{1234}(z,\zb) &= \frac{1}{2} \sum_{J} \int_{\frac d 2 - i \oo}^{\frac d 2 + i\oo} \frac{d\De}{2\pi i}\, 
I^{1234}(\De,J)\, \psi^{1234}_{\text{good},\cO} (z,\zb) \,.
\label{eq:partialwaveexpansion_s}
\ee
Let 
$\psi^{1234}(z,\zb)$ be the partial wave $\Psi^{1234}(y_i)$ with the prefactor $T^{1234}$ stripped off. $\psi^{1234}_{\text{good},\cO} (z,\zb)$ is the conformal partial wave with an additional term that 
vanishes for integer spin but ensures favorable properties for non-integer spin  \cite{Caron-Huot:2020nem}. The new partial wave is given by
\bea
\psi^{1234}_{\text{good},\cO}(z,\zb)  =  \psi^{1234}_\cO(z,\zb)   +  2\pi \, S(\cO_3 \cO_4 [\tl\cO^\dag]) \, K_{J+d-1,1-\Delta} \, \xi_{\De,J}^{(a,b)} \, g^{1234}_{J+d-1,1 - \Delta}(z,\zb) \,,
\eea{eq:psigood}
where $g^{1234}_{\De,J}(z,\zb)$ is the usual conformal block, the constants $a,b$ are defined below \eqref{eq:dDisc_conventional} and 
\bea
\xi_{\De,J}^{(a,b)} = & \left(s^{(a,b)}_{\Delta+J}-s^{(a,b)}_{\Delta+2-d-J}\right) \frac{\Gamma\big(-J-\tfrac{d-2}{2}\big)}{\Gamma(-J)}   \,, \\
s^{(a,b)}_{\beta} = &  \, \frac{\sin\big(\pi(a+\beta/2 )\big) \, \sin\big(\pi(b+\beta/2)\big)}{\sin(\pi\beta)}  \,, \\
K_{\De,J} = &\, \frac{\Gamma(\De - 1)}{\Gamma\big(\De - \frac{d}{2}\big)}\, \kappa^{(a,b)}_{\De+J}  \,,\\
\kappa^{(a,b)}_{\beta} = &\, \frac{\Gamma\big(\frac{\beta}{2}-a\big)\Gamma\big(\frac{\beta}{2}+a\big)\Gamma\big(\frac{\beta}{2}-b\big)\Gamma\big(\frac{\beta}{2}+b\big)}{2\pi^{2}\Gamma(\beta-1)\Gamma(\beta)}  \,. 
\eea{eq:shortening} 
With this conformal partial wave it is possible to compute the discontinuity exactly \cite{Caron-Huot:2020nem}
\beq
\frac{\Disc_{14} \, \psi^{1234}_{\text{good},\cO}(z,\zb)}{S(\cO_3 \cO_4 [\tl\cO^\dag])} = \frac{R^{1234}_{\cO}(z,\zb)}{\pi i \kappa_{\De,J}^{(a,b)}}\,.
\label{eq:Disc14Psi}
\eeq
Here $R$ is the so-called Regge block
\bea
 R^{1234}_{\cO}(z,\zb) &= g^{1234}_{1-J,1-\Delta}
 -\kappa'^{(a,b)}_{\Delta+J} g^{1234}_{\Delta,J}
-\frac{\Gamma(d-\Delta-1)\Gamma\big(\Delta-\tfrac{d}{2}\big)}{\Gamma(\Delta-1)\Gamma\big(\tfrac{d}{2}-\Delta\big)}\,\kappa'^{(a,b)}_{d-\Delta+J}\ g^{1234}_{d-\Delta,J}+
\\ &\phantom{=}+ 
\frac{\Gamma(J+d-2)\Gamma\big(-J-\tfrac{d-2}{2}\big)}{\Gamma\big(J+\tfrac{d-2}{2}\big)\Gamma(-J)}
\,\kappa'^{(a,b)}_{\Delta+J}\ \kappa'^{(a,b)}_{d-\Delta+J}\ g^{1234}_{J+d-1,1-\Delta}\,,
\eea{eq:Regge_block}
with $\kappa'^{(a,b)}_{\beta}$ defined as
\be
\label{eq:kappaprime}
\kappa'^{(a,b)}_{\beta} = \frac{r^{(a,b)}_{\beta}}{r^{(a,b)}_{2-\beta}}\,, 
\qquad 
r^{(a,b)}_{\beta} = \frac{\Gamma(\frac{\beta}{2}+a)\Gamma(\frac{\beta}{2}+b)}{\Gamma(\beta)}    \, . 
\ee
$\Disc_{23}$ in the $3412$ OPE channel can be obtained by using \eqref{eq:disc_relation} on \eqref{eq:Disc14Psi}
\beq
\frac{\Disc_{23} \, \psi^{3412}_{\text{good},\cO}(z,\zb)}{S(\cO_{1}\cO_{2}[\tl\cO^\dag])}= \frac{R^{3412}_{\cO}(z,\zb)}{\pi i \kappa_{\De,J}^{(-b,-a)}}\,,
\eeq
which was the reason to consider this channel for the correlators on the right hand side of \eqref{eq:cft_optical_theorem_s}.
The Regge block is dominated in the Regge limit by \cite{Caron-Huot:2020nem,Caron_Huot_2017,Cornalba:2007fs}
\beq
g^{1234}_{1-J,1-\De}(z,\zb) = \frac{4 \pi^{\frac{d}{2}} \Gamma(\De-\frac{d}{2})}{\Gamma(\De-1)} \sigma^{1-J} \left( \Omega_{\De-\frac{d}{2}}(\rho)
+ O(\sigma) \right)\,.
\eeq
The $\sigma,\rho$ cross-ratios introduced here are defined as 
\beq
\sigma = \sqrt{z \zb} = \sqrt{x^2 \xb^2}\,, \quad
\cosh(\rho) = \frac{z+\zb}{2 \sqrt{z \zb}} = -\frac{x \cdot \bar{x}}{\sqrt{x^2 \xb^2}} \,.
\label{eq:sigma_rho}
\eeq
$\Omega_{i\nu}(\rho)$ is the harmonic function on $d-1$ dimensional hyperbolic space $H_{d-1}$ transverse to the scattering plane in $\text{AdS}_{d+1}$ \cite{Cornalba:2006xk}
\bea
\Omega_{i \nu} (\rho) 
={}&-\frac{i \nu \sin(\pi i \nu) \Gamma(h-1+i \nu) \Gamma(h-1-i \nu) }{2^{2h-1}\pi^{h+\frac{1}{2}} \Gamma\big(h-\frac{1}{2}\big)} \\
& {}_2F_1 \left(h-1+i \nu, h-1-i\nu,h-\frac{1}{2},\frac{1-\cosh(\rho)}{2}\right)\,.
\eea{eq:Omega}
Inserting everything into \eqref{eq:partialwaveexpansion_s}, we find the following expression for the discontinuity of the correlator in the Regge limit
\beq
\label{eq:sameoldregge}
\Disc_{14}  A^{1234}(z,\bar{z}) = 2\pi i\sum\limits_J \int\limits_{-\oo}^{\oo} d\nu \ \a(\nu,J)\, \sigma^{1-J} \Omega_{i\nu} (\rho)\,,
\eeq
with
\beq
\label{eq:alphaus}
\a(\nu,J) = - \, \frac{\pi^{\frac d 2 -2} \, S\big(\cO_{3}\cO_{4}\big[\big(-i\nu-\frac{d}{2}\big)^\dag\big]\big) \, \Gamma(i \nu)}{2\pi \kappa_{i\nu+\frac{d}{2},J}^{(a,b)}\, \Gamma\big(i \nu + \frac d 2 -1\big)} \,I^{1234}\!\left(i \nu + \frac{d}{2},J\right) .
\eeq

As in flat space the sum in \eqref{eq:sameoldregge} is dominated by the large $J$ contributions in the Regge limit and only finite due to a conspiration of the coefficients to ensure Regge boundedness.
The next step is therefore to perform a  Sommerfeld-Watson resummation over $J$ to evaluate \eqref{eq:sameoldregge}. Also, note that the spectral function, as given by the Lorentzian inversion formula \cite{Caron_Huot_2017}, is of the form
\be
I^{1234}(\nu) = I^{1234,t}(\nu) + (-1)^{J}I^{1234,u}(\nu)   \,.        \label{eq:spectral_break}
\ee
Let us first consider the case of a correlator with pairwise equal external operators i.e.\ $a=b=0$, $I^{1234,t} = I^{1234,u}$, where only even spins are exchanged.
Now for the resummation we replace the sum by an integral,
\beq
2\sum\limits_{J \text{ even}} \to
\int_C dJ \frac{e^{i\pi J}}{1-e^{i\pi J}}  \,.
\label{eq:sommerfeld-watson}
\eeq
The contour $C$ encloses all poles on the positive real axis (at even integers) in a clockwise direction. 
The leading Regge trajectory is given by the operators with the lowest dimension $\De(J)$ for every even spin $J$ and $\a(\nu,J)$ has poles at $i \nu = \pm (\De(J)- d/2)$.
Defining the inverse function $j=j(\nu)$ of the spectral function $\De(J)$ by
\beq
\nu^2 + \big(\De(j(\nu))-\tfrac{d}{2}\big)^2 = 0\,,
\eeq
we see that the poles in $\nu$ translate into a single pole at $J=j(\nu)$.
By deforming the $J$ contour to the left one sees that the $J$ integral is given by the residue at $J=j(\nu)$, i.e.
\begin{equation}
\Disc_{14}  A^{1234}(z,\bar{z}) = 2\pi i \int\limits_{-\oo}^{\oo} d\nu \ \a(\nu)\, \sigma^{1-j(\nu)} \Omega_{i\nu} (\rho)\,, 
\label{eq:discA_regge} 
\end{equation}
where
\begin{equation}
\a(\nu)  = - \underset{J=j(\nu)}{\Res} i \, \frac{e^{i\pi J}}{1-e^{-i\pi J}} \frac{\pi^{\frac d 2 -2} \, S\big(\cO_{3}\cO_{4}\big[\big(-i\nu-\frac{d}{2}\big)^\dag\big]\big) \, \Gamma(i \nu)}{\kappa_{i\nu+\frac{d}{2},J}^{(a,b)}\, \Gamma\!\big(i \nu + \frac d 2 -1\big)} \,
 I^{1234,t}\left(i \nu + \frac{d}{2},J\right) .
\end{equation}
When the operators are not pairwise equal, the even and odd spin operators organize into two analytic families as evident from the Lorentzian inversion formula \cite{Caron_Huot_2017}. To obtain the contribution of the leading Regge trajectory we still sum over the even spin exchanges. The result is of the same form as in \eqref{eq:discA_regge} with $I^{1234,t}$ replaced by $ \frac{1}{2}I^{1234}$ in $\alpha(\nu)$.
For the other correlator we can use \eqref{eq:disc_relation} to see that we get an analogous result with the complex conjugate spectral function
\beq
\Disc_{23} A^{3412}(z,\bar{z}) = 2\pi i \int\limits_{-\oo}^{\oo} d\nu \ \a(\nu)^*\, \sigma^{1-j(\nu)} \Omega_{i\nu} (\rho)\,.
\eeq
One can show that the corresponding impact parameter representation is given in general
by the same spectral function times a multiplicative factor which cancels poles for the external double-trace operators \cite{Cornalba:2006xm}
\beq
\cB(p,\pb) = 2\pi i \int\limits_{-\oo}^{\oo} d\nu \ \b(\nu)\, S^{j(\nu)-1} \Omega_{i\nu} (L)\,,
\label{eq:B}
\eeq
where 
\beq
\label{eq:betaexpr}
\b(\nu) = \frac{ 4 \pi^{2-d}(\sqrt{\cM_{\cO_1} \cM_{\cO_2} \cM_{\cO_3} \cM_{\cO_4}})^{-1} \; \a(\nu)}
{\chi_{j(\nu)}(\nu)  \,\chi_{j(\nu)}(-\nu) }\,,
\eeq
with the definition
\begin{equation}
\chi_{j(\nu)}(\nu) = \Gamma\left( \frac{\Delta_1+\Delta_2+j(\nu)-d/2 + i \nu}{2}\right)
 \Gamma\left( \frac{\Delta_3+\Delta_4+j(\nu)-d/2 + i \nu}{2}\right).
 \label{eq:chi_definition}
\end{equation}
The impact parameter space cross-ratios, analogous to \eqref{eq:sigma_rho}, are
\beq
S= \sqrt{p^2 \pb^2}, \qquad \cosh L=-\frac{p \cdot \bar{p}}{\sqrt{p^2 \pb^2}}\,.    \label{eq:impactCR}
\eeq
In the dual AdS scattering process  these cross ratios are interpreted  as the squared of the energy with respect to global time and as 
the impact parameter in the transverse space $H_{d-1}$.

\subsection{Spinning particles and the vertex function}
\label{sec:vertex_function}

In this section we will introduce concrete expressions for the tree amplitudes with spinning external legs and show that the contributions of the contracted spinning legs can be expressed in terms of a scalar function of three spectral parameters which we call the vertex function, analogous to \eqref{eq:V_flat_def} in flat space.
We construct tensor structures in terms of differential operators, which are a Regge limit version of weight-shifting operators that generate spinning conformal blocks from the scalar ones \cite{Costa:2011dw,Karateev:2017jgd}. 
It is convenient to work with tensor structures which are homogeneous in $p$ and $\bar{p}$, i.e.\ independent of the cross-ratio $S$ in \eqref{eq:impactCR}, such that all tensor structures have the same large $S$ behavior in the Regge limit. These differential operators can be constructed from the covariant derivative on the hyperboloid $H_{d-1}$ and from $\hat{p}= p/|p|, \hat{\bar{p}}=\bar{p}/|\bar{p}|$ 
\cite{Cornalba:2009ax,Costa:2017twz}.
The possible differential operators that generate spin for a single particle are
	\beq
		\mathcal{D}^{\rho,k}_{\mathbf{m}} (p) = 
		\hat{p}_{m_1} \ldots \hat{p}_{m_k} {\nabla_{p}}_{m_{k+1}} \ldots {\nabla_{p}}_{m_{|\rho|}} \,, \qquad k = 0,\ldots, |\rho|\,.
		\label{eq:ts_operators}
	\eeq
Tree diagrams for exchange of the Pomeron   then have the form
	\beq
		\cB^{(\De_5,\rho_5),(\De_6,\rho_6)}_{\mathbf{m} \mathbf{n}}  (p,\bar{p}) 
		= 2\pi i\int\limits_{-\infty}^\infty d\nu \, S^{j(\nu)-1} 
		\mathfrak{D}^{(\De_5,\rho_5),(\De_6,\rho_6)}_{\mathbf{m} \mathbf{n}} (\nu)\,
		\Omega_{i \nu} (L)\,.
		\label{eq:Btree_differential}
	\eeq
	Here
$\cB^{(\De_5,\rho_5),(\De_6,\rho_6)}_{\mathbf{m}\mathbf{n}}  (p,\bar{p})$ is defined just as in \eqref{eq:BtoA} and \eqref{eq:strip}, but with tensor structures constructed from $\hat{p}$ and $\hat{\bar{p}}$.
In \eqref{eq:Btree_differential} we introduced the following definition for the combination of spectral functions $\b(\nu)$ and differential operators that generate different tensor structures
	\beq
		\mathfrak{D}^{(\De_5,\rho_5),(\De_6,\rho_6)}_{\mathbf{m} \mathbf{n}} (\nu)
		= \sum\limits_{k_5=0}^{|\rho_5|}  \sum\limits_{k_6=0}^{|\rho_6|} \beta^{k_5,k_6}_{(\De_5,\rho_5),(\De_6,\rho_6)} (\nu)
		 \mathcal{D}^{\rho_5,k_5}_{\mathbf{m}} (p) \mathcal{D}^{\rho_6,k_6}_{\mathbf{n}} (\bar{p})\,.
\label{eq:Dfrak}
	\eeq
Notice that, in contrast to flat space, we do not impose a full factorization into three-point structures but rather allow for a separate spectral function for each combination of three-point structures.	

The next step is to derive the general functional form of \eqref{eq:gluing_stripped} after the contractions and sums have been done. We begin by inserting \eqref{eq:Btree_differential} into \eqref{eq:gluing_stripped},
	\begin{align}
- \Re \cB_{\text{1-loop}} (p,\pb)={}& 2\pi^{2}  \sum\limits_{\De_5,\De_6,\rho_5,\rho_6} \int\limits_{-\infty}^\infty d\nu_1 d\nu_2 \, S^{j(\nu_1) + j(\nu_2)-2} 
		\label{eq:Btilde_neater_fs} \\
		&\mathfrak{D}^{(\De_5,\rho_5),(\De_6,\rho_6)}_{\mathbf{m} \mathbf{n}} (\nu_1)^*\, \Omega_{i \nu_1} (L)
		\,\pi_{\rho_{5}}^{\mathbf{m}; \mathbf{p}}
		\pi_{\rho_{6}}^{\mathbf{n}; \mathbf{q}}
		\,\mathfrak{D}^{(\De_5,\rho_5),(\De_6,\rho_6)}_{\mathbf{p} \mathbf{q}} (\nu_2)
		 \,\Omega_{i \nu_2} (L)\,.
		\nonumber
	\end{align}
Here $\pi_\rho$ is the projector to the irreducible representation $\rho$ of $SO(d)$, which is necessary because the operators \eqref{eq:ts_operators} do not ensure the properties of irreducible representations such as tracelessness and Young symmetrization.
Next we will show how one can replace the contractions and derivatives in the previous equation by spectral parameters.
Note first that due to $p \cdot \nabla_p = 0$, all contractions involving $\hat{p}$ or $\hat{\pb}$ give factors of their norm $-1$. The remaining contractions involve only covariant derivatives. These contracted derivatives can all be replaced by functions of the spectral parameters by using the Laplace equation for the harmonic function
	\beq
		\left( \nabla_{H_{d-1}}^2 + \nu^2 + (d/2-1)^2 \right) \Omega_{i \nu} (L) = 0 \,.
		\label{eq:nabla}
	\eeq
Using this equation, factors of $\nabla_p^2$ can directly be replaced.
To evaluate contractions between derivatives acting on different harmonic functions we expand the product of two scalar harmonic functions as follows,
	\beq
		\Omega_{i \nu_1} (L) \Omega_{i \nu_2} (L) = \int\limits_{-\infty}^\infty d\nu \, \Phi(\nu_1,\nu_2,\nu) \Omega_{i \nu} (L)\,,
		\label{eq:Omega_prod}
	\eeq
where $\Phi(\nu_1,\nu_2,\nu)$ was computed (for the similar case of harmonic functions on AdS$_{d+1}$) in appendix D of \cite{Penedones:2010ue}.\footnote{Note that $\Phi_{\text{here}} \Omega_{i \nu}(0)=\Phi_{\text{there}}$.}
By acting repeatedly with \eqref{eq:nabla} on this equation, one can determine the function $W_k$ that appears in
	\beq
		{\nabla_{p}}_{m_1} \ldots {\nabla_{p}}_{m_k}   \Omega_{i \nu_1} (L)
		\nabla_p^{m_1} \ldots \nabla_p^{m_k} \Omega_{i \nu_2} (L) =
		\int\limits_{-\infty}^\infty d\nu \, W_{k} \big(\nu_1^2,\nu_2^2,\nu^2\big)\, \Phi(\nu_1,\nu_2,\nu) \,\Omega_{i \nu} (L)\,.
		\label{eq:Wk}
	\eeq
$W_{k}$ is a fixed kinematical  polynomial of maximal degree $k$ in its arguments.
For example, the first non-trivial case is
	\beq
		\int\limits_{-\infty}^\infty \!\! d\nu \, \Phi(\nu_1,\nu_2,\nu) \nu^2 \,\Omega_{i \nu} (L)
		= \left( \nu_1^2 + \nu_2^2 +(\tfrac{d}{2}-1)^2 \right) \Omega_{i \nu_1} (L)\, \Omega_{i \nu_2} (L)
		-2 \nabla_\mu \Omega_{i \nu_1} (L) \nabla^\mu \Omega_{i \nu_2} (L),
	\eeq
from which one can read off $W_{0}$ and $W_1$ to be 
	\beq
		W_{0} \big(\nu_1^2,\nu_2^2,\nu^2\big) = 1\,, \qquad
		W_{1} \big(\nu_1^2,\nu_2^2,\nu^2\big)  = \frac{1}{2} \Big(
		\nu_1^2 + \nu_2^2 - \nu^2 + (d/2-1)^2
		\Big) .
	\eeq
More generally, by acting with the Laplacian on both sides of (\ref{eq:Wk}) one can derive a recursion relation of the form
	\bea
{}&		\int d\nu \,W_{k+1}(\nu_i) \, \Phi(\nu_i)\, \Omega_{i \nu} (L) = \int d\nu \,W_{k}(\nu_i)\,W_{1}(\nu_i)\,\Phi(\nu_i) \,\Omega_{i \nu} (L) \\
		& + \frac{1}{2} \left([\nabla^2,\nabla_{m_1}\dots\nabla_{m_{k}}] \Omega_{i \nu_1} (L) \nabla^{m_1}\dots\nabla^{m_{k}} \Omega_{i \nu_2} (L) +(\nu_1 \leftrightarrow \nu_2)\right) \,.
	\eea{eq:Wrecrer}
The terms with commutators, which will vanish in the flat space limit, can be evaluated using the fact that the commutators of covariant derivatives can be replaced by Riemann tensors, which for the hyperboloid can be written in terms of the metric. This means that these terms have two derivatives less than the other terms, and will therefore produce less than maximal powers of $\nu_i$. This shows that the maximal power of $\nu_i$ in $W_k$ is just given by repeatedly multiplying $W_1$. Therefore we have
	\beq
		W_{k} \big(\nu_1^2,\nu_2^2,\nu^2\big)  = \left(\frac{\nu_1^2 + \nu_2^2 - \nu^2}{2} 
		 \right)^k + O\!\left(\nu_i^{2(k-1)}\right) .
\label{eq:Wk_leading}
	\eeq
Having shown that all derivatives can be replaced by polynomials of the spectral parameters, we can define
	\bea
		&\mathfrak{D}^{(\De_5,\rho_5),(\De_6,\rho_6)}_{\mathbf{m} \mathbf{n}} (\nu_1)^*\, \Omega_{i \nu_1} (L)
		\,\pi_{\rho_5}^{\mathbf{m};\mathbf{p}}
		\pi_{\rho_6}^{\mathbf{n};\mathbf{q}}\,
		\mathfrak{D}^{(\De_5,\rho_5),(\De_6,\rho_6)}_{\mathbf{p} \mathbf{q}} (\nu_2)
		\, \Omega_{i \nu_2} (L)\\
		&=\int\limits_{-\infty}^\infty d\nu \, W_{(\De_5,\rho_5),(\De_6,\rho_6)} \big(\nu_1^2,\nu_2^2,\nu^2\big)\, \Phi(\nu_1,\nu_2,\nu) \,\Omega_{i \nu} (L)\,.
	\eea{eq:W_llb}
This gives the contribution of a given pair of intermediate states labeled by $(\Delta_5,\rho_5)$ and $(\Delta_6,\rho_6)$ to $- \Re \cB_{\text{1-loop}}(p,\pb)$.
Now we can define the vertex function  $V (\nu_1,\nu_2,\nu)$, which is even in all its arguments, in analogy to \eqref{eq:V_flat_def} as the sum over all such contributions in \eqref{eq:Btilde_neater_fs}
	\bea
		\sum\limits_{\De_5,\De_6,\rho_5,\rho_6,}
		W_{(\De_5,\rho_5),(\De_6,\rho_6)} \big(\nu_1^2,\nu_2^2,\nu^2\big)
		= \beta (\nu_1)^* \beta (\nu_2) V (\nu_1,\nu_2,\nu)^2\,,
	\eea{eq:vertex_ansatz}
and reach the following representation for the 1-loop amplitude
	\bea
		- \Re \cB_{\text{1-loop}} (p,\pb) = 2\pi^2  \int\limits_{-\infty}^\infty  d\nu d\nu_1  d\nu_2 \, \beta(\nu_1)^* \beta(\nu_2)
		 \, V(\nu_1,\nu_2,\nu)^2
		  \\ S^{j(\nu_1)+j(\nu_2)-2} \Phi(\nu_1,\nu_2,\nu)\, \Omega_{i \nu} (L)\,.
			\eea{eq:Bt_SL_vertex}
All the information about the spinning tree-level correlators and their contractions is encoded in the vertex function $V(\nu_1,\nu_2,\nu)$ which  mirrors the role of its flat space analogue.

However, in order to compute the full impact parameter representation rather than just its real part, we have to go through a detour via the Lorentzian inversion formula, as described in \cite{Meltzer:2019pyl}.
We first Fourier transform back to $\dDisc_t \, A_{\text{1-loop}}$ from which we obtain the $s$-channel OPE coefficients.
Then we can compute $\Disc_{14} A_{\text{1-loop}}$ which we can finally Fourier transform to obtain $\cB_{\text{1-loop}}(p,\pb)$. Since in the Regge limit the difference between $\dDisc_t \, A_{\text{1-loop}}$ and $\Disc_{14} A_{\text{1-loop}}$ is just a phase factor (see \cite{Meltzer:2019pyl}), the same happens for the impact parameter representation
	\begin{align}
		\cB_{\text{1-loop}} (p,\pb) =  -4\pi^2 \int\limits_{-\infty}^\infty d\nu d\nu_1 d\nu_2 \, & \frac{1 + e^{-i \pi (j(\nu_1)+j(\nu_2)-1)}}{1 - e^{-2 \pi i (j(\nu_1)+j(\nu_2)-1)}} \, \beta(\nu_1)^* \beta(\nu_2) \,V(\nu_1,\nu_2,\nu)^2
		       \nonumber \\ &
S^{j(\nu_1)+j(\nu_2)-2} \; \Phi(\nu_1,\nu_2,\nu)\, \Omega_{i \nu} (L)  \,.
	\label{eq:B_SL_vertex}
	\end{align}
It is important to emphasize that this provides a finite $\Delta_{gap}$ description for the one-loop correlator in the Regge limit up to the knowledge of the vertex function $V(\nu_1,\nu_2,\nu)^2$. For CFTs that admit a flat space limit, 
we will see in  sections \ref{sec:flat_space_limit} and \ref{sec:IIB_AdS_flat} how one can fix part of this vertex function from the knowledge of its flat space analogue. In section \ref{sec:Appendix_tchannel} below, we make a comparison with the large $\Delta_{gap}$ limit studied in reference \cite{Meltzer:2019pyl}, and also describe the implications of \eqref{eq:Bt_SL_vertex} for t-channel CFT data.

\section{Constraints on CFT data}
\label{sec:Appendix_tchannel}

\subsection{Comparison with the large $\De_{\text{gap}}$ limit}
\label{sec:Dav_match}

In \cite{Meltzer:2019pyl} the Regge limit of the four-point correlator of pairwise identical scalars was studied in an expansion in $1/N$ in the limit of large $\De_{\text{gap}}$. The specific limit considered was $S\gg \De_{\text{gap}}^{2}\gg 1$, so the result is sensitive to all the higher spin interactions in the leading Regge trajectory, but tidal excitations are ignored. Since we have kept $\De_{\text{gap}}$ finite, we should be able to obtain a match between the result for the one-loop correlator in \cite{Meltzer:2019pyl} with our result \eqref{eq:gluing_stripped} after dropping the tidal excitations $\cO_5 \neq \cO_1$ and $\cO_6\neq \cO_3$. 

We pick the term $\De_5=\De_1$ and $\De_6=\De_3$ that is the sole contribution to \eqref{eq:gluing_stripped} in the large $\De_\text{gap}$ limit, and use \eqref{eq:B} to obtain
\be
- \Re \cB_{\text{1-loop}}(S,L) = 2\pi^2  \int d\nu_1 d\nu_2 d\nu\, \beta^{*}(\nu_1)\beta(\nu_2) \,\Phi(\nu_1,\nu_2,\nu) \, S^{j(\nu_1)+j(\nu_2)-2}\Omega_{i\nu}(L) \, \Bigg|_{\left[\cO_5\cO_6\right]}  \,.
\label{eq:tnt1}
\ee
Now let us extract the corresponding result from \cite{Meltzer:2019pyl}. Equation (3.15) of  \cite{Meltzer:2019pyl}  
gives the double discontinuity of the one-loop correlator $\mathcal{G}^{(2)}$ as follows (with $\Delta_\phi = \De_1$ and $\Delta_\psi = \De_3$)
\bea
\dDisc_t [\mathcal{G}^{(2)}(z,\zb)] =\ & \frac{\pi^4}{8} \int d\nu_1 d\nu_2 d\nu \, \chi_{j(\nu_1)+j(\nu_2)-1}(\nu) \,\chi_{j(\nu_1)+j(\nu_2)-1}(-\nu) \, \mathcal{N} 
  \\ 					
  	&\widehat{\gamma}^{(1)}(\nu_1)\widehat{\gamma}^{(1)}(\nu_2) \, \Phi(\nu_1,\nu_2,\nu) (z\bar{z})^{\frac{2-j(\nu_1)-j(\nu_2)}{2}}\Omega_{i\nu}\left(\frac{1}{2}\log(z/\bar{z})\right)    \,,
\eea{eq:david1}
where  $\chi_{j(\nu_1)+j(\nu_2)-1}(\nu)$ is defined in \eqref{eq:chi_definition} but with $j(\nu)$ replaced by $j(\nu_1)+j(\nu_2)-1$, and with $\Delta_2=\Delta_1$ and $\Delta_4=\Delta_3$. It accounts for the double-trace exchanges $[\cO_1 \cO_1]$ and $[\cO_3 \cO_3]$ analytically continued to spin $j(\nu_1)+j(\nu_2)-1$. The operators contributing to the $t$-channel expansion in the large $\De_{\text{gap}}$ limit are the double-traces $[\cO_1 \cO_3]_{n,\ell}$ with dimensions and OPE coefficients given by
\bea
\Delta_{h,\bar{h}}&= \Delta^{(0)}_{h,\bar{h}} +\frac{1}{N^{2}} \, \gamma^{(1)}_{h,\bar{h}}+\frac{1}{N^{4}} \,\gamma^{(2)}_{h,\bar{h}}+\cdots\,,
\qquad \Delta^{(0)}_{h,\bar{h}}&= \De_1 +\De_3 +2n +\ell \,,
\\ 
P_{h,\bar{h}}&= P^{MFT}_{h,\bar{h}}\left(1+\frac{1}{N^{2}} \,\delta P^{(1)}_{h,\bar{h}}+\frac{1}{N^{4}}\,\delta P^{(2)}_{h,\bar{h}}+\cdots\right)\,,
\eea{eq:CFT_data_expansion}
where $h,\bar{h} = \De\mp\ell$.
The tree-level anomalous dimensions $\gamma^{(1)}_{h,\bar{h}}$ and tree-level corrections to OPE coefficients $\delta P^{(1)}_{h,\bar{h}}$ can be extracted respectively from $\widehat{\gamma}^{(1)}(\nu)$ and $\widehat{\delta P}^{(1)}(\nu)$  by
\bea
\gamma^{(1)}_{h,\bar{h}} &\approx \int\limits_{-\infty}^{\infty}d\nu \, \widehat{\gamma}^{(1)}(\nu) \, (h\bar{h})^{j(\nu)-1}\,\Omega_{i\nu}\big(\log(h/\bar{h})\big)  \,, \\
\delta P^{(1)}_{h,\bar{h}} &\approx \int\limits_{-\infty}^{\infty}d\nu \, \widehat{\delta P}^{(1)}(\nu) \, (h\bar{h})^{j(\nu)-1}\,\Omega_{i\nu}\big(\log(h/\bar{h})\big)   \,.
\eea{eq:extract_anom}
$\widehat{\gamma}^{(1)}(\nu)$ and $\widehat{\delta P}^{(1)}(\nu)$ can be obtained respectively from the real and imaginary parts of the phase shift, and are related to $\beta$ by\footnote{Note that due to difference in conventions, $\mathcal{N}\beta$ for us is equal to $\beta$, as defined in \cite{Meltzer:2019pyl}.}
\bea
\widehat{\gamma}^{(1)}(\nu) = &\ 2\, \text{Re} \beta(\nu) \,, \\
\widehat{\delta P}^{(1)}(\nu) = &-2\pi\, \text{Im} \beta(\nu) \,.
\eea{eq:gamma_beta}
Taking the Fourier transform to impact parameter space on \eqref{eq:david1} and then using \eqref{eq:gamma_beta} gives\footnote{In our conventions the Fourier transform takes $\dDisc_t [\mathcal{G}^{(2)}(z,\zb)]$ to $- \mathcal{N} \Re \cB$ upto scaling factors.}
\bea
- \Re \cB_{\text{1-loop}}(S,L) = 2\pi^2   \int d\nu_1 d\nu_2 d\nu \, \text{Re}\beta(\nu_1) \, \text{Re}\beta(\nu_2) \, \Phi(\nu_1,\nu_2,\nu) S^{j(\nu_1)+j(\nu_2)-2} \, \Omega_{i\nu}(L) \,.
\eea{eq:david7}
We need to compare \eqref{eq:tnt1} with \eqref{eq:david7}. The only difference are the real parts in \eqref{eq:david7}, however $\text{Im} \beta$ in \eqref{eq:tnt1} is related to tree-level corrections to the OPE coefficients and these are suppressed at large $\De_{\text{gap}}$ \cite{Meltzer:2019pyl}. This can be seen for example from \eqref{eq:gamma_beta} and using in it the explicit expression for $\alpha(\nu)$ from \cite{Meltzer:2019pyl}. The result is 
\beq
\widehat{\delta P}^{(1)}(\nu) = \frac{- \pi \Im \left( \frac{i e^{i\pi j(\nu)}}{1-e^{i\pi j(\nu)}} \right) }{\Re \left( \frac{i e^{i\pi j(\nu)}}{1-e^{i\pi j(\nu)}} \right)} \,\widehat{\gamma}^{(1)}(\nu) 
= - \pi \tan\Big( \frac{\pi}{2} j(\nu) \Big)\, \widehat{\gamma}^{(1)}(\nu) \,.
\label{eq:OPE_supp}
\eeq
The suppression is due to the tan factor, since for large $N$ theories it is known that \cite{Brower:2006ea,Cornalba:2007fs,Costa:2012cb}
\be
\label{eq:jnu}
j(\nu) = 2 - 2\,\frac{ \nu^{2} + (d/2)^2}{\De_{\text{gap}}^{2}} + O\big(\De_{\text{gap}}^{-4}\big)    \,.
\ee
The anomalous dimensions $\gamma^{(1)}_{h,\bar{h}}$ are order 1, while the the corrections to the OPE coefficients $\delta P^{(1)}_{h,\bar{h}}$ are at order $\De_{\text{gap}}^{-2}$.

Thus we have matched our result for the Regge limit of the dDisc of a one-loop correlator at large $\De_{\text{gap}}$ to that of \cite{Meltzer:2019pyl}. Note that we managed to reproduce the result without the need for any projections to the physical double-traces. Therefore it is reasonable to assume that the gluing of tree-level correlators in the Regge limit does not receive contributions from the double-traces of shadows and we can use the optical theorem \eqref{eq:gluing_stripped} without the projections onto $[\cO_5 \cO_6]$.

\subsection{Extracting t-channel CFT data}
\label{sec:extract_data}

Next we shall see how we can extract the CFT data for the double-trace operators exchanged in the $t$-channel to order $1/N^4$ from the vertex function $V(\nu_1^2,\nu_2^2,\nu^2)$. To this end we follow section 3.2 of \cite{Meltzer:2019pyl} and extend the results therein by including tidal excitations, which make our statements valid at finite $\De_{\text{gap}}$.
As discussed in the previous section, the only operators contributing to the $t$-channel expansion in the large $\De_{\text{gap}}$  limit are the double-traces $[\cO_1 \cO_3]$ \cite{Li:2017lmh}. The three-point function of these double-traces with  their constituent operators $\cO_1$ and  $\cO_3$
 has the large $N$ behavior
\beq
\< \cO_1 \cO_3 [\cO_1 \cO_3] \> \sim 1\,.
\eeq
By including tidal excitations we have to include also double-traces $[\cO_5 \cO_6]$ corresponding to additional double-traces coupling to 
$\cO_1$ and  $\cO_3$. These satisfy
\beq
\< \cO_1 \cO_3 [\cO_5 \cO_6] \> \sim \frac{1}{N^2}\,, \qquad
[\cO_5 \cO_6] \neq [\cO_1 \cO_3]\,,
\eeq
so that only their classical dimension and leading OPE coefficient squared\footnote{We are free to insert $P^{\text{MFT}}$ here, defining $\delta P$ accordingly. This will be useful below in \eqref{eq:very_long2}.}
\beq
\Delta_{h',\bar{h}'} = \De_{\cO_5} + \De_{\cO_6} +2n + \ell\,, \qquad
P_{h',\bar{h}'} =
\frac{1}{N^{4}} P^{\text{MFT}}_{h',\bar{h}'} \delta P^{(56)}_{h',\bar{h}'}+\ldots\,,
\eeq
appear in the one-loop correlator. This is compatible with the large $N$ behavior for single-trace exchange in the direct channel,
\beq
\langle \cO_1 \mathcal{O}_5 \mathcal{O}_{\Delta(J)} \rangle \sim \frac{1}{N} ~,~\langle  \mathcal{O}_{\Delta(J)} \mathcal{O}_6 \cO_3 \rangle \sim \frac{1}{N}  \,,
\eeq
which justifies that the OPE coefficients $c_{\cO_1 \cO_3 [\mathcal{O}_5\mathcal{O}_6]}$ start at order $1/N^2$.
As explained in \cite{Meltzer:2019pyl}, the cross channel expansion of the correlator is then dominated by the terms
\begin{align}
 \frac{\mathcal{A}_{\text{1-loop}}^{\circlearrowleft}(z,\bar{z})}{(z\bar{z})^{\Delta_{\f}}}
& \approx  \sum_{h,\bar{h}}
P^{MFT}_{h,\bar{h}}\bigg[i\pi \gamma^{(2)}_{h,\bar{h}}+\delta P^{(2)}_{h,\bar{h}}+i\pi \gamma^{(1)}_{h,\bar{h}} \, \delta P^{(1)}_{h,\bar{h}}-\frac{\pi^{2}}{2}\left(\gamma^{(1)}_{h,\bar{h}}\right)^{2}\bigg] g_{h,\bar{h}}(1-z,1-\bar{z}) \nonumber\\
&+\sum_{h',\bar{h}'}
P^{MFT}_{h',\bar{h}'} \delta P^{(56)}_{h',\bar{h}'} \,g_{h',\bar{h}'}(1-z,1-\bar{z})\,.
\label{eq:loopCrossingV2}
\end{align}

We shall now compare with our result for the one-loop correlator in the Regge limit and use it in the light of \eqref{eq:loopCrossingV2} to extract CFT data. We start with the dDisc of the correlator in the impact parameter representation as in \eqref{eq:Bt_SL_vertex}. Doing an inverse Fourier transform on this and taking out the appropriate scale factors gives us the dDisc of the one-loop correlator in the Regge limit
	\bea
	\dDisc_t \mathcal{A}_{\text{1-loop}}(z,\zb) = \frac{\pi^4 \mathcal{N}}{4} \int\limits_{-\infty}^\infty d\nu d\nu_1 d\nu_2 \, \big(\beta^{*}(\nu_1) \beta(\nu_2) + \beta(\nu_1) \beta^{*}(\nu_2)\big)
		  		V(\nu_1,\nu_2,\nu)^2 \\ 
			\Phi(\nu_1,\nu_2,\nu) \, \chi_{j(\nu_1)+j(\nu_2)-1}(\nu) \, \chi_{j(\nu_1)+j(\nu_2)-1}(-\nu) \, \sigma^{2-j(\nu_1)-j(\nu_2)}\, \Omega_{i\nu}(\rho)    \,,
	\eea{eq:dDisc_invFT2} 
where we  symmetrized the product of $\beta$'s by using the symmetry of the expression under $\nu_1 \leftrightarrow \nu_2$. 
We can now use the Lorentzian inversion formula \cite{Caron_Huot_2017} on \eqref{eq:dDisc_invFT2}, as shown in \cite{Meltzer:2019pyl}, to obtain the one-loop correlator in the Regge limit, and then use \eqref{eq:gamma_beta} to express it as
	\bea
	\mathcal{A}_{\text{1-loop}}^{\circlearrowleft}(z,\bar{z})\approx -\frac{\pi^4 \mathcal{N}}{4}\int\limits_{-\infty}^{\infty} d\nu_1 d\nu_2 d\nu \,
		\frac{1+e^{-i\pi (j(\nu_1)+j(\nu_2)-1)}}{1-e^{-2\pi i (j(\nu_1)+j(\nu_2)-1)}} 	\,	V(\nu_1,\nu_2,\nu)^2  \\
		\Phi(\nu_1,\nu_2,\nu)\, \chi_{j(\nu_1)+j(\nu_2)-1}(\nu) \, \chi_{j(\nu_1)+j(\nu_2)-1}(-\nu) \,
 			\sigma^{2-j(\nu_1)-j(\nu_2)}\, \Omega_{i\nu}(\rho)  \\
		\left[\widehat{\gamma}^{(1)}(\nu_1)\,\widehat{\gamma}^{(1)}(\nu_2)  + \frac{1}{\pi^2}\widehat{\delta P}^{(1)}(\nu_1)\, \widehat{\delta P}^{(1)}(\nu_2)\right] 	.		
	\eea{eq:A_1_loop}
We now take the $t$-channel expansion \eqref{eq:loopCrossingV2}, and use in it \eqref{eq:extract_anom}, \eqref{eq:Omega_prod}, and the following ansatz,
\bea
\gamma^{(2)}_{h,\bar{h}}&\approx\int\limits_{-\infty}^{\infty} d\nu_{1}d\nu_2d\nu \, \widehat{\gamma}^{(2)}(\nu_1,\nu_2,\nu) \, (h\bar{h})^{j(\nu_1)+j(\nu_2)-2}\,\Omega_{i\nu}\big(\log(h/\bar{h})\big)\,,\\
\delta P^{(2)/(56)}_{h,\bar{h}}&\approx\int\limits_{-\infty}^{\infty} d\nu_{1}d\nu_2d\nu \, \widehat{\delta P}^{(2)/(56)}(\nu_1,\nu_2,\nu) \, (h\bar{h})^{j(\nu_1)+j(\nu_2)-2}\,\Omega_{i\nu}\big(\log(h/\bar{h})\big)\,,
\eea{eq:2loop_CFT_data_spectral}
to obtain
\begin{align}
(z\bar{z})^{-\Delta_{\f}}& \mathcal{A}_{\text{1-loop}}^{\circlearrowleft}(z,\bar{z}) \approx \int\limits_{-\infty}^{\infty} d\nu_{1}d\nu_2d\nu \Bigg[\sum_{h,\bar{h}}  (h\bar{h})^{j(\nu_1) +j(\nu_2) -2}\,\Omega_{i\nu}(\log h/\bar{h})\, P^{MFT}_{h,\bar{h}}  
\label{eq:very_long}    \\
		& \bigg[i\pi \widehat{\gamma}^{(2)}(\nu_1,\nu_2,\nu)  + \frac{i\pi}{2}\left(\widehat{\gamma}^{(1)}(\nu_1)\,\widehat{\delta P}^{(1)}(\nu_2) + \widehat{\gamma}^{(1)}(\nu_2)\,\widehat{\delta P}^{(1)}(\nu_1)\right)\Phi(\nu_1,\nu_2,\nu)   \nonumber\\
		&  \qquad  \qquad -\frac{\pi^2}{2}\,\widehat{\gamma}^{(1)}(\nu_1)\,\widehat{\gamma}^{(1)}(\nu_2)\Phi(\nu_1,\nu_2,\nu) + \widehat{\delta P}^{(2)}(\nu_1,\nu_2,\nu)   \bigg]\, g_{h,\bar{h}}(1-z,1-\bar{z}) \nonumber \\
		&  + \sum_{h',\bar{h}'} P^{MFT}_{h',\bar{h}'} \,\widehat{\delta P}^{(56)}(\nu_1,\nu_2,\nu) (h'\bar{h}')^{j(\nu_1) +j(\nu_2) -2)}\,\Omega_{i\nu}(\log h'/\bar{h}') \,g_{h',\bar{h}'}(1-z,1-\bar{z}) \Bigg]\,.
		 \nonumber
\end{align}
We can approximate the $h,\bar{h}$ and $h',\bar{h}'$ sums with integrals, $\sum_{h,\bar{h}} \rightarrow \frac{1}{2}\int_{0}^{\infty}dh\, d\bar{h}$, and evaluate them using Bessel function integrals (see section 2.2 of \cite{Meltzer:2019pyl}) to arrive at the result
\begin{align}
\mathcal{A}_{\text{1-loop}}^{\circlearrowleft}(z,\bar{z})&\approx \frac{\pi^2 \mathcal{N}}{4}\int\limits_{-\infty}^{\infty} d\nu_{1}d\nu_2d\nu \, \chi_{j(\nu_1)+j(\nu_2)-1}(\nu)\,\chi_{j(\nu_1)+j(\nu_2)-1}(-\nu) \,
 			\sigma^{2-j(\nu_1)-j(\nu_2)} \,\Omega_{i\nu}(\rho) \nonumber  \\
		& \left[ i\pi \widehat{\gamma}^{(2)}(\nu_1,\nu_2,\nu)  -\frac{\pi^2}{2}\,\widehat{\gamma}^{(1)}(\nu_1)\,\widehat{\gamma}^{(1)}(\nu_2)\,\Phi(\nu_1,\nu_2,\nu) + \widehat{\delta P}^{(2)}(\nu_1,\nu_2,\nu) \right.  
\label{eq:very_long2}  \\
		&  \left. + \frac{i\pi}{2}\left(\widehat{\gamma}^{(1)}(\nu_1)\,\widehat{\delta P}^{(1)}(\nu_2) + \widehat{\gamma}^{(1)}(\nu_2)\,\widehat{\delta P}^{(1)}(\nu_1)\right)\Phi(\nu_1,\nu_2,\nu) + \widehat{\delta P}^{(56)}(\nu_1,\nu_2,\nu) \right] .
\nonumber
\end{align}
Comparing the real parts of the coefficient of $\chi(\nu)\chi(-\nu)\sigma^{2-j(\nu_1)-j(\nu_2)} \Omega_{i\nu}(\rho)$ in the integrands of \eqref{eq:A_1_loop} and \eqref{eq:very_long2}, and using
\beq
\frac{1+e^{-i\pi (j(\nu_1)+j(\nu_2)-1)}}{1-e^{-2\pi i (j(\nu_1)+j(\nu_2)-1)}}
= \frac{1}{2} + \frac{i}{2} \tan \left( \frac{\pi}{2} \big(j(\nu_1)+j(\nu_2) \big) \right) ,
\label{eq:trig_ID}
\eeq
we conclude that
\begin{align}
\widehat{\delta P}^{(2)}(\nu_1,\nu_2;\nu) + \widehat{\delta P}^{(56)}(\nu_1,\nu_2;\nu) & = 
 -\frac{1}{2}\bigg[\pi^2 \Big(V(\nu_1,\nu_2,\nu)^2 -1\Big) \widehat{\gamma}^{(1)} (\nu_1)\, \widehat{\gamma}^{(1)} (\nu_2)
\label{eq:deltaP2}\\
& 
+ V(\nu_1,\nu_2,\nu)^2\, \widehat{\delta P}^{(1)}(\nu_1)\,\widehat{\delta P}^{(1)}(\nu_2) \bigg] \Phi(\nu_1,\nu_2,\nu )  \,.
\nonumber
\end{align}
This is the general result for fixed $\De_{\text{gap}}$ that extracts  OPE data from the AdS vertex function. 
Let us now take the large $\De_{\text{gap}}$ limit to make contact with \cite{Meltzer:2019pyl}.
In this limit, $V(\nu_1,\nu_2,\nu)^2 =1$ and $\widehat{\delta P}^{(1)}$, $\widehat{\delta P}^{(56)}$ are suppressed with respect to 
$\widehat{\gamma}^{(1)}$. Therefore $\delta{P}^{(2)}_{h,\bar{h}}=0$, as was obtained in \cite{Meltzer:2019pyl}. 

Similarly, comparing the imaginary parts 
 we have
\begin{align}
\widehat{\gamma}^{(2)}(\nu_1,\nu_2;\nu)& =-\frac{1}{2}
\bigg[ \left(\widehat{\gamma}^{(1)} (\nu_1) \widehat{\delta P}^{(1)}(\nu_2)
+ \widehat{\delta P}^{(1)}(\nu_1) \widehat{\gamma}^{(1)} (\nu_2) \right)  + \pi \tan \left( \frac{\pi}{2} (j(\nu_1)+j(\nu_2)) \right)  
\nonumber\\
&
 \!\!
 V(\nu_1,\nu_2,\nu)^2 \Big( \widehat{\gamma}^{(1)} (\nu_1) \widehat{\gamma}^{(1)} (\nu_2)  
 +\frac{1}{\pi^2}\widehat{\delta P}^{(1)}(\nu_1)\widehat{\delta P}^{(1)}(\nu_2) \Big) \bigg] \Phi(\nu_1,\nu_2,\nu )\,.
 \label{eq:gamma2}
\end{align}
The term $\widehat{\delta P}^{(1)}(\nu_1)\,\widehat{\delta P}^{(1)}(\nu_2)$ is suppressed by $\De_{\text{gap}}^{-4}$ with respect to 
the other terms. At leading order in $\De_{\text{gap}}$, we can discard this term and set $V(\nu_1,\nu_2,\nu)^2 =1$. We can then use \eqref{eq:OPE_supp} to simplify the expression to,
\bea
\widehat{\gamma}^{(2)}(\nu_1,\nu_2;\nu)&=-\frac{1}{2} \pi \tan \left(\frac{1}{2} \pi  j(\nu_1)\right) \tan \left(\frac{1}{2} \pi  j(\nu_2)\right) \tan \left(\frac{1}{2} \pi (j(\nu_1)+j(\nu_2))\right) 
\\ & \hspace{2.45in}\times\widehat{\gamma}^{(1)}(\nu_1)   \widehat{\gamma}^{(1)}(\nu_2) \Phi(\nu_1,\nu_2,\nu ) \,.
\eea{eq:gamma2_meltzer}
This is the same result as obtained in \cite{Meltzer:2019pyl} for $\widehat{\gamma}^{(2)}(\nu_1,\nu_2;\nu)$.
More generally, knowledge of the vertex function $V(\nu_1,\nu_2,\nu)^2$ and of the $\langle\cO_1 \cO_3 [\mathcal{O}_5\mathcal{O}_6]\rangle$ OPE coefficients gives additional information about the one-loop CFT data of the $[\cO_1 \cO_3]$ double-trace operators. It would be interesting to 
analyze these equations order by order in the $1/\De_{\text{gap}}^2$ expansion.


\section{Flat space limit}
\label{sec:flat_space_limit}

Having fixed the general form of the impact parameter representation of the one-loop correlator from first principles in section \ref{sec:ads},
we now want to fix part of the dynamical data by taking the flat space limit,
which relates it to the known flat space amplitudes.
The prescription to achieve this was discovered in
\cite{Cornalba:2007fs}, where it was applied to scalar tree-level amplitudes.
This limit is taken by sending the AdS radius $R$ to infinity while scaling the relevant quantities in order to match them to flat space quantities in a sensible way.
The dimensionless quantities $S$ and $L$ are sent to dimensionless combinations of $R$ with the flat space center of mass energy $s$ and impact parameter $b$ as
\beq
 S = \frac{R^2 s}{4}\,, \qquad  L=\frac{b}{R}\,.
\eeq
Note that $L$ is the AdS impact parameter, as it describes the geodesic distance on $H_{d-1}$ between the impact points in transverse space.
If we impose the identification of Casimir eigenvalues
\beq
\De (\De-d) = R^2 m^2\,,
\label{eq:casimirdeltam}
\eeq
for the states on the leading Regge trajectory and take this equation off-shell,
it becomes
\beq
\nu^2 + \left( \frac{d}{2} \right)^2 =  R^2 q^2\,,
\eeq
so for large $R$ we further impose
\beq
\nu^2 = R^2 q^2\,.
\eeq
Our expressions in AdS are integrals in $\nu$, while in flat space we have vector integrals in $q$,
where we recall that $q$ is a vector in the transverse space $\mathbb{R}^{D-2}$.
In order to compare the expressions, it is instructive to do the flat space angular integrals and keep only the integral over the modulus $|q|$.
In this way, the exponential is replaced by the harmonic function $\omega_q(b)$ according to 
\beq
\int\limits_{\mathbb{R}^{D-2}} \frac{d q}{(2\pi)^{D-2}} \, e^{i b \cdot q}
= 2 \int\limits_0^{\infty} d|q| \, \omega_q(b)\,,
\label{eq:exp_to_harmonic}
\eeq
so that \cite{Costa:2014kfa}
	\bea
\omega_{q} (b) = q \int\limits_{\mathbb{R}^{D-2}} \frac{d p}{(2\pi)^{D-2}} \, e^{i b \cdot p}\, \delta(p^2 - q^2) = \frac{1}{2 (2\pi)^{\frac{D-2}{2}}}\frac{|q|^\frac{D-2}{2}}{|b|^{\frac{D-4}{2}}} J_{\frac{D-4}{2}}(|q| |b|)\,,
	\eea{eq:harmonic_flat}
where $J$ denotes the Bessel $J$-function and we recall that $\omega_{q} (b)$ only depends on the modulus of the vectors $q$ and $b$.
One can check that the flat space limit of the $H_{d-1}$ harmonic function \eqref{eq:Omega} yields the flat space harmonic function
\beq
R^{3-D} \Omega_{i \nu} (L) \to \omega_{q} (b)\,, \qquad \nu \geq 0\,.
\label{eq:Omega_fsl}
\eeq
For even $d$ this can be checked directly, while for general $d$ it is convenient to use an integral representation for the hypergeometric function which under the limit is related to an integral representation for the Bessel function \cite{Carmi:2018qzm}.
For even $d$ the relation is also valid for $\nu < 0$.

In the context of string theory we further have the dimensionless coupling $\lambda$ which is expressed in terms of $\alpha'$ and $R$ as
\beq
\sqrt{\l} = \frac{R^2}{\a'}\,,
\eeq
meaning we can also express $S$ as
\beq
S= \frac{\sqrt{\lambda}\,  \alpha's}{4}\,.
\label{eq:Stos}
\eeq
To summarize, the flat space limit is taken by
sending the AdS radius $R$ to infinity while replacing
\beq
S = \frac{\sqrt{\lambda} \alpha' s}{4}, \quad L=\frac{b}{R}, \quad	\nu^2 = R^2 q^2, \quad \nu_1^2 = R^2 q_1^2, \quad  \nu_2^2 = R^2 q_2^2, \quad \sqrt{\l} = \frac{R^2}{\a'}  \,,
		\label{eq:flat_space_limit}
\eeq
and impact parameter representations can be compared by using \eqref{eq:Omega_fsl}.
We can also use these relations to relate $\Delta_{gap}$ to $\lambda$ taking as reference a string state of mass $m^2=4/\alpha'$, therefore
\beq
\Delta^2_{\text{gap}}=\frac{4 R^2}{\alpha'} = 4 \sqrt{\lambda}\,.
\eeq

\subsection{Matching in impact parameter space}
\label{sec:matching_impact_parameter}
Let us now see what we can learn when we apply the flat space limit to the impact parameter representations studied in section \ref{sec:ads}. We begin with the tree-level correlator of four scalars for which the limit was originally imposed in \cite{Cornalba:2007fs}.
The flat space limit of the AdS result $\cB$ in \eqref{eq:B} should match the flat space impact parameter representation $i \de$ from \eqref{eq:impact_flat} of the amplitude \eqref{eq:A_tree_regge}\footnote{The relative factor $i$ in $\cB \to i \de$ can be determined by matching the exponents in the eikonal approximation for $\lambda \to \infty$.}
\beq
\cB_{\text{tree}}(p,\pb) = 4 \pi i \int\limits_{0}^{\oo} d\nu \, \b(\nu)\,S^{j(\nu)-1}\, \Omega_{i\nu} (L)
\ \to \ 
i\de_{\text{tree}} (s,b) = 2i \int\limits_0^{\oo} d|q|\, \b(t) \left(\frac{\alpha's}{4}\right)^{j(t)-1} \omega_q(b) \,.
\label{eq:flat_space_limit_dilatons}
\eeq
Here and below we do not always write the overall factors of $R$ as in \eqref{eq:Omega_fsl}, but they do work out correctly when including the expansion parameters from
\eqref{eq:loop_expansion} and \eqref{eq:amplitude_loop_expansion} and using the relation
\beq
\frac{1}{N^2} = \frac{1}{R^{D-2}} \frac{2 G_N}{\pi}\,.
\eeq
From \eqref{eq:Omega_fsl} and \eqref{eq:Stos} we see that
this does indeed match, provided  the flat space limit of the AdS Regge trajectory and spectral function are sent to the flat space Regge trajectory and Pomeron propagator%
\footnote{$j(\nu)$, $j(t)$ and $\b(\nu)$, $\b(t)$ are different functions and not the same function with different arguments.}
\beq
j(\nu)\, \to\,  j(t)\,, \qquad \lambda^{\frac{j(\nu)-1}{2}}\b(\nu)\, \to\,  \frac{1}{2 \pi} \,\b(t)\,.
\label{eq:lim_beta}
\eeq
The power of $\lambda$ in the relation of $\beta$'s is necessary to cancel the powers of $\lambda$ in the relation between $S$ and $\alpha' s$.
It is compatible with the expectation that each derivative in the couplings of the spin $J$ operators forming the Pomeron comes at least with a power of $\lambda^{-\frac{1}{4}}$.

Next we consider the optical theorem in AdS \eqref{eq:gluing_stripped} and flat space \eqref{eq:impact_optical_theorem}
\beq
- \Re \cB_{\text{1-loop}} = \frac{1}{2}
\sum\limits_{\substack{\De_5,\rho_5\\\De_6,\rho_6}}
\cB^{3652\, *}_\text{tree}  \cB^{1564}_\text{tree}
\ \to \ 
		\Im \de_{\text{1-loop}} (b) 
		= \frac{1}{2} \sum\limits_{\substack{m_5,\rho_5,\e_5\\m_6,\rho_6,\e_6}}
		\de_\text{tree}^{3652\,*} \de_\text{tree}^{1564}\,.
\eeq
The similarity is striking, however we have to make sure  the sums and summands are in fact related by the flat space limit.
The additional sums over polarizations can be evaluated using completeness relations such as 
\eqref{eq:completeness_relation}, which evaluate to contractions just as in the AdS equation.
We also have to make sure that the labels $\rho$ on both sides are irreducible representations of the same group $SO(d)$. This is indeed the case for massive particles if we consider the flat space limit $AdS_{d+1} \to \mathbb{R}^{1,d}$, which has the massive Little group $SO(d)$.

The next step is to match the tree-level correlators \eqref{eq:Btree_differential} and amplitudes \eqref{eq:A_tree_regge} that involve spinning particles 5 and 6. In this case the flat space limit gives
\begin{align}
&\cB^{(\De_5,\rho_5),(\De_6,\rho_6)}_{\mathbf{m} \mathbf{n}}  (p,\bar{p}) 
		= 4 \pi i \int\limits_{0}^\infty d\nu \, S^{j(\nu)-1} 
		\,\mathfrak{D}^{(\De_5,\rho_5),(\De_6,\rho_6)}_{\mathbf{m} \mathbf{n}} (\nu)
		\,\Omega_{i \nu} (L)\ \to
		\label{eq:flat_space_limit_spinning}
\\
\to\ &
i \de^{(m_5,\rho_5),(m_6,\rho_6)}_{\mathbf{m} \mathbf{n}} (s,b) 
= i \int\limits_{\mathbb{R}^{D-2}} \frac{dq}{(2\pi)^{D-2}} \left(\frac{\alpha's}{4}\right)^{j(t)-1} A^{12P}_{m_5,\rho_5,\mathbf{m}} (q,v)\, \beta(t) A^{34P}_{m_6,\rho_6,\mathbf{n}}(q,v)\, e^{i q\cdot b}
\nonumber \\
& \qquad \qquad = 2i \int\limits_0^{\oo} d|q| \left(\frac{\alpha's}{4}\right)^{j(t)-1} A^{12P}_{m_5,\rho_5,\mathbf{m}} (-i\partial_{b},v)\, \beta(t) A^{34P}_{m_6,\rho_6,\mathbf{n}}(-i\partial_{b},v)\, \omega_q(b)\,,
\nonumber
\end{align}
where the derivative $\partial_{b}$ is with respect to the components of the transverse vector $b$.
The difference compared to \eqref{eq:flat_space_limit_dilatons} is that in AdS we  have differential operators that generate tensor structures, while  in flat space the tensor structures are the ones of the on-shell three-point amplitudes. As discussed in section \ref{sec:3pt_amplitudes}, these three-point amplitudes are given in terms of the Pomeron momentum $q$ and, for massive particles, the longitudinal polarization vector $v$, which is transverse to $q$.
We will now study the relation of these two kinds of tensor structures to the flat space limit.

We begin with the covariant derivatives and will argue that they become derivatives in  impact parameter in flat space, i.e.
	\beq
		\nabla_p^{m} \Omega_{i \nu} (L)  \ \to\  
		R \partial_b^m e^{i b \cdot q} = R i q^m e^{i b \cdot q}\,.
		\label{eq:nabla_fsl}
	\eeq
In order to show this, we will act with two contracted covariant derivatives either on a single harmonic function or on two different ones, covering all situations that can occur.
Acting on a single harmonic function we obtain, from \eqref{eq:nabla} and \eqref{eq:flat_space_limit},
	\beq
		\frac{1}{\sqrt{\l}}\, \nabla_p^2 \, \Omega_{i \nu_{1,2}} (L) \ \to\  - \alpha' q_{1,2}^2\, \omega_{i \nu_{1,2}} (b)\,.
\label{eq:laplace_flat_space_limit}
	\eeq
The action of contracted covariant derivatives on two different harmonic functions is captured by the functions $W_k$ in \eqref{eq:Wk}, which is given in the flat space limit by the leading term \eqref{eq:Wk_leading}
	\bea
		\frac{W_{k} \big(\nu_1^2,\nu_2^2,\nu^2\big)}{\big(\sqrt{\l}\big)^k}\  \to\ 
		\left( \frac{\nu_1^2+\nu_2^2 - \nu^2}{2\sqrt{\l}} \right)^k
		\ \to \ \left(  \alpha'\, \frac{q_1^2+q_2^2 - q^2}{2} \right)^k
		= (\alpha')^{k} (- q_{1}\cdot q_2)^k\,, 
	\eea{eq:Wk_fs}
where we used that $q = q_1 + q_2$. This implies that the flat space limit of \eqref{eq:Wk} is
	\begin{align}
	&
\frac{1}{\big(\sqrt{\l}\big)^k}
		{\nabla_{p}}_{m_1} \ldots {\nabla_{p}}_{m_k}   \Omega_{i \nu_1} (L)
		\nabla_p^{m_1} \ldots \nabla_p^{m_k} \Omega_{i \nu_2} (L)
		\ \to
		\\
		\to \ & 
		(-\a')^k q_{1 \, m_1} \ldots q_{1 \, m_k}   e^{i b \cdot q_1}
		q_2^{m_1} \ldots q_2^{m_k} e^{i b \cdot q_2}\,.
\label{eq:cov_flat_12}
	\end{align}
We conclude that both \eqref{eq:laplace_flat_space_limit} and \eqref{eq:cov_flat_12} are compatible with \eqref{eq:nabla_fsl}.
Apart from the covariant derivative, tensor structures depend also on the direction $\hat{p}$, which is normal to the transverse space $H_{d-1}$ and satisfies $\hat{p}^2=-1$. In flat space the only possible direction for polarizations that is normal to the transverse space is the unit vector $v$, hence we have to require that in the flat space limit
	\beq
		\hat{p}^m \,\to\, i v^{m}\,.
		\label{eq:phat_fsl}
	\eeq

With the identifications \eqref{eq:nabla_fsl} and \eqref{eq:phat_fsl}, the matching in \eqref{eq:flat_space_limit_spinning} works provided that the spectral functions $\beta^{k_5,k_6}_{(\De_5,\rho_5),(\De_6,\rho_6)} (\nu)$ in \eqref{eq:Dfrak} are such that
	\beq
		\lambda^{\frac{j(\nu)-1}{2}}\mathfrak{D}^{(\De_5,\rho_5),(\De_6,\rho_6)}_{\mathbf{m} \mathbf{n}} (\nu)\, \Omega_{i \nu} (L) \, \to\, 
\frac{1}{2 \pi}
		A^{12P}_{m_5,\rho_5,\mathbf{m}} (-i\partial_{b},v) \,\beta(t)\, A^{34P}_{m_6,\rho_6,\mathbf{n}} (-i\partial_{b},v)
		\,\omega_{q}(b)\,.
\label{eq:flat_space_limit_ts}
	\eeq
Using the explicit tensor structures for three-point amplitudes in \eqref{eq:3pt_spinning}, the matching \eqref{eq:flat_space_limit_ts} can also be expressed for the spectral function for any given tensor structure
\beq
 \lambda^{\frac{j(\nu)-1}{2}} \lambda^{\frac{|\rho_5|-k_5}{4}}\lambda^{\frac{|\rho_6|-k_6}{4}}\beta^{k_5,k_6}_{(\De_5,\rho_5),(\De_6,\rho_6)} (\nu)
\ \to\ 
\frac{1}{2 \pi}\,
a_{m_5,\rho_5}^{k_5}(t) \,a_{m_6,\rho_6}^{k_6}(t) \,\b(t)\,.
\label{eq:flat_space_limit_beta_spinning}
\eeq
The powers of $\lambda$ are again compatible with a factor of $\lambda^{-\frac{1}{4}}$ in $\beta^{k_5,k_6}_{(\De_5,\rho_5),(\De_6,\rho_6)} (\nu)$ for each derivative in the coupling.
Such a scaling is expected from the general arguments of \cite{Costa:2017twz,Meltzer:2017rtf}.
We have now shown that all tree-level phase shifts appearing in the AdS and flat space optical theorems can be related by the flat space limit. 

Let us also compare the vertex functions that appear in both  AdS and flat space impact parameter representations of the one-loop amplitudes. 
The flat space limit of   \eqref{eq:Bt_SL_vertex}  is given by the impact parameter transform of \eqref{eq:optical_theorem_V_flat}, i.e.
\begin{equation}
- \Re \cB_{\text{1-loop}} (p,\pb) \ \to\  \Im \delta_{\text{1-loop}} (s, b)\,,
\end{equation}
becomes
\bea
		& 2 \pi^2  \int\limits_{-\infty}^\infty d\nu d\nu_1 d\nu_2 \, \beta(\nu_1)^* \beta(\nu_2)
		  \, V(\nu_1,\nu_2,\nu)^2 S^{j(\nu_1)+j(\nu_2)-2} \Phi(\nu_1,\nu_2,\nu) \, \Omega_{i \nu} (L) \ \to\\
\to \ & \frac{1}{2}
\int\limits_{\mathbb{R}^{D-2}} \frac{d q d q_1  d q_2}{(2\pi)^{2(D-2)}} 
\, \b(t_1)^* \b(t_2) \, V(q_1,q_2)^2 \left(\frac{\alpha's}{4}\right)^{j(t_1) + j(t_2)-2} 
\de(q-q_1-q_2) \, e^{i q\cdot b}\,.
\eea{eq:flat_space_limit_with_V}
In this case we can use the delta function to write all the other scalar functions in terms of $q_1^2$, $q_2^2$ and $q^2$, however we need to do the angular integral over the delta function itself. To this end we can define
\beq
\int\limits_{\mathbb{R}^{D-2}}  \frac{dq_1 dq_2}{(2\pi)^{D-2}} \, \de(q-q_1-q_2)
= 4\int\limits_{0}^{\oo} d|q_1| d|q_2|\, \phi(q_1, q_2, q)\,.
\label{eq:flat_phi}
\eeq
Using this and \eqref{eq:exp_to_harmonic}, we can compute the angular integrals in
	\beq
\int\limits_{\mathbb{R}^{D-2}}  \frac{dq_1 dq_2}{(2\pi)^{2(D-2)}} \, e^{i b \cdot (q_1 + q_2)} = \int\limits_{\mathbb{R}^{D-2}}  \frac{dq_1 dq_2 dq}{(2\pi)^{2(D-2)}} \, \de(q-q_1 -q_2)\, e^{i b \cdot q} \,,		\label{eq:Wk_flat}
	\eeq
to find the flat space version of \eqref{eq:Omega_prod}
\beq
\omega_{q_1}(b)\,  \omega_{q_2}(b) = 2 \int\limits_{0}^{\infty} d|q| \, \phi(q_1, q_2, q) \, \omega_{q}(b)\,.
\eeq
Using the explicit expressions for $\Phi$ and $\phi$ (which can be found for instance in appendix E of  \cite{Penedones:2010ue}), one can further check that under the flat space limit
\beq
R^{4-D} \Phi(\nu_1, \nu_2, \nu) \ \to\  \phi(q_1, q_2, q)\,.
\eeq
With this relation is clear that the expressions in
\eqref{eq:flat_space_limit_with_V} are indeed related by the flat space limit provided that the vertex functions are related by
\beq
V(\nu_1,\nu_2,\nu)\  \to\  V(q_1,q_2) = V(t_1, t_2, t)\,.
\eeq

\subsection{Constraining AdS quantities}
\label{eq:constraining_from_flat_space_limit}

We saw above that all elements of the impact parameter optical theorems in AdS and flat space are related by the flat space limit provided that $j(\nu)$, $\beta(\nu)$, $\beta^{k_5,k_6}_{(\De_5,\rho_5),(\De_6,\rho_6)} (\nu)$ and $V\big(\nu_1^2,\nu_2^2,\nu^2\big)$ are given by their flat space counterparts in the limit.
In this subsection, we briefly review how the limit actually constrains these functions.  All of these objects depend on two dimensionless quantities, the spectral parameters $\nu$ and the t'Hooft coupling $\lambda$.  Let us discuss this for a generic function $f(\nu)$ that is required to satisfy the flat space limit
\beq
f(\nu,\lambda) \, \to \,  f(t)\,.
\eeq
Existence of the  gravity limit requires the function to have an expansion in negative powers of $\sqrt{\lambda}$ of the form
\beq
f(\nu,\lambda) = \sum\limits_{n=0}^{\infty} \frac{f_n(\nu)}{\lambda^{n/2}}\,.
\eeq
In order for the flat space limit \eqref{eq:flat_space_limit} of the function to be finite, the functions $f_n(\nu)$ must  have an expansion in large $\nu$ with leading power not larger than $2n$,
\beq
f_n(\nu) = a_{n,n} \nu^{2n} + a_{n,n-1} \nu^{2(n-1)} + a_{n,n-2} \nu^{2(n-2)} + \ldots\,,
\eeq
which ensures finiteness of the limit order by order in the large $\lambda$ expansion.
The flat space limit of $f(\nu,\lambda)$ is then
\beq
f(\nu,\lambda)\  \to\  \sum\limits_{n=0}^{\infty} a_{n,n} \left(\frac{\nu^{2}}{\sqrt{\lambda}}\right)^n \to\
\sum\limits_{n=0}^{\infty} a_{n,n} (\a' q^2)^n\,.
\eeq
At every order in $1/\sqrt{\l}$ the leading power of $\nu$ survives and is fixed by the flat space limit, while all the other powers are subleading, and cannot be determined from this condition. 
These considerations hold for $j(\nu)$, $\beta(\nu)$, $\beta^{k_5,k_6}_{(\De_5,\rho_5),(\De_6,\rho_6)} (\nu)$ and $V(\nu_1,\nu_2,\nu)$, fixing part of these functions. These facts have been explored in detail for the functions  $j(\nu)$ and $\beta(\nu)$ in \cite{Cornalba:2008qf,Costa:2012cb}.


\section{Relating type IIB string theory in AdS and flat space}
\label{sec:IIB_AdS_flat}

Let us now apply our general ideas to a concrete example, the scattering of four dilatons in type IIB superstring theory on $AdS_5 \times S^5$. In the flat space limit this is related to
type IIB superstring theory on 10-dimensional flat space where the kinematics is restricted to the five dimensions arising from AdS. This happens since both the dilatons and the Pomerons are $R$-symmetry singlets, meaning the tidal excitations they couple to also have to be singlets.
As a consequence, the Regge limit does not probe the 10 dimensional nature of the string scattering process, as we consider only states with the vacuum quantum numbers associated to the compact manifold $S^5$.
 For this case the discontinuity of the (finite $\alpha'$) one-loop amplitude in the Regge limit was computed in \cite{Amati:1987uf} and is precisely of the form \eqref{eq:optical_theorem_V_flat} with $D=5$. The regime of validity of this description was discussed in detail in \cite{Amati:1987uf}.
All we need to specify are the four dynamic quantities that we already discussed in the previous section. For the Regge trajectory and Pomeron propagator we have
\beq
j(t) = 2 + \frac{\a'}{2}\,t\,, \qquad
\beta(t) = 2 \pi^2 \frac{\G\big(- \frac{\a'}{4} t\big)}{\G\big(1+\frac{\a'}{4} t\big)}
		\,e^{- \frac{i \pi \a'}{4} t}\,.
\eeq
As discussed in section \ref{sec:vertex_function_flat} the vertex function can be obtained from the scattering amplitude of two dilatons and two Pomerons. This amplitude was computed in 
\cite{Amati:1987uf} and reads
	\beq
		A^{12P_1P_2} (k, q_1, q_2) = - \frac{\Gamma(1+\alpha'q_{12}/2) \,\Gamma\big(-\alpha'\frac{k^2}{4}-\alpha'q_{12}/2\big)\, \Gamma\big(\alpha'\frac{k^2}{4}\big)}{2 \Gamma(-\alpha'q_{12}/2) \,\Gamma\big(1+\alpha'q_{12}/2+\alpha'\frac{k^2}{4}\big) \,\Gamma\big(1-\alpha'\frac{k^2}{4}\big)}\,,
		\label{eq:Aa1a2}
	\eeq
where $q_{12} = q_1\cdot q_2$.
This amplitude has poles at the masses ($m^2=4n/\alpha'$) of the string states with residues
	\beq
		\underset{k^2=-4n/\alpha'}{\Res} A^{12P_1P_2} (k, q_1, q_2)
		= \left( \frac{(-\alpha'q_{12}/2)_n}{n!} \right)^2\,, \qquad n=0,1,2,\ldots\,,
		\label{eq:3pt_A_calculated}
	\eeq
and the resulting vertex function is given by \eqref{eq:V_diagrams}
\beq
V(q_1,q_2) = 
\sum\limits_{n=0}^\infty \left( \frac{(-\alpha'q_{12}/2)_n}{n!} \right)^2
=\frac{\Gamma\big(1+\alpha'\frac{t_1+t_2-t}{2}\big)}{\Gamma\big(1+\alpha'\frac{t_1+t_2-t}{4}\big)^2}\,.
\label{eq:V_flat}
\eeq
Using the reasoning of section \ref{eq:constraining_from_flat_space_limit}, this immediately fixes the leading terms in $\nu, \nu_i$ of the AdS vertex function at every order in $\lambda$
	\beq
		V(\nu_1,\nu_2,\nu) = \frac{\Gamma\Big(1-\frac{\nu_1^2+\nu_2^2-\nu^2}{2 \sqrt{\l}}\Big)}{\Gamma\Big(1-\frac{\nu_1^2+\nu_2^2-\nu^2}{4 \sqrt{\l}}\Big)^2}
		+ \text{vanishing in flat space limit}\,.
		\label{eq:V_from_flat_space}
	\eeq
Thus, the first two corrections from expanding \eqref{eq:V_from_flat_space} at large $\lambda$ are
	\begin{align}
V(\nu_1,\nu_2,\nu) &= 1+\Big(0\cdot\big( \nu_1^2+\nu_2^2-\nu^2 \big) + c_{1,0}\Big) \frac{1}{\sqrt{\l}} 
\label{eq:V_polynomial}\\
&+\left( \frac{\pi^2}{96} \left( \nu_1^2+\nu_2^2-\nu^2 \right)^2 +c_{2,1} \big(\nu_1^2  + \nu_2^2\big)  + c'_{2,1} \nu^2 + c_{2,0} \right) \frac{1}{\l} + \ldots\,,
		\nonumber
	\end{align}
where we also included the constants  that are not fixed by the flat space limit. We note that the constants multiplying the leading power of $\nu$ at order $\big(\sqrt{\lambda}\big)^{-n}$ have a uniform transcendentality of weight $n$, which can be seen by explicitly expanding \eqref{eq:V_from_flat_space}. It would be interesting to understand the relation of this property with features of maximal transcendentality in $\mathcal{N}=4$ SYM \cite{Kotikov:2002ab,Kotikov:2007cy}.

Finally, all the spectral functions $\beta^{k_5,k_6}_{(\De_5,\rho_5),(\De_6,\rho_6)} (\nu)$ of tree-level correlators that contribute to the optical theorem are constrained 
by the flat space limit. Their flat space limit \eqref{eq:flat_space_limit_beta_spinning} is parameterized by on-shell three-point amplitudes in flat space.
These amplitudes are in principle encoded in the result \eqref{eq:3pt_A_calculated},
which separates the contributions of particles with different masses, but not the ones of particles in different representations $\rho$. 
The attempt to expand \eqref{eq:3pt_A_calculated} into products of three-point amplitudes for different $\r$ and tensor structures $k$ using \eqref{eq:residue_generic} shows that this does not fully fix the $a_{m_5,\rho_5}^{k}(t)$ in \eqref{eq:3pt_spinning} because the equations are quadratic.
However the three-point amplitudes can of course be computed in string theory, which is what the next subsection is about.
We will start with the 10D open superstring amplitudes of a massless vector, a Pomeron and an open string state up to mass level 2 which were computed in \cite{DAppollonio:2013mgj} by studying string-brane scattering. These amplitudes have to be squared to obtain closed string amplitudes. Then the irreducible representations of the 10D massive Little group $SO(9)$ have to be branched into irreducible representations of $SO(4)$ to match with the CFT irreps and account for the fact that we have five non-compact dimensions.

\subsection{Massive tree amplitudes in flat space}
\label{sec:massive_tree_flat}

Now we will discuss the flat space three-point amplitudes that take part in the process and that will fix part of the tree-level correlators with external spinning legs in AdS via the flat space limit. 
The goal is to derive the three-point amplitudes that appear in the unitarity cut \eqref{eq:3pt_A_calculated} of the
four-point amplitude of two dilatons and two Pomerons \eqref{eq:Aa1a2}. It will be convenient to consider the more general case of two gravitons instead of dilatons, with polarizations $\epsilon_i^{\mu\nu}=\epsilon_i^{\mu}\epsilon_i^{\nu}$, and obtain the dilaton amplitudes in the very end by replacing $\epsilon_i^{\mu\nu}$ with $\eta^{\mu\nu}$.
By using explicitly transverse three-point amplitudes and the completeness relation \eqref{eq:completeness_relation}, we can write tree-level unitarity \eqref{eq:residue_generic}
in the form
	\beq
		\underset{k^2=-4n/\alpha'}{\Res} A^{12P_1 P_2} (k, q_{12})
		= \left( \frac{(-\alpha'q_{12}/2)_n}{n!} \right)^2 (\epsilon_1\cdot \epsilon_2)^2
		= \sum\limits_{\rho,i} A^{15P_1}_{n,\r,i,\mathbf{m}}
\pi^{\mathbf{m},\mathbf{n}}_{\rho}
		A^{52P_2}_{n,\r,i,\mathbf{n}} \,,
		\label{eq:A12_cut}
	\eeq
where for the massive levels, on which we will mostly focus, $\rho$ is summed over irreducible representations of $SO(4)$ and $i$ is summed over degenerate states in the same representation.

Our starting point will be the open string three-point amplitudes of a massless vector, a Pomeron and an arbitrary massive state up to mass level 2 (we give some simpler explicit examples in Appendix \ref{sec:examples}). These amplitudes were computed in \cite{DAppollonio:2013mgj} by studying string-brane scattering.
Since in flat space there is no interaction between the left- and right-moving string modes, the closed string amplitudes factorize into products of open string amplitudes.
We can indeed check that the square root of the residues \eqref{eq:A12_cut} matches the expansion in terms of the open string three-point amplitudes of \cite{DAppollonio:2013mgj}
	\beq
		\sqrt{\underset{k^2=-4n/\alpha'}{\Res} A^{12P_1P_2} (k, q_{12})} = \frac{(-\alpha'q_{12}/2)_n}{n!} (\epsilon_1\cdot \epsilon_2)
		= \sum\limits_{\rho_L} A^{15P_1}_{n,\r_L,\balpha} 
\pi^{\balpha,\bgamma}_{\rho_L}
		A^{52P_2}_{n,\r_L,\bgamma} \,.
		\label{eq:A12_cut_chiral}
	\eeq
We did this consistency check for the first three mass levels, for which $\r_L$ is summed over
the bosonic part (NS sector) of the chiral superstring spectrum in 10 dimensions, given by \cite{Hanany:2010da}
	\bea
		n&=0:&&\qquad\ydiagram{1}_{\,8}\,,\\
		n&=1:&&\qquad\ydiagram{2}_{\,9} \oplus \, \ydiagram{1,1,1}_{\,9}\,,\\
		n&=2:&&\qquad\ydiagram{3}_{\,9} \oplus \, \ydiagram{2,1,1}_{\,9}
		\oplus \, \ydiagram{2,1}_{\,9}
		\oplus \, \ydiagram{1,1}_{\,9}
		\oplus \, \ydiagram{1}_{\,9}\,.
	\eea{eq:chiral_spectrum_10d}
In order to obtain three-point amplitudes for closed strings in $10D$, we need to square \eqref{eq:A12_cut_chiral} and expand again in irreducible representations.
The first step is trivial
	\beq
		\underset{k^2=-4n/\alpha'}{\Res} A^{12P_1 P_2} (k, q_{12})
		= \sum\limits_{\rho_L,\rho_R} 
A^{15P_1}_{n,\r_L,\balpha} \ A^{15P_1}_{n,\r_R,\bbeta} \
\pi^{\balpha,\bgamma}_{\rho_L} \ \pi^{\bbeta,\bdelta}_{\rho_R} \ 
		A^{52P_2}_{n,\r_L,\bgamma} \ A^{52P_2}_{n,\r_R,\bdelta}\,,
		\label{eq:A12_cut_chiral_squared}
	\eeq
however expanding this into irreducible representations requires some more work.
On an abstract level this is easily done in terms of the tensor product
\beq
\r_L \otimes \r_R = \bigoplus\limits_{\rho_C} \rho_C\,,
\label{eq:tensor_product}
\eeq
which can be computed explicitly in terms of characters using e.g.\ the WeylCharacterRing implementation in SageMath \cite{sagemath}. For example, the closed string spectrum for the first two mass levels is
	\begin{align}
		&n=0:\ \ydiagram{1}_{\,8}\otimes \, \ydiagram{1}_{\,8} = \ydiagram{2}_{\,8}\oplus \ydiagram{1,1}_{\,8}\oplus \bullet \,,\label{eq:closed_string_spectrum_10d}\\
		&n=1:\ \left(\ydiagram{2}_{\,9} \oplus \, \ydiagram{1,1,1}_{\,9}\right)^{2} = 
		     \ydiagram{2,2,2}_{\,9}
		\oplus \,      \ydiagram{2,2,1,1}_{\,9}
		\oplus \, 2 \, \ydiagram{3,1,1}_{\,9}
		\oplus \, 3 \, \ydiagram{2,1,1,1}_{\,9}
		\oplus \,      \ydiagram{4}_{\,9}\nonumber\\
		&\oplus \,      \ydiagram{3,1}_{\,9}
		\oplus \, 2 \, \ydiagram{2,2}_{\,9}
		\oplus \,      \ydiagram{2,1,1}_{\,9}
		\oplus \,      \ydiagram{1,1,1,1}_{\,9}
		\oplus \, 2 \, \ydiagram{2,1}_{\,9}
		\oplus \, 3 \, \ydiagram{1,1,1}_{\,9}
		\oplus \, 2 \, \ydiagram{2}_{\,9}
		\oplus \, 2 \, \ydiagram{1,1}_{\,9}
		\oplus \, 2 \, \bullet
		\,.\nonumber
	\end{align}
	
To use the tensor product in explicit calculations requires considerably more work and can be done by formulating \eqref{eq:tensor_product} as an equation in terms of projectors to irreducible representations
\beq
\pi_{\rho_L}^{\balpha;\bgamma}\pi_{\rho_R}^{\bbeta;\bdelta}
= \sum_{\rho_C \subset \rho_L \otimes \rho_R}
p^{\balpha \bbeta}_{\rho_L \otimes \rho_R \to \rho_C,\bmu}
\pi_{\rho_C}^{\bmu;\bnu}
p^{\bgamma \bdelta}_{\rho_L \otimes \rho_R \to \rho_C,\bnu}\,.
\label{eq:general_tensorproduct_identity}
\eeq
The tensors $p_{\rho_L \otimes \rho_R \to \rho_C}$ are constructed from Kronecker deltas and are uniquely determined by this equation. By inserting \eqref{eq:general_tensorproduct_identity} into \eqref{eq:A12_cut_chiral_squared}
we find the expansion of the residue
	\beq
		\underset{k^2=-4n/\alpha'}{\Res} A^{12P_1 P_2} (k, q_{12})
		= \sum\limits_{\rho_L,\rho_R,\rho_C} A^{15P_1}_{n,\rho_L \otimes \rho_R \to \rho_C,\bmu}
\pi^{\bmu,\bnu}_{\rho_C}
		A^{52P_2}_{n,\rho_L \otimes \rho_R \to \rho_C,\bnu} \,,
		\label{eq:A12_cut_closed_string}
	\eeq
in terms of the closed string amplitudes
\beq
A^{15P_1}_{n,\rho_L \otimes \rho_R \to \rho_C,\bmu}
= A^{15P_1}_{n,\r_L,\balpha} \ A^{15P_1}_{n,\r_R,\bbeta} \
p^{\balpha \bbeta}_{\rho_L \otimes \rho_R \to \rho_C,\bmu}\,.
\label{eq:closed_string_3pt_10d}
\eeq

The final step is to restrict the indices of the amplitudes to five dimensions and expand once again into irreducible representations, this times for the massive Little group $SO(4)$.
In terms of representation theory, this is done by using branching rules to expand the $SO(9)$ representations in terms of irreps of the product $SO(4) \times SO(5)$,
\beq
\r_C = \bigoplus\limits_{(\r,\s) \subset \r_C} (\r,\s)\,,
\eeq
where, for massive levels, $\r$ is an irreducible representation of $SO(4)$ and $\s$ of $SO(5)$. Since we consider Pomeron exchange, which caries the vacuum quantum numbers, 
we project onto the singlets of $SO(5)$
\beq
\r_C |_{\bullet_5} = \bigoplus\limits_{(\r,\bullet) \subset \r_C} (\r,\bullet)\,.
\label{eq:branching}
\eeq
 This step is also abstractly implemented in SageMath. For example, we have
\beq
\ydiagram{2}_{\,9} = 
\Big(\ydiagram{2}_{\,4} , \bullet_{5}  \Big) \oplus
\Big(\ydiagram{1}_{\,4} , \ydiagram{1}_{\,5}  \Big) \oplus
\Big(\bullet_{4} , \ydiagram{2}_{\,5}  \Big) \oplus
\Big(\bullet_{4} ,  \bullet_{5} \Big)\,.
\eeq
and after projection to $SO(5)$ singlets
\beq
\ydiagram{2}_{\,9} \Big|_{\bullet_{5}} = \ydiagram{2}_{\,4} \oplus \bullet_{4}\,.
\label{eq:branching_example_spin2}
\eeq
In this way we find the  $SO(5)$ singlets for the closed string spectrum in terms of $SO(3)$ or $SO(4)$ irreps for the first two levels
	\begin{align}
		n&=0:&\ & \ydiagram{2}_{\,3}\oplus \, \ydiagram{1}_{\,3} \, \oplus \,\,2\,\bullet \,,
		\nonumber\\
		n&=1:&\ &
		\ydiagram{4}_{\,4}
		\oplus \,      \ydiagram{3,1}_{\,4}
		\oplus \, 2 \, \ydiagram{2,2}_{\,4}
		\oplus \, 2 \, \ydiagram{3}_{\,4}
		\oplus \, 4 \, \ydiagram{2,1}_{\,4}
		\oplus \, 8 \, \ydiagram{2}_{\,4}
		\nonumber\\
		&&&\oplus \, 5 \, \ydiagram{1,1}_{\,4}
		\oplus \, 10 \, \ydiagram{1}_{\,4}
		\oplus \, 9 \, \bullet_{\,4}\,.
		\label{eq:closed_string_spectrum_5d}
	\end{align}

As for the tensor product, we can rephrase \eqref{eq:branching} as an equation in terms of projectors.
In this case we get an equation for the $SO(9)$ projector with indices restricted to the $SO(4)$ directions $a=1,\ldots,4$
\beq
\pi_{\rho_C}^{\mathbf{a};\mathbf{b}}
= \sum_{\rho \subset \rho_C |_{\bullet_5}}
b^{\mathbf{a}}_{\rho_C \to \rho,\mathbf{m}} \ 
\pi_{\rho}^{\mathbf{m};\mathbf{n}} \ 
b^{\mathbf{b}}_{\rho_C \to \rho,\mathbf{n}}\,,
\label{eq:general_branching_identity}
\eeq
where the tensors $b_{\rho_C \to \rho}$ are uniquely determined by this equation and can be expressed in terms of Kronecker deltas and the $SO(4)$ Levi-Civita symbol.
Since we are assuming the flat space limit kinematics to be restricted to five dimensions, we can simply insert this into \eqref{eq:A12_cut_closed_string} and obtain the residue in the anticipated form \eqref{eq:A12_cut}
	\beq
		\underset{k^2=-4n/\alpha'}{\Res} A^{12P_1 P_2} (k, q_{12})
		= \sum\limits_{\rho_L,\rho_R,\rho_C,\rho} A^{15P_1}_{n,\rho_L \otimes \rho_R \to \rho_C \to \rho,\mathbf{m}}
\pi^{\mathbf{m},\mathbf{n}}_{\rho}
		A^{52P_2}_{n,\rho_L \otimes \rho_R \to \rho_C \to \rho,\mathbf{n}} \,,
		\label{eq:A12_cut_closed_string_5d}
	\eeq
with the $5D$ closed string amplitudes given by
\beq
A^{15P_1}_{n,\rho_L \otimes \rho_R \to \rho_C \to \rho,\mathbf{m}} =
A^{15P_1}_{n,\rho_L \otimes \rho_R \to \rho_C,\mathbf{a}}  
b^{\mathbf{a}}_{\rho_C \to \rho,\mathbf{m}}\,.
\label{eq:closed_string_3pt_5d}
\eeq

\subsubsection{Example}
\label{sec:3pt_example}

Let us now give a specific example of the procedure outlined above. We will consider the following chain of expansions at mass level 1, starting from the product of two open string massive spin 2 fields that 
give rise to a $5D$ scalar 
\beq
\ydiagram{2}_{\,9} \otimes \ydiagram{2}_{\,9} \to \ydiagram{2}_{\,9} \to \bullet_{4}\,.
\eeq
In this example we discard the $5D$ massive spin 2 field that also appears in the projection (\ref{eq:branching_example_spin2}).
We alert the reader that whenever we write explicit amplitudes they are neither appropriately symmetrized nor traceless in order to write them more compactly. All explicit amplitudes should be understood as objects to be contracted with the projector for the associated representation.

We start with the open string amplitude for the state $(1,\ydiagram{2}_{\,9})$ from \cite{DAppollonio:2013mgj}
	\beq
A^{15P_1}_{1,[2]_9,\a_1 \a_2}
		=-\sqrt{\frac{\alpha'}{2}}\left(\epsilon_{1 \alpha_1} q_{1 \alpha_2} + \frac{1}{2} (q_1 \cdot \epsilon_1) \, v_{\alpha_1}v_{\alpha_2} \right) \,.
	\eeq
Squaring this amplitude produces the following closed string states
\beq
\ydiagram{2}_{\,9} \otimes \ydiagram{2}_{\,9} =
\ydiagram{4}_{\,9} \oplus
\ydiagram{3,1}_{\,9} \oplus
\ydiagram{2,2}_{\,9} \oplus
\ydiagram{1,1}_{\,9} \oplus
\ydiagram{2}_{\,9} \oplus
\bullet_{\,9}\,.
\eeq
To construct the relation \eqref{eq:general_tensorproduct_identity} we need the projectors for all the representations in this list. They can be found in \cite{Costa:2016hju,Costa:2018mcg}, however here we just remind the reader of two of the most familiar ones
\beq
\pi_{[2]_d}^{\mu_1 \mu_2,\nu_1 \nu_2}
= \frac{1}{2}\big(\eta^{\mu_1 \nu_1}\eta^{\mu_2 \nu_2}+\eta^{\mu_1 \nu_2}\eta^{\mu_2 \nu_1} \big)-\frac{1}{d} \, \eta^{\mu_1 \mu_2}\eta^{\nu_1 \nu_2}\,, \qquad
\pi_{\bullet} = 1\,,
\label{eq:projectors_example}
\eeq
and state that the closed string state $\ydiagram{2}_{\,9}$ comes in \eqref{eq:general_tensorproduct_identity} with the tensor
\beq
p^{\a_1 \a_2 \b_1 \b_2}_{[2]_9 \otimes [2]_9 \to [2]_9,\mu_1 \mu_2} = \sqrt{\frac{36}{91}} \, \pi_{[2]_9}^{\alpha_1 \alpha_2;\gamma_1 \gamma_2}\pi_{[2]_9}^{\beta_1 \beta_2;\delta_1 \delta_2}\eta_{\gamma_2 \delta_1} \eta_{\gamma_1\mu_1}\eta_{\delta_1\mu_2} \,,
\eeq
where we introduced additional projectors contracted with metrics in order to have the correct index properties. This determines the following closed string amplitude via \eqref{eq:closed_string_3pt_10d}
\bea
A^{15P_1}_{1,[2]_9 \otimes [2]_9 \to [2]_9,\mu_1 \mu_2} &= 
\frac{\alpha '}{4 \sqrt{91}}
\Big[ \big(q_{1,\mu_1} (\epsilon _{1,\mu_2} q_1\cdot \epsilon _1+3 q_{1,\mu_2} )
\\
&
+\epsilon
_{1,\mu_1} (q_{1,\mu_2} q_1\cdot \epsilon _1-3 t_1 \epsilon _{1,\mu_2} )+v_{\mu_1} v_{\mu_2}
(q_1\cdot \epsilon _1){}^2\big) \Big]\,.
\eea{eq:closed_string_3pt_10d_ex}
The branching rule for $\ydiagram{2}_{\,9}$ was already considered in \eqref{eq:branching_example_spin2}. Using the projectors \eqref{eq:projectors_example} it is easy to see that we can write explicitly
\beq
\pi_{[2]_9}^{a_1 a_2,b_1 b_2} = 
\de^{a_1}_{m_1} \de^{a_2}_{m_2} \pi_{[2]_4}^{m_1 m_2,n_1 n_2} \de^{b_1}_{n_1} \de^{b_2}_{n_2}
+ \frac{5}{36} \, \de^{a_1 a_2} \pi_{\bullet} \de^{b_1 b_2}\,,
\eeq
from which we read off
\beq
b^{a_1 a_2}_{[2]_9 \to [2]_4,m_1 m_2} = \de^{a_1}_{m_1} \de^{a_2}_{m_2}\,, \qquad
b^{a_1 a_2}_{[2]_9 \to \bullet_4} = \sqrt{\frac{5}{36}} \, \de^{a_1 a_2}\,.
\label{eq:b_tensor_ex}
\eeq
Finally, inserting \eqref{eq:closed_string_3pt_10d_ex} and \eqref{eq:b_tensor_ex} into \eqref{eq:closed_string_3pt_5d} we compute the $5D$ closed string amplitude
\beq
A^{15P_1}_{1,[2]_9 \otimes [2]_9 \to [2]_9 \to \bullet_4} = \frac{1}{8} \sqrt{\frac{5}{91}} \alpha ' \left( (q_1\cdot \epsilon _1)^2-2 t_1\right)
\label{eq:3pt_example_graviton}
\eeq
We derive the complete list of such level 1  three-point amplitudes  in appendices \ref{sec:tensorprodprojs} and \ref{sec:branchingprojs}.

 \subsection{Constraints on spinning AdS amplitudes}
 \label{sec:constraints_spinning_amplitude}
 In this section we use the flat-space string amplitudes to constrain the high-energy, tree-level AdS$_5$ amplitudes with two dilatons and two spinning operators $\langle \phi \phi \mathcal{O}_5 \mathcal{O}_6\rangle$. Since the operators in question are of stringy nature (i.e. the bulk fields have $m^2 \sim 4n/\alpha'$ for a very large AdS radius),  their dimensions grow with the 't Hooft coupling ($\Delta\sim \lambda^{1/4}$) and transform in the $SO(4)$ representations discussed above in the flat space case. This means that generically $\mathcal{O}_5$ and $\mathcal{O}_6$ are in bosonic mixed-symmetry representations.

As discussed in section \ref{sec:matching_impact_parameter}, the spectral functions that determine the spinning AdS correlators \eqref{eq:Btree_differential} via \eqref{eq:Dfrak} are determined in the flat space limit by the three-point amplitudes \eqref{eq:3pt_spinning}
\beq
 \lambda^{\frac{j(\nu)-1}{2}} \lambda^{\frac{|\rho_5|-k_5}{4}}\lambda^{\frac{|\rho_6|-k_6}{4}}\beta^{k_5,k_6}_{(\De_5,\rho_5),(\De_6,\rho_6)} (\nu)
 \ \to\ 
 \frac{1}{2 \pi}\,
 a_{m_5,\rho_5}^{k_5}(t)  \, a_{m_6,\rho_6}^{k_6}(t) \b(t)\,.
\eeq
In other words, the leading term in $\nu$ at each order in $\lambda$ is fixed by the flat space expression (see section \ref{eq:constraining_from_flat_space_limit})
\beq
\lambda^{\frac{j(\nu)-1}{2}} \lambda^{\frac{|\rho_5|-k_5}{4}}\lambda^{\frac{|\rho_6|-k_6}{4}}\beta^{k_5,k_6}_{(\De_5,\rho_5),(\De_6,\rho_6)} (\nu)
= \frac{1}{2 \pi} a_{m_5,\rho_5}^{k_5}(t)  \,a_{m_6,\rho_6}^{k_6}(t) \b(t) \Big|_{\a't = - \frac{\nu^2}{\sqrt{\lambda}}} + \ldots
\,,
\eeq
where $\ldots$ are terms that vanish in the flat space limit.
Let us now use this to study some specific examples. We begin with the example of section \ref{sec:3pt_example} where we computed an amplitude involving a graviton, a Pomeron and a particular scalar at mass level 1 in \eqref{eq:3pt_example_graviton}.
In this result $\epsilon_{1,\mu}$ is such that $\epsilon_{\mu\nu}=\epsilon_{1,\mu}\epsilon_{1,\nu}$ parametrizes a general graviton polarization which must be replaced by the metric $\epsilon_{\mu \nu}\rightarrow \eta_{\mu\nu}$ to obtain the dilaton amplitude
 \beq
A_{1,[2]_9\otimes[2]_9 \rightarrow[2]_9\rightarrow \bullet_4}^{D 5 P_1}=
 a_{4/\alpha',[2]_9\otimes[2]_9 \rightarrow[2]_9\rightarrow \bullet_4}^{0}(t)= -\frac{3}{8}\sqrt{\frac{5}{91}} \, \a't \,,
 \eeq
 where we used that $(q_1\cdot\epsilon_1)^2=-t_1$ for the dilaton case.
 Thus, for the AdS correlator of this particular scalar and three dilatons we have
 \beq 
 \beta^{0,0}_{(2 \l^{\frac{1}{4}} + \ldots,\bullet),(4,\bullet)}(\nu) =
\frac{3}{8}\sqrt{\frac{5}{91}} \frac{\nu^2}{\sqrt{\lambda}}  \, \beta(\nu)
+ \text{vanishing in the flat space limit}\,.
 \eeq
Note that with respect to the case of four dilaton scattering we have an extra power of $1/\sqrt{\lambda}$, so the term of order $\l^0$ is absent, confirming that tidal excitations are suppressed at large $\lambda$, which in turn agrees with the considerations of \cite{Meltzer:2019pyl} (we verified this fact for all amplitudes at level 1).
  In particular, this is consistent with the large $\lambda$ suppression of $c_{\phi_1 \phi_2 j(\nu)}$ for non-identical scalars, since our stringy mode is certainly different from the dilaton. Such a suppression is not a priori obvious from writing a bulk interaction between two different scalars and a spin $J$ field (to be Sommerfeld-Watson transformed into a Pomeron), which makes this a non-trivial realization of the bounds derived in \cite{Costa:2017twz,Meltzer:2017rtf}.
This example is particularly simple, since there is a unique three-point structure in the case of the scalar.

More generally, we can consider amplitudes with several tensor structures constructed from $v_a$'s and $q_a$'s (equivalently, $\hat{p}$ and $\nabla_p$ in AdS). Let us take as a representative example, the case of the spin 4 operator at level 1. This operator is typically used to define $\Delta_{\text{gap}}$ and sits in the leading Regge trajectory. The corresponding graviton-Pomeron-spin 4 amplitude was worked out in Appendices \ref{sec:tensorprodprojs} and \ref{sec:branchingprojs}, and reads
\beq
A_{[2]_9\otimes[2]_9 \rightarrow[4]_9 \rightarrow[4]_4}^{a_1 a_2 a_3 a_4}(\epsilon_1)=\frac{1}{8} \alpha ' \big(2 \epsilon _{1,a _1} q_{1,a _2}+v_{a _1} v_{a _2} q_1\cdot \epsilon _1\big)
\big(2 \epsilon _{1,a _3} q_{1,a _4}+v_{a _3} v_{a _4} q_1\cdot \epsilon _1\big)\,.
\eeq 
We again emphasize that it is understood that the amplitude should be contracted with $\pi_{[4]_4}^{\balpha;\bbeta}$. Furthermore, we are using the simplifying transverse kinematics discussed above. This means that upon doing the dilaton replacement, we have
\beq
A_{[2]_9\otimes[2]_9 \rightarrow[4]_9 \rightarrow[4]_4}^{a_1 a_2 a_3 a_4}=\frac{\alpha '}{8} \big(-t_1 v_{a_1}v_{a_2}v_{a_3}v_{a_4} +4v_{a_1}v_{a_2}q_{1,a_3}q_{1,a_4} \big)\,, 
\eeq
where we used the symmetry of the indices and note that transverse kinematics ensures that the term proportional to $\epsilon_{1,a_1}\epsilon_{1,a_2}$ gets mapped to a transverse metric which is annihilated by the projector to $[4]_4$.
In this case we can directly use \eqref{eq:flat_space_limit_ts} to match both tensor structures at once
\bea
\mathfrak{D}^{(2 \l^{\frac{1}{4}} + \ldots,[4]_4),(4,\bullet)}_{a_1 a_2 a_3 a_4} (\nu)
={}& \frac{\beta(\nu)}{8\sqrt{\l}} \big( \nu^2 \hat{p}_{a_1} \hat{p}_{a_2} \hat{p}_{a_3} \hat{p}_{a_4} 
+ 4 \hat{p}_{a_1} \hat{p}_{a_2} \nabla_{p\, a_3} \nabla_{p\, a_4}\big)\\
&+  \text{vanishing in flat space limit}
\,.
\eea{eq:Dfrac_example}
We again note that these corrections are suppressed at large $\lambda$.


\section{Conclusions}
\label{sec:conclusions}
In this work, we  derived a perturbative CFT optical theorem which computes the dDisc of a correlator in the $1/N$ expansion in terms of single discontinuities of lower order correlators. Notably, this allows the determination of double-trace contributions to a given one-loop holographic correlator, even when the intermediate fields have spin, which makes them much harder to handle  using unitarity formulas in terms of the CFT data. This also clarifies the underlying CFT principles behind cutting formulas for AdS Witten diagrams, which so far used bulk quantities \cite{Meltzer:2019nbs,Meltzer:2020qbr}.

Using the perturbative CFT optical theorem we fixed the form of the AdS one-loop four-point scattering amplitude in the high-energy limit, accounting for the physical effect of tidal excitations.  This corresponds to box Witten diagrams with two-Pomeron exchange and general string fields as intermediate states. To do this, we transformed the optical theorem to CFT impact parameter space, in which the loop level phase shift is obtained as a contraction of tree-level phase shifts.
Using the general structure of spinning correlators in the s-channel Regge limit, we rewrote all the tidal excitations in terms of a single scalar function, the AdS vertex function.

For the case of ${\cal N}=4$ SYM, dual to type IIB strings, we fixed part of the answer by relating our expression to the flat space results of \cite{Amati:1987wq,Amati:1987uf,Amati:1988tn} for high energy string scattering, requiring consistency with the flat space limit in impact parameter space. This procedure fixes part of the AdS vertex function and therefore also part of the CFT correlation function at one-loop in the Regge limit.
Additionally, interpreting the previous result in terms of unitarity, we used the flat-space behavior  to constrain the spectral function for certain spinning CFT correlators  at tree level in the Regge limit.

There are several open directions and applications of this work.
First, we emphasize that the CFT optical theorem is quite general and does not rely on AdS ingredients.
Moreover,   it works directly at the level of correlators instead of having to extract the CFT data, which is very difficult to resum into correlators. It would be interesting to test and use this formula for more general holographic correlators and, since the expansion parameter does not necessarily need to be $1/N$, in weakly coupled CFTs such as $\phi^4$ theory at the Wilson-Fisher fixed point in $4-\epsilon$ dimensions. 

Another playground to apply our gluing formula  is $\mathcal{N}=4$ SYM at weak t'Hooft coupling in the Regge limit. One could try to derive the order $1/N^4$ correlator
explicitly  at leading order in $\lambda$, using the techniques introduced in \cite{Cornalba:2008qf}. The corresponding double discontinuity should be the square 
of order  $1/N^2$ correlators with impact factors that include the intermediate states. 

In the Regge limit there are kinematical conditions in the CFT optical theorem that simplified the integrations over Lorentzian configurations. An interesting generalization would be to systematically study kinematic corrections to the Regge limit. In fact, in the recent work \cite{Caron-Huot:2020nem} the authors derived a Regge expansion for the correlator valid for any boost. It would be interesting to see how to incorporate this into our analysis, both in a general structural way, but also potentially to impose specific constraints from the flat space limit in a more general kinematic setup. More generally, it would be interesting to understand the Regge limit integrations in terms of light-ray operators \cite{Kravchuk:2018htv}, and to use the more general Lorentzian machinery of \cite{Simmons_Duffin_2018,Kravchuk:2018htv} to write an intrinsically Lorentzian gluing formula in general kinematics.

A possible extension of this work is to consider a higher number of bulk loops. This was analyzed in the large $\Delta_{\text{gap}}$ limit in \cite{Meltzer:2019pyl}. Let us give a few concrete ideas for the stringy generalization of that analysis.
The leading contribution in the Regge limit at $k-1$ loops is expected to be $k$-Pomeron exchange, related to a $k$-fold product of tree-level phase shifts.
By repeatedly using \eqref{eq:Omega_prod} one can define a generalization of the function $\Phi$ for such a product, so we expect that the contribution of intermediate states can again be expressed by a vertex function\footnote{This corresponds to eikonalization in the operator sense of \cite{Amati:1987uf} where the phase shift is an operator in the string Hilbert space, with matrix elements between all possible string states.}
\bea
-\Re \mathcal{B}_{k-1}(S,L) = \int\limits_{-\infty}^\infty d\nu  \left(\prod\limits_{n=1}^k d\nu_n \, \beta^{(*)}(\nu_n)  \right)
&V(\nu_1,\ldots,\nu_k,\nu)^2 \,S^{\sum_m j(\nu_m)-k}\\
&\Phi(\nu_1,\ldots,\nu_k,\nu) \,\Omega_{i \nu} (L)\,,
\eea{eq:Bk_result}
where the product of $\beta(\nu_n)$ must be real, which means that the answer is slightly different depending on whether the number of loops is even or odd \cite{Meltzer:2019pyl}.
In order to find the flat space limit of this $(k-1)$-loop vertex function we would need to consider the flat space result for higher loops, which is known at least in integral form \cite{Amati:1987wq,Amati:1987uf,Amati:1988tn} 
\begin{equation}
\label{eq:nloopVfunctn}
V_k(q_1,\dots,q_k)= \int \prod_{i=1}^{k} \frac{d \sigma_i}{2 \pi} \prod_{1\leq j <l \leq k} |e^{i \sigma_j}- e^{i \sigma_l}|^{ \alpha' q_{l}\cdot q_j}\,.
\end{equation}
Note that the symmetry of the integrand is such that only $k-1$ integrals are non-trivial. For example, the one-loop case just gives
\beq
V_2(q_1,q_2)= \int \frac{d \sigma_1 d \sigma_2}{(2\pi)^2} \,  |e^{i \sigma_1}- e^{i \sigma_2}|^{\alpha'q_{12}}=
 \int \frac{d \sigma}{2\pi}|1- e^{i \sigma} \, |^{\alpha'q_{12}}= \frac{\Gamma\big(1+\alpha'\frac{t_1+t_2-t}{2}\big)}{\Gamma \big(1+\alpha'\frac{t_1+t_2-t}{4}\big)^2},
\eeq
recovering \eqref{eq:V_flat}.


\section{Acknowledgements}
\label{sec:ack}

We would like to thank David Meltzer for discussions on conformal Regge theory and Fernando Alday for comments on the draft. 
This research received funding from the Simons Foundation grants 488637  (Simons collaboration on the Non-perturbative bootstrap).
Centro de F\'\i sica do Porto is partially funded by Funda\c c\~ao para a Ci\^encia e a Tecnologia (FCT) under the grant
UID-04650-FCUP.
AA is funded by FCT under the IDPASC doctoral program with the fellowship  PD/BD/\allowbreak 135436/2017.
TH received funding from the Knut and Alice Wallenberg Foundation grant KAW 2016.0129, the VR grant 2018-04438 and the European Research Council (ERC) under the European Union’s Horizon
2020 research and innovation programme (Grant No.\ 787185).

\appendix

\section{Additional examples of string amplitudes}
\label{sec:examples}
In this appendix we provide some additional examples and comments on the chiral and closed string amplitudes.
\subsection{Chiral Amplitudes}

In the chiral case it is trivial to reproduce level 0. Here we have only the massless particles with residue
	\beq
		\sqrt{\underset{k^2=0}{\Res} A^{12P_1P_2} (k, q_{12})}  = \epsilon_1^{\mu} A^{15P_1}_{\mu \alpha} \pi_{[1]_8}^{\alpha;\beta} A^{52P_2}_{\beta \nu}\epsilon_2^{\nu}=\epsilon_1^{\mu} \eta_{\mu \alpha} \eta^{\alpha \beta}\eta_{\beta \nu} \epsilon_2^\nu= \epsilon_1\cdot \epsilon_2\,,
	\eeq
where we have used $A^{15P_1}_{\mu \alpha}= \eta_{\mu \alpha}$ and $\pi_{[1]_8}^{\alpha;\beta}=\eta^{\alpha \beta}$. At higher levels we will have non-trivial three-point functions. It will be convenient to absorb the external polarization into the amplitude
	\beq
		\epsilon_1^{\mu} A^{15P_1}_{n,\, \mu,\rho_5, \bnu} \equiv A^{15P_1}_{n,\rho_5, \bnu}\,,
	\eeq
to be more compact in writing the amplitudes (we are using the integer $n$ to label the mass level of the state). 

From the spectrum described above, we will have two amplitudes at level 1 which are $A^{15P_1}_{1,[1,1,1]_9,\balpha}$ and $A^{15P_1}_{1,[2]_9,\balpha}$.\\
 Here we 
will  keep in mind the Young diagrams explained above, along with the index symmetrization that comes with them, packaged in our boldface multi-index notation.
The explicit level 1 three point amplitudes in the IIB superstring are
	\beq
		A^{15P_1}_{1,[1,1,1]_9,\balpha}= \frac{\sqrt{6}}{m_1} \epsilon_{1 \alpha_1}q_{1 \alpha_2} v_{\alpha_3}~,~ A^{15P_1}_{1,[2]_9,\balpha}
		=-\sqrt{\frac{\alpha'}{2}}\left(\epsilon_{1 \alpha_1} q_{1 \alpha_2} + \frac{1}{2} (q_1 \cdot \epsilon_1) v_{\alpha_1}v_{\alpha_2} \right) \,,
	\eeq
with $m_1=2/\sqrt{\alpha'}$ being the mass at level 1, and $q_1$ is the transverse momentum carried by the Pomeron $P_1$ (similarly for $q_2$ and the Pomeron $P_2$). All the relative factors between the different tensor structures and the overall normalization are fixed by computing the three-point amplitudes with the correctly normalized vertex operators for the excited NS states in IIB super string theory \cite{DAppollonio:2013mgj}. We can now contract the three-point amplitudes on each side using the projector for the appropriate representation and check the residue
	\bea
		\sqrt{\underset{k^2=-4/\alpha'}{\Res} A^{12P_1P_2} (k, q_{12})}&= A^{15P_1}_{1,\balpha} 
		\pi_{[1,1,1]_9}^{\balpha;\bbeta} A^{52P_1}_{1,\bbeta}
		+A^{15P_2}_{1,\balpha} \pi_{[2]_9}^{\alpha;\bbeta} A^{52P_2}_{1,\bbeta}\\
		&=\frac{\alpha'}{2} (-q_1\cdot q_2)(\epsilon_1\cdot \epsilon_2),
	\eea{eq:level1_res_flat}
where we refrained from writing the representation labels in the amplitudes since they are contracted with a projector with the appropriate label.	
This matches what we extracted from $A^{\alpha_1 \alpha_2}(q_{12})$, or equivalently from the vertex function. We can continue this procedure to the second level, where mixed symmetry tensors appear for the first time. For example, the $[2,1,1]_9$ tensor has the amplitude
	\beq
		A^{15P_1}_{2,[2,1,1]_9,\balpha}= \sqrt{\frac{3}{8}} \sqrt{\frac{\alpha'}{2}} 
		\left(\frac{4}{m_2}q_{1 \alpha_2} +2 v_{\alpha_2}\right) \epsilon_{1 \alpha_1} q_{1 \alpha_3}v_{\alpha_4},
	\eeq
where $m_2=\sqrt{8/\alpha'}$ is the mass at level 2. It is important to emphasize that the level 2 amplitude contains a term with more powers of $\alpha'$ than any of the level 1 amplitudes. This would lead to further suppression in $1/\sqrt{\lambda}$ in the AdS theory.\footnote{Clearly, states with higher spin, which can only appear at higher levels, can have higher powers of $\alpha'$ leading to a spin-dependent suppression of couplings, as is expected from the general arguments of \cite{Costa:2017twz,Meltzer:2017rtf}.}

The remaining amplitudes can be found in section 5 of \cite{DAppollonio:2013mgj}. For our purposes it is just important to know that the amplitudes satisfy
	\bea
		\sqrt{\underset{k^2=-8/\alpha'}{\Res} A^{12P_1P_2} (k, q_{12})}&= 
		\sum_{\rho \in S}A^{15P_1}_{2,\rho,\balpha}  \pi_{\rho}^{\balpha;\bbeta}A^{52P_2}_{2,\rho,\bbeta} = \frac{(-\frac{\alpha'}{2}q_1\cdot q_2)_2}{2!} (\epsilon_1\cdot \epsilon_2)\,,\\ 
		S & =\left\lbrace[3]_9 ~,~ [2,1,1]_9 ~,~[2,1]_9 ~,~[1,1]_9 ~,~[1]_9 \right\rbrace  ,
	\eea{eq:level2_res_flat}
which we explicitly checked. More generally, we can conclude that the square root of the residue at mass level $n$ of $A^{12P_1P_2}(k,q_{12})$ can be recovered by unitarity if we account for all the covariant SO(9) representations corresponding to the massive NS states. This gives a microscopic interpretation for the vertex function at a given mass level. As already mentioned, summing over all these mass levels reconstructs the full vertex function.

\subsection{Closed string amplitudes}
Here we consider the simple but instructive level 0 case for the closed string amplitudes, where the little group is SO(8). The square of the residue reads
	\begin{align}
\label{eq:level0_closed_so8}
		{}&A^{15P_1}_{L1, \alpha}A^{15P_1}_{R1,\beta} \; \left( \pi_{[1]_8}^{\alpha ;\gamma}\pi_{[1]_8}^{\beta ;\delta} \right) \;  A^{52P_2}_{L1,\gamma}A^{52P_2}_{R1\delta}     \\
		&= \hspace{-.4cm} \sum_{\rho_C=[2]_8,[1,1]_8,\bullet_8} \hspace{-.4cm}   A^{15P_1}_{L1, \alpha}A^{15P_1}_{R1,\beta}(
		p^{\alpha \beta}_{[1]_8 \otimes [1]_8 \to \rho_C,\mu_1 \mu_2}
		\pi_{\rho_C}^{\mu_1 \mu_2;\nu_1 \nu_2}
		p^{\gamma \delta}_{[1]_8 \otimes [1]_8 \to \rho_C,\nu_1 \nu_2})A^{52P_2}_{L1,\gamma}A^{52P_2}_{R1\delta}
		= (\epsilon_1\cdot \epsilon_2)^2  \,,
\nonumber
\end{align}
where we used the group theoretical tensor product identity for projectors
	\beq
	 	\pi_{[1]_8}^{\alpha ;\gamma}\pi_{[1]_8}^{\beta ;\delta} \; = \sum_{\rho_C=[2]_8,[1,1]_8,\bullet_8} p^{\alpha \beta}_{[1]_8 \otimes [1]_8 \to \rho_C,\mu_1 \mu_2}
	 	\pi_{\rho_C}^{\mu_1 \mu_2;\nu_1 \nu_2}
	 	p^{\gamma \delta}_{[1]_8 \otimes [1]_8 \to \rho_C,\nu_1 \nu_2}\,.
	\eeq
We can solve this equation for the tensors $p$ by contracting with polarization vectors for the left and right modes on both sides of the projector ($z_L,z_R$ and $\zb_L,\zb_R$) and equating the polynomials in scalar products of $z$'s. In practice we will always use this procedure, or a similar one where we contract with amplitudes to fix coefficients. In this case it is trivial to directly check that
	\begin{align}
		\pi_{[1]_8}^{\alpha ;\gamma}\pi_{[1]_8}^{\beta ;\delta} \; 
		&= \; \eta^{\alpha \gamma}\eta^{\beta \delta}	\;\equiv \;  \pi_{[2]_8}^{\alpha \beta;\gamma \delta} + \pi_{[1,1]_8}^{\alpha \beta;\gamma \delta} + \frac{1}{8} \eta^{\alpha \beta} \eta^{\gamma \delta}
	\label{eq:level0_closed_check}
	       \\
		& = \left(\frac{1}{2}( \eta^{\alpha \gamma}\eta^{\beta \delta}+\eta^{\alpha \delta}\eta^{\beta \gamma})-\frac{1}{8} \eta^{\alpha \beta}\eta^{\gamma \delta}\right) 
		+ \frac{1}{2}(\eta^{\alpha \gamma}\eta^{\beta \delta}-\eta^{\alpha \delta}\eta^{\beta \gamma}) + \frac{1}{8}(\eta^{\alpha \beta}\eta^{\gamma \delta})\,,
	\nonumber
\end{align}
where, obviously $p^{\alpha \beta}_{[1]_8 \otimes [1]_8 \to [2]_8,\mu_1 \mu_2}=\delta^\alpha_{\mu_1}\delta^\beta_{\mu_2} ~,~p^{\alpha \beta}_{[1]_8 \otimes [1]_8 \to [1,1]_8,\mu_1 \mu_2}=\delta^\alpha_{\mu_1}\delta^\beta_{\mu_2}$ and $p^{\alpha \beta}_{[1]_8 \otimes [1]_8 \to \bullet_8}= \sqrt{\frac{1 }{8}}\eta^{\alpha\beta}$ extracts traces, thereby projecting to a singlet state.


\section{Tensor products for projectors}
\label{sec:tensorprodprojs}

In this appendix we explain how to realize the tensor product of open string states into closed string states in terms of the corresponding projectors/tensors. We will consider mass level $n=1$. The chiral spectrum at this level is
\beq
n=1:\quad \ydiagram{2}_{\,9} \oplus \, \ydiagram{1,1,1}_{\,9}\,.
\eeq
We square the irreps using the tensor product as in the main text and analyze the decomposition term by term. For example, taking $\rho_L= \rho_R = [2]_9$ corresponds to the tensor product
\beq
\ydiagram{2}_{\,9} \otimes \, \ydiagram{2}_{\,9}= 
\ydiagram{4}_{\,9}
\oplus \,      \ydiagram{3,1}_{\,9}
\oplus \, \ydiagram{2,2}_{\,9}
\oplus \,  \ydiagram{2}_{\,9}
\oplus \,      \ydiagram{1,1}_{\,9}
\oplus \,      \bullet_{\,9}\,,
\eeq
which we want to write in terms of SO(9) tensors as
\bea
\pi_{[2]_9}^{\balpha;\bgamma}\pi_{[2]_9}^{\bbeta;\bdelta}&= \sum_{\rho_C \in S}p^{\balpha \bbeta}_{[2]_9 \otimes [2]_9 \to \rho_C,\bmu}
\pi_{\rho_C}^{\bmu;\bnu}
p^{\bgamma \bdelta}_{[2]_9 \otimes [2]_9 \to \rho_C,\bnu}\,,\\
S&=\left\lbrace [4]_9 \,, [3,1]_9 \,, [2,2]_9\,, [2]_9\,, [1,1]_9\,, \bullet_9\right\rbrace\,. 
\eea{eq:level1_squared_proj}
It will be useful to manifestly symmetrize the $\balpha$ and $\bbeta$ indices of the tensors $p$, in order to write down these tensors more compactly. Therefore, we will use
\beq
p^{\balpha \bbeta}_{\rho_L \otimes \rho_R \to \rho_C,\bmu} \equiv \pi^{\balpha}_{\rho_L;\balpha'} \pi^{\bbeta}_{\rho_R;\bbeta'} \tilde{p}^{\balpha' \bbeta'}_{\rho_L \otimes \rho_R \to \rho_C,\bmu},
\label{eq:symmetrize_factors}
\eeq
and we will present the simpler trial tensors $\tilde{p}$ for each case.
 In this case $\rho_L=\rho_R=[2]_9$ and we used the trial tensors\footnote{For irreps with the same number of indices as the tensor product we will always take $\tilde{p}^{\balpha \bbeta}_{\bmu} \propto \delta^{\balpha \bbeta}_{\bmu}$.}
\begin{align}
\label{eq:tensorprods_2t2}
\tilde{p}^{\alpha_1\alpha_2\beta_1 \beta_2}_{[2]_9 \otimes [2]_9 \to [4]_9,\mu_1 \mu_2\mu_3\mu_4}&= \delta^{\alpha_1}_{\mu_1} \delta^{\alpha_2}_{\mu_2}\delta^{\beta_1}_{\mu_3}\delta^{\beta_2}_{\mu_4} \equiv \delta^{\balpha \bbeta}_{\bmu}
\,,&&&
\tilde{p}^{\balpha\bbeta}_{[2]_9 \otimes [2]_9 \to [3,1]_9,\bmu}&= \delta^{\balpha \bbeta}_{\bmu}\,,\\
\tilde{p}^{\balpha\bbeta}_{[2]_9 \otimes [2]_9 \to [2,2]_9,\bmu}&= \sqrt{3} \delta^{\balpha \bbeta}_{\bmu}\,, &&&
\tilde{p}^{\alpha_1\alpha_2\beta_1 \beta_2}_{[2]_9 \otimes [2]_9 \to [2]_9,\mu_1 \mu_2}&    = \sqrt{\frac{36}{91}} \delta^{\alpha_1}_{\mu_1} \delta^{\beta_2}_{\mu_2} \eta^{\alpha_2 \beta_1}\,,
\nonumber\\
\tilde{p}^{\alpha_1\alpha_2\beta_1 \beta_2}_{[2]_9 \otimes [2]_9 \to [1,1]_9,\mu_1 \mu_2}&= \sqrt{\frac{4}{11}} \delta^{\alpha_1}_{\mu_1} \delta^{\beta_2}_{\mu_2} \eta^{\alpha_2 \beta_1}\,, &&& \tilde{p}^{\alpha_1\alpha_2\beta_1 \beta_2}_{[2]_9 \otimes [2]_9 \to \bullet_9}&= \sqrt{\frac{1}{44}} \eta^{\alpha_1\beta_1} \eta^{\alpha_2 \beta_2}\,.
\nonumber
\end{align}
The remaining tensor products have some additional subtleties. Taking the cross term in the tensor product
\beq
\ydiagram{2}_{\,9} \otimes \, \ydiagram{1,1,1}_{\,9}= 
\ydiagram{1,1,1}_{\,9}
\oplus \,      \ydiagram{2,1}_{\,9}
\oplus \, \ydiagram{2,1,1,1}_{\,9}
\oplus \,  \ydiagram{3,1,1}_{\,9}\,,
\eeq
we note that there is a 4 row tensor appearing. When contracted with the amplitudes, this contribution will vanish, because we have only 3 independent vectors. However, from the point of view of the projector equation, we must still have 
\bea
\pi_{[2]_9}^{\balpha;\bgamma}\pi_{[1,1,1]_9}^{\bbeta;\bdelta}&= \sum_{\rho_C \in S'}p^{\balpha \bbeta}_{[2]_9 \otimes [1,1,1]_9 \to \rho_C,\bmu}
\pi_{\rho_C}^{\bmu;\bnu}
p^{\bgamma \bdelta}_{[2]_9 \otimes [1,1,1]_9 \to \rho_C,\bnu}\,,\\
S'&=\left\lbrace [3,1,1]_9 \,, [1,1,1]_9 \,, [2,1]_9\,, [2,1,1,1]_9\right\rbrace\,. 
\eea{eq:tensorprod_2_111_proj}
with a non-vanishing $p^{\balpha \bbeta}_{[2]_9 \otimes [1,1,1]_9 \to [2,1,1,1]_9,\bmu}$. However, by contracting 
directly with the amplitudes $A^{\balpha}_{[2]_9}A^{\bbeta}_{[1,1,1]_9}A^{\bgamma}_{[2]_9}A^{\bdelta}_{[1,1,1]_9}$ we automatically eliminate the contribution of this tensor (this also avoids the computation of a complicated 4 row projector). With this in mind, we use again \eqref{eq:symmetrize_factors}, with the trial projectors
\bea
\tilde{p}^{\balpha \bbeta}_{[2]_9 \otimes [1,1,1]_9 \to [3,1,1],\bmu} &= \sqrt{\frac{5}{4}} \delta^{\alpha_1}_{\mu_1} \delta^{\alpha_2}_{\mu_2}\delta^{\beta_1}_{\mu_3}\delta^{\beta_2}_{\mu_4}\delta^{\beta_3}_{\mu_5} \equiv \sqrt{\frac{5}{4}} \delta^{\balpha \bbeta}_{\bmu}\,, \\ \tilde{p}^{\balpha \bbeta}_{[2]_9 \otimes [1,1,1]_9 \to [1,1,1],\bmu} &= \sqrt{\frac{585}{352}} \eta^{\alpha_2 \beta_1} \delta^{\alpha_1}_{\mu_1}\delta^{\beta_2}_{\mu_2}\delta^{\beta_3}_{\mu_3}\,,\,
\tilde{p}^{\balpha \bbeta}_{[2]_9 \otimes [1,1,1]_9 \to [2,1],\bmu} =  \eta^{\alpha_2 \beta_1} \delta^{\alpha_1}_{\mu_1}\delta^{\beta_2}_{\mu_2}\delta^{\beta_3}_{\mu_3}\,.
\eea{eq:prod_2_111_trial}
With these tensors and setting $p^{\balpha \bbeta}_{[2]_9 \otimes [1,1,1]_9 = [2,1,1,1]_9,\bmu}\to0$, which suffices for our purposes, we have that the identity (\ref{eq:tensorprod_2_111_proj}) holds, but only when inserted between the amplitudes.

The remaining tensor product
\bea
\ydiagram{1,1,1}_{\,9} \otimes \, \ydiagram{1,1,1}_{\,9}={}& 
\ydiagram{2,2,1,1}_{\,9}
\oplus \,      \ydiagram{2,1,1,1}_{\,9}
\oplus \, \ydiagram{1,1,1,1}_{\,9}
\oplus \,  \ydiagram{2,2,2}_{\,9}\\
&\oplus \,  \ydiagram{2,1,1}_{\,9}
\oplus \,  \ydiagram{2,2}_{\,9}
\oplus \,  \ydiagram{1,1,1}_{\,9}
\oplus \,  \ydiagram{2}_{\,9}
\oplus \,  \ydiagram{1,1}_{\,9}
\oplus \,  \bullet_{\,9}\,,
\eea{eq:tensor_111_111}
can be dealt with similarly, by contracting with only 3 independent polarizations instead of six, eliminating the contributions of the complicated 4 row tensors. The only non-vanishing contributions to projectors turn out to be
\bea
\pi_{[1,1,1]_9}^{\balpha;\bgamma}\pi_{[1,1,1]_9}^{\bbeta;\bdelta}&= \sum_{\rho_C \in S''}p^{\balpha \bbeta}_{[1,1,1]_9 \otimes [1,1,1]_9 \to \rho_C,\bmu}
\pi_{\rho_C}^{\bmu;\bnu}
p^{\bgamma \bdelta}_{[1,1,1]_9 \otimes [1,1,1]_9 \to \rho_C,\bnu}\,,\\
S''&=\left\lbrace [2,2,2]_9 \,, [2,2]_9 \,, [2]_9\,, \bullet_9\right\rbrace\,.
\eea{eq:tensor_111_111_proj}
where we used \eqref{eq:symmetrize_factors} and the trial projectors
\bea
\tilde{p}^{\balpha \bbeta}_{[1,1,1]_9 \otimes [1,1,1]_9 \to [2,2,2]_9,\bmu} &= \sqrt{8} \delta^{\alpha_1}_{\mu_1} \delta^{\alpha_2}_{\mu_2}\delta^{\alpha_3}_{\mu_3}\delta^{\beta_1}_{\mu_4}\delta^{\beta_2}_{\mu_5}\delta^{\beta_3}_{\mu_6} \equiv \sqrt{8} \delta^{\balpha \bbeta}_{\bmu}\,,\\
\tilde{p}^{\balpha \bbeta}_{[1,1,1]_9 \otimes [1,1,1]_9 \to [2,2]_9,\bmu}& = \sqrt{\frac{12}{5}} \eta^{\alpha_1 \beta_1} \delta^{\alpha_2}_{\mu_1} \delta^{\beta_3}_{\mu_2}\delta^{\alpha_3}_{\mu_3}\delta^{\beta_2}_{\mu_4} \,,\\
\tilde{p}^{\balpha \bbeta}_{[1,1,1]_9 \otimes [1,1,1]_9 \to [2]_9,\bmu} &= \sqrt{\frac{3}{7}} \eta^{\alpha_1 \beta_1} \eta^{\alpha_2 \beta_3} \delta^{\alpha_3}_{\mu_1}\delta^{\beta_2}_{\mu_2}\,,\\
\tilde{p}^{\balpha \bbeta}_{[1,1,1]_9 \otimes [1,1,1]_9 \to \bullet_9} &= \sqrt{\frac{1}{84}} \eta^{\alpha_1 \beta_1} \eta^{\alpha_2 \beta_2}\eta^{\alpha_3 \beta_3} \,,
\eea{eq:naive_proj_111_111}
where the unusual index ordering is to ensure that the resulting tensor doesn't vanish when we act with $\pi_{[1,1,1]_9}$ on the trial projectors $\tilde{p}$. 
This turns (\ref{eq:tensor_111_111_proj}) into an identity which holds for two identical $[1,1,1]_9$ tensors, as will be the case for our amplitudes. 


\section{Branching relations for projectors}
\label{sec:branchingprojs}

In this Appendix we provide a detailed account of all the branching relations for closed string state projectors utilized in section \ref{sec:massive_tree_flat}.
Let us start by recalling the SO(9) closed string states at level 1
\begin{align}
&\left(\ydiagram{2}_{\,9} \oplus \, \ydiagram{1,1,1}_{\,9}\right)^{2} = 
      \ydiagram{2,2,2}_{\,9}
\oplus \,      \ydiagram{2,2,1,1}_{\,9}
\oplus \, 2 \, \ydiagram{3,1,1}_{\,9}
\oplus \, 3 \, \ydiagram{2,1,1,1}_{\,9}
\oplus \,      \ydiagram{4}_{\,9}
\oplus \,      \ydiagram{3,1}_{\,9}
\nonumber\\
&\oplus \, 2 \, \ydiagram{2,2}_{\,9}
\oplus \,      \ydiagram{2,1,1}_{\,9}
\oplus \,      \ydiagram{1,1,1,1}_{\,9}
\oplus \, 2 \, \ydiagram{2,1}_{\,9}
\oplus \, 3 \, \ydiagram{1,1,1}_{\,9}
\oplus \, 2 \, \ydiagram{2}_{\,9}
\oplus \, 2 \, \ydiagram{1,1}_{\,9}
\oplus \, 2 \, \bullet
\,.
\label{eq:closed_states_level_1}
\end{align}
We recall that states with more than 3 columns will not contribute as we only have 3 different vectors to anti-symmetrize. We are going to perform the branching
\beq
\text{SO(9)} ~\rightarrow~ \text{SO(4)}\times \text{SO(5)}|_\bullet \,,
\eeq
where we denote the projection to singlets of SO(5) by $|_\bullet$. Note that certain representations can naturally be restricted to SO(4) by simply taking the 5d subset of 10d indices. It is obvious that
\beq
\ydiagram{1,1}_{\,9}\,\,\rightarrow\,\, \ydiagram{1,1}_{\,4}\,,\qquad \bullet_{\,9} \,\, \rightarrow \,\, \bullet_{\,4}\,.
\eeq
Additionally, the projector for these representations is identical for SO(9) and SO(4) (up to restriction of the indices $\alpha \to a \,,\, \beta \to b \,,\, \mu \to m \,,\, \dots$)
\beq
\pi_{[1,1]_9}^{a_1a_2;b_1b_2} = \pi_{[1,1]_4}^{a_1a_2;b_1b_2}= \frac{1}{2}\left(\eta^{a_1 b_1}\eta^{a_2 b_2}-\eta^{a_1 b_2}\eta^{a_2 b_1} \right) \,. 
\eeq
Other representations admit a direct restriction, but can also give additional irreps, by the creation of SO(5) singlets, through the contraction of indices with legs on the compact manifold. For example the spin 2 states branch as
\beq
\ydiagram{2}_{\,9}\,\,\rightarrow\,\, \ydiagram{2}_{\,4} \,\, \oplus \,\, \bullet_{\,4}\,.
\eeq
The spin 2 on the RHS is interpreted as the restriction of indices to the SO(4) and the singlet as a trace over the compact space indices. In terms of projectors the statement is simply
\beq
\pi_{[2]_9}^{a_1a_2;b_1b_2} = \pi_{[2]_4}^{a_1a_2;b_1b_2}+ \frac{5}{36}\eta^{a_1 a_2}\eta^{b_1 b_2}\,.
\eeq
Similarly, for the spin 4 case
\beq
\ydiagram{4}_{\,9}\,\,\rightarrow\,\, \ydiagram{4}_{\,4} \,\, \oplus \,\, \ydiagram{2}_{\,4} \,\, \oplus \,\,\bullet_{\,4},
\eeq
and the projector equation is
\beq
\pi_{[4]_9}^{\mathbf{a};\mathbf{b}}
= \sum_{\rho =[4]_4,[2]_4,\bullet_4}
b^{\mathbf{a}}_{[4]_9 \to \rho,\mathbf{m}} \ 
\pi_{\rho}^{\mathbf{m};\mathbf{n}} \ 
b^{\mathbf{b}}_{[4]_9 \to \rho,\mathbf{n}}\,.
\eeq
It will be again convenient to manifestly symmetrize the tensors, in order to present them more compactly. We define
\beq
b^{\mathbf{a}}_{\rho_C \to \rho,\mathbf{m}}\equiv \pi^{\mathbf{a}}_{\rho_C, \mathbf{a}'} \tilde{b}^{\mathbf{a'}}_{\rho_C \to \rho,\mathbf{m}}\,,
\label{eq:branching_tensors_symmetrization}
\eeq
and then present a list of the simpler $\tilde{b}$. In this case we have
\bea
\tilde{b}^{a_1 a_2 a_3 a_4}_{[4]_9 \to [4]_4,m_1 m_2 m_3 m_4}&= \delta^{a_1}_{m_1}\delta^{a_2}_{m_2}\delta^{a_3}_{m_3}\delta^{a_4}_{m_4}\equiv \delta^ {\mathbf{a}}_{\mathbf{m}}\\
\tilde{b}^{a_1 a_2 a_3 a_4}_{[4]_9 \to [2]_4,m_1 m_2 }&= \sqrt{\frac{39}{20}}\delta^{a_1}_ {m_1}\delta^{a_2}_ {m_2}\eta^{a_3 a_4} ~,~ \tilde{b}^{a_1 a_2 a_3 a_4}_{[4]_9 \to \bullet_4}
  = \sqrt{\frac{143}{280}}\eta^{a_1a_2}\eta^{a_3 a_4}\,.
\eea{eq:branch4}
The fact that the direct restriction of the irrep $[4]_9 \rightarrow[4]_4$ comes with coefficient 1 is a non-trivial consistency check of the previous procedure.

There are other irreps that don't admit a direct restriction, because they have more than two columns (SO(4) Young tableaux have at most two columns, and traces can vanish by antisymmetry). For this we need to use the SO(4) Levi-Civita tensor. We will simply write it as $\varepsilon^{a_1a_2a_3a_4}$. The simplest case is the 3-form
\beq
\ydiagram{1,1,1}_{\,9}\,\,\rightarrow\,\, \ydiagram{1}_{\,4}\,,
\eeq
and the corresponding projector equation is
\beq
\pi_{[1,1,1]_9}^{\mathbf{a};\mathbf{b}} =\frac{1}{6} \varepsilon^{a_1 a_2 a_3}_{\,\,\,\,\qquad m} \pi_{[1]_4}^{m;n} \, \varepsilon^{\,\,\,b_1 b_2 b_3}_{n}= 4\, \pi_{[1,1,1,1]_4}^{\mathbf
{a} m;\mathbf{b} n}\,\pi_{[1]_4,m;n}\,,
\label{eq:111_to_1}
\eeq
From the first equation we can read off
\beq
b^{a_1 a_2 a_3}_{[1,1,1]_9\to[1]_4,m}= \sqrt{\frac{1}{6}}\varepsilon^{a_1 a_2 a_3}_{\,\,\,\,\qquad m}\,.
\eeq
For the second equality in (\ref{eq:111_to_1}) we used the standard identity
\beq
\epsilon^{a_1a_2a_3a_4}\epsilon^{b_1b_2b_3b_4}= 4!\, \pi_{[1,1,1,1]_4}^{\mathbf{a};\mathbf{b}}\,.
\eeq
This is convenient to square the three-point amplitudes when computing the residue of the four-point function $A^{12P_1P_2}$.
Using trace subtractions and products of epsilon tensors we can now derive branching identities for all the relevant irreps. Let us list the remaining identities, where we write the trace subtractions and the Levi-Civita tensors using the trial projectors $\tilde{b}$, but then appropriately symmetrize them through (\ref{eq:branching_tensors_symmetrization})
\bea
\ydiagram{2,1,1}_{\,9}\,\,&\rightarrow\,\, \ydiagram{2}_{\,4} \,\, \oplus \,\,  \ydiagram{1,1}_{\,4}\,,\\
\tilde{b}^{a_1 a_2 a_3 a_4}_{[2,1,1]_9\to[2]_4,m_1 m_2}&=\sqrt{\frac{6}{4!}}  \varepsilon^{a_1a_3a_4}_{\qquad m_1}\delta^{a_2}_{m_2} ~,~ \tilde{b}^{a_1 a_2 a_3 a_4}_{[2,1,1]_9\to[1,1]_4,m_1 m_2}=\sqrt{\frac{42}{4!5}} \varepsilon^{a_1a_3a_4}_{\,\,\,\,\qquad m_1} \delta^{a_2}_{m_2}\,,\\
\ydiagram{2,1}_{\,9}\,\,&\rightarrow\,\, \ydiagram{2,1}_{\,4} \,\, \oplus \,\,  \ydiagram{1}_{\,4}\,,\\
\tilde{b}^{\mathbf{a
}}_{[2,1]_9\to[2,1]_4,\mathbf{m}}&= \delta^{\mathbf{a}}_{\mathbf{m}} ~,~ \tilde{b}^{a_1 a_2 a_3}_{[2,1]_9\to[1]_4,m} = \sqrt{\frac{16}{5}} \delta^{a_1}_{m_1} \eta^{a_2 a_3}\,,\\
\eea{eq:branching1}

\bea
\ydiagram{3,1,1}_{\,9}\,\,&\rightarrow\,\, \ydiagram{3}_{\,4} \,\, \oplus \,\,  \ydiagram{2,1}_{\,4}\,\, \oplus \,\,  \ydiagram{1}_{\,4}\,,\\
\tilde{b}^{\mathbf{a}}_{[3,1,1]_9\to[3]_4,\mathbf{m}} &= \sqrt{\frac{36}{4!5}} \varepsilon^{a_1a_4a_5}_{\,\,\,\, \qquad m_1} \delta^ {a_2}_{m_2} \delta^ {a_3}_{m_3} ~,~ \tilde{b}^{\mathbf{a}}_{[3,1,1]_9\to[2,1]_4,\mathbf{m}} = \sqrt{\frac{1152}{4!25}} \varepsilon^{a_1a_4a_5}_{\,\,\,\, \qquad m_1} \delta^ {a_2}_{m_2} \delta^ {a_3}_{m_3}\,,\\
\tilde{b}^{a_1 a_2 a_3 a_4 a_5}_{[3,1,1]_9\to[1]_4,m} &= \sqrt{\frac{22}{4! 5}} \varepsilon^{a_1a_4a_5}_{\,\,\,\, \qquad m} \eta^{a_2 a_3}\,,\\
\ydiagram{3,1}_{\,9}\,\,&\rightarrow\,\, \ydiagram{3,1}_{\,4} \,\, \oplus \,\,  \ydiagram{1,1}_{\,4}\,\, \oplus \,\,  \ydiagram{2}_{\,4}\,,\\
\tilde{b}^{\mathbf{a
	}}_{[3,1]_9\to[3,1]_4,\mathbf{m}}&= \delta^{\mathbf{a}}_{\mathbf{m}}, \tilde{b}^{\mathbf{a
}}_{[3,1]_9\to[1,1]_4,\mathbf{m}}= \sqrt{\frac{11}{10}} \delta^{a_1}_{m_1} \eta^{a_2 a_3} \delta^{a_4}_{m_2}, \tilde{b}^{\mathbf{a
}}_{[3,1]_9\to[2]_4,\mathbf{m}}= \sqrt{\frac{27}{10}} \delta^{a_1}_{m_1} \eta^{a_2 a_4} \delta^{a_3}_{m_2},\\
\ydiagram{2,2}_{\,9}\,\,&\rightarrow\,\, \ydiagram{2,2}_{\,4} \,\, \oplus \,\,  \ydiagram{2}_{\,4}\,\, \oplus \,\, \bullet_{\,4}\,,\\
\tilde{b}^{\mathbf{a
	}}_{[2,2]_9\to[2,2]_4,\mathbf{m}}&= \delta^{\mathbf{a}}_{\mathbf{m}} ~,~
\tilde{b}^{\mathbf{a
	}}_{[2,2]_9\to[2]_4,\mathbf{m}}= \sqrt{\frac{42}{5}}  \delta^{a_1}_{m_1} \eta^{a_2 a_3} \delta^{a_4}_{m_2} ~,~\tilde{b}^{\mathbf{a
}}_{[2,2]_9\to \bullet_4}= \sqrt{\frac{7}{5}}\eta^{a_1 a_4}\eta^{a_2 a_3} \,,\\
\ydiagram{2,2,2}_{\,9}\,\,&\rightarrow\,\, \ydiagram{2}_{\,4} \,\, \oplus \,\, \bullet_{\,4}\,,\\
\tilde{b}^{\mathbf{a
	}}_{[2,2,2]_9\to[2]_4,m_1 m_2}&= \frac{\sqrt{48}}{4!} \varepsilon^ {a_1 a_5 a_3}_{\,\,\,\,\qquad m_1} \varepsilon^ {a_4 a_2 a_6}_{\,\,\,\,\qquad m_2} ~,~
\tilde{b}^{\mathbf{a
		}}_{[2,2,2]_9\to\bullet_4}= \frac{\sqrt{28}}{4!} \varepsilon^ {a_1 a_5 a_3}_{\,\,\,\,\qquad m} \varepsilon^ {a_4 a_2 a_6 m}  \,.
\eea{eq:listofbranchings}
Note that for the last diagram, which has more than 2 boxes in both columns, we are forced to use two pairs of epsilon tensors. Tensors with more than three rows aren't allowed by the 10d kinematics, but even if they were, their branchings do not contain singlets of SO(5) so we can simply discard them.
\subsection{All 5d closed string amplitudes}
\label{sec:all_5d_closed}
 Let us first write down in generality the 5d amplitudes using the relations derived in the main text. We have
\beq
A^{15P_1}_{n,\rho_L \otimes \rho_R \to \rho_C \to \rho,\mathbf{m}} = 
 A^{15P_1}_{\r_L,\balpha} \ A^{15P_1}_{\r_R,\bbeta} \
p^{\balpha \bbeta}_{\rho_L \otimes \rho_R \to \rho_C,\mathbf{a}}
b^{\mathbf{a}}_{\rho_C \to \rho,\mathbf{m}}\,.
\eeq
With all the group theory identities established, we can enumerate all the amplitudes used in the main text to reproduce the residue at the cut with mass level 1. However, we again emphasize that we have not explicitly symmetrized the amplitudes by contracting with the respective projector, in order to maintain some compactness of the tables below. Namely, all amplitudes are to be contracted with the projector to the SO(4) irrep, and furthermore, amplitudes where $\text{rank}(\rho)=\text{rank}(\rho_L)+\text{rank}(\rho_R)$ are also not explicitly symmetrized with respect to $\rho_L$ and $\rho_R$ as in equation (\ref{eq:symmetrize_factors}). 
Additionally, for amplitudes where the $b$ tensors contain a Levi-Civita symbol, we write the square of the amplitude
\beq
\left( A_{\mathbf{m}}^{\rho_L\otimes \rho_R \to \rho_C \rightarrow \rho}\right) ^2 \equiv
A_{\mathbf{m}}^{\rho_L\otimes \rho_R \to \rho_C \rightarrow \rho}\pi^ {\mathbf{m};\mathbf{n}}_{\rho}A_{\mathbf{n}}^{\rho_L\otimes \rho_R \to \rho_C \rightarrow \rho}\,,
\eeq
since the amplitude itself cannot be written nicely in terms of $v_\alpha, q_\alpha,\epsilon_\alpha$. 
With these caveats in mind, we list the amplitudes starting by the ones with the most indices\\
\bea
\ydiagram{4}_{\,4}\\
A^{[2]_9\otimes[2]_9 \rightarrow[4]_9 \rightarrow[4]_4}_{m_1 m_2 m_3 m_4}&=
\frac{1}{8} \alpha ' \left(2 \epsilon _{1,m _1} q_{1,m _2}+v_{m _1} v_{m _2} q_1\cdot \epsilon _1\right)\\ &\times
\left(2 \epsilon _{1,m _3} q_{1,m _4}+v_{m _3} v_{m _4} q_1\cdot \epsilon _1\right) \,,\\
&&\\
2 \,\ydiagram{2,2}_{\,4}\\
A^{[2]_9\otimes[2]_9 \rightarrow[2,2]_9 \rightarrow[2,2]_4}_{m_1 m_2 m_3 m_4}&=
\frac{1}{8} \alpha ' \left(2 \epsilon _{1,m _1} q_{1,m _2}+v_{m _1} v_{m _2} q_1\cdot \epsilon _1\right)\\ &\times
\left(2 \epsilon _{1,m _3} q_{1,m _4}+v_{m _3} v_{m _4} q_1\cdot \epsilon _1\right)\,,\\
&&\\
A^{[1,1,1]_9\otimes[1,1,1]_9 \rightarrow[2,2]_9 \rightarrow[2,2]_4}_{m_1 m_2 m_3 m_4}&=\frac{\alpha'}{4 \sqrt{15}} [ \left(v_{m _1} \epsilon _{1,m _3}-v_{m _3} \epsilon _{1,m _1}\right) \left(q_1\cdot
\epsilon _1 \left(v_{m _2} q_{1,m _4}-v_{m _4} q_{1,m _2}\right)\right. \\
&+t_1 \left.\left(v_{m _2} \epsilon
_{1,m _4}-v_{m _4} \epsilon _{1,m _2}\right)\right)\\&+q_{1,m _1} \left(\epsilon _{1,m _3} \left(\epsilon
_{1,m _2} q_{1,m _4}-\epsilon _{1,m _4} q_{1,m _2}\right)\right.\\
&+\left.v_{m _3} \left(v_{m _4} \left(\epsilon
_{1,m _2} q_1\cdot \epsilon _1-q_{1,m _2}\right)+v_{m _2} \left(q_{1,m _4}-\epsilon _{1,m _4} q_1\cdot
\epsilon _1\right)\right)\right)\\
&+q_{1,m _3} \left(\epsilon _{1,m _1} \left(\epsilon _{1,m _4} q_{1,m
	_2}-\epsilon _{1,m _2} q_{1,m _4}\right)\right.\\
&+\left.v_{m _1} \left(v_{m _4} \left(q_{1,m _2}-\epsilon _{1,m
	_2} q_1\cdot \epsilon _1\right)+v_{m _2} \left(\epsilon _{1,m _4} q_1\cdot \epsilon _1-q_{1,m
_4}\right)\right)\right)]\,,\\
&\\
\ydiagram{3}_{\,4}\\
\left( A_{\mathbf{m}}^{[2]_9\otimes[111]_9 \rightarrow[311]_9 \rightarrow[3]_4}\right)^2 &=-\frac{\left(\alpha '\right)^2}{1152}\left(48 \left(q_1\cdot \epsilon _2\right){}^2 \left(3
	\left(q_2\cdot \epsilon _1\right){}^2+t_2\right)\right.\\&+48 t_1 \left(\left(q_2\cdot \epsilon
	_1\right){}^2+t_2 \left(\epsilon _1\cdot \epsilon _2\right){}^2\right)+48 \left(q_1\cdot
	q_2\right){}^2 \left(1-3 \left(\epsilon _1\cdot \epsilon _2\right){}^2\right)\\&+115
	(q_1\cdot q_2) (\epsilon _1\cdot \epsilon _2) (q_1\cdot \epsilon _1) (q_2\cdot \epsilon _2)\\ &\left.-115
	(q_1\cdot \epsilon _1) (q_1\cdot \epsilon _2) (q_2\cdot \epsilon _1) (q_2\cdot \epsilon
	_2)\right)\,,\\
&\\
2\,\ydiagram{1}_{\,4}\\
\left( A_{\mathbf{m}}^{[2]_9\otimes[111]_9 \rightarrow[311]_9 \rightarrow[1]_4}\right)^2&=-\frac{625 \left(\alpha '\right)^2}{12672} (q_1\cdot \epsilon _1) (q_2\cdot \epsilon _2)\\&\times \left((q_1\cdot
	\epsilon _2) (q_2\cdot \epsilon _1)-(q_1\cdot q_2) (\epsilon _1\cdot \epsilon
	_2)\right)\,,\\
&\\
\left( A_{\mathbf{m}}^{[2]_9\otimes[111]_9 \rightarrow[111]_9 \rightarrow[1]_4}\right)^2&=\frac{65 \left(\alpha '\right)^2}{1408} (q_1\cdot \epsilon _1) (q_2\cdot \epsilon _2)\\&\times \left((q_1\cdot
	q_2) (\epsilon _1\cdot \epsilon _2)-(q_1\cdot \epsilon _2) (q_2\cdot \epsilon _1)\right)\,.
\eea{eq:list_amplitudes_flat}
And
\bea
2\,\ydiagram{2,1}_{\,4}\\
A_{m_1 m_2 m_3}^{[2]_9\otimes[111]_9 \rightarrow[21]_9 \rightarrow[21]_4}&=
\frac{\alpha'}{12
	\sqrt{6}}[ \left(q_{1,m _1} \left(v_{m _3} \left(2 \epsilon _{1,m _2} q_1\cdot \epsilon _1-3 q_{1,m
	_2}\right)\right.\right.\\&+\left.v_{m _2} \left(3 q_{1,m _3}-2 \epsilon _{1,m _3} q_1\cdot \epsilon _1\right)\right)\\&+2 q_1\cdot
\epsilon _1 \left(q_{1,m _3} \left(v_{m _2} \epsilon _{1,m _1}-v_{m _1} \epsilon _{1,m
	_2}\right)\right.\\&+\left.q_{1,m _2} \left(v_{m _1} \epsilon _{1,m _3}-v_{m _3} \epsilon _{1,m _1}\right)\right)\\&+\left.3
t_1 \epsilon _{1,m _1} \left(v_{m _2} \epsilon _{1,m _3}-v_{m _3} \epsilon _{1,\alpha _2}\right)\right)]\,,\\
&\\
\left( A_{\mathbf{m}}^{[2]_9\otimes[111]_9 \rightarrow[311]_9 \rightarrow[21]_4}\right)^2&=\frac{5 \left(\alpha '\right)^2 }{1152}\left(6 t_2 \left(q_1\cdot \epsilon _2\right){}^2+6 t_1
\left(\left(q_2\cdot \epsilon _1\right){}^2+t_2 \left(\epsilon _1\cdot \epsilon
_2\right){}^2\right)\right.\\&+(q_1\cdot q_2) (\epsilon _1\cdot \epsilon _2) (q_1\cdot \epsilon _1)
(q_2\cdot \epsilon _2)\\&-\left.(q_1\cdot \epsilon _1) (q_1\cdot \epsilon _2) (q_2\cdot \epsilon _1)
(q_2\cdot \epsilon _2)+6 \left(q_1\cdot q_2\right){}^2\right)\,,\\
&\\
\eea{eq:21diagrams}
We also have the sixfold degenerate spin 2 states
\bea
6\,\ydiagram{2}_{\,4}\\
A_{m_1 m_2}^{[2]_9\otimes[2]_9\rightarrow[4]_9\rightarrow[2]_4}&= \frac{\alpha '}{16} \sqrt{\frac{5}{39}}  \left[q_{1,m _1} \left(5 \epsilon _{1,m _2} (q_1\cdot \epsilon _1)+2 q_{1,m
	_2}\right)\right.\\ &+\epsilon _{1,m _1} \left(5 q_{1,m _2} (q_1\cdot \epsilon _1)-2 t_1 \epsilon _{1,m _2}\right)\\&\left.+5
v_{m _1} v_{m _2} \left(q_1\cdot \epsilon _1\right){}^2\right]\,,\\
&\\
A_{m_1 m_2}^{[2]_9\otimes[2]_9\rightarrow[2,2]_9\rightarrow[2]_4}&= \frac{\alpha '}{4} \sqrt{\frac{5}{42}}  \left[-q_{1,m _1} \left(q_{1,m _2}-2 \epsilon _{1,m _2} (q_1\cdot \epsilon
_1)\right)\right.\\&+\epsilon _{1,m _1} \left(2 q_{1,m _2} (q_1\cdot \epsilon _1)+t_1 \epsilon _{1,m _2}\right)\\&+\left.2 v_{m
	_1} v_{m _2} \left(q_1\cdot \epsilon _1\right){}^2\right]\,,\\
&\\
A_{m_1 m_2}^{[2]_9\otimes[2]_9\rightarrow[2]_9\rightarrow[2]_4}&= \frac{\alpha '}{4 \sqrt{91}}[ \left(q_{1,m _1} \left(\epsilon _{1,m _2} (q_1\cdot \epsilon _1)+3 q_{1,m _2}\right)\right.\\&+\epsilon
_{1,m _1} \left(q_{1,m _2} (q_1\cdot \epsilon _1)-3 t_1 \epsilon _{1,m _2}\right)\\&+\left.v_{m _1} v_{m _2}
\left(q_1\cdot \epsilon _1\right){}^2\right)]\,,\\
&\\
\eea{spin2part1}
\bea
A_{m_1m_2}^{[1,1,1]_9\otimes[1,1,1]_9\rightarrow[2,2]_9\rightarrow[2]_4}&= \frac{\alpha ' }{2 \sqrt{14}}[\left(q_{1,m _1} \left(q_{1,m _2}-\epsilon _{1,m _2} (q_1\cdot \epsilon _1)\right)\right.\\&+\epsilon _{1,m
	_1} \left(-q_{1,m _2} \left(q_1\cdot \epsilon _1\right)-t_1 \epsilon _{1,m _2}\right)\\&+\left.v_{m _1}
v_{m _2} \left(-\left(q_1\cdot \epsilon _1\right){}^2-t_1\right)\right)]\,,\\
&\\
A_{m_1 m_2}^{[1,1,1]_9\otimes[1,1,1]_9\rightarrow[2]_9\rightarrow[2]_4}&=\frac{\alpha '}{4 \sqrt{21}}[ \left(q_{1,m _1} \left(\epsilon _{1,m _2} (q_1\cdot \epsilon _1)-q_{1,m _2}\right)\right.\\&+\epsilon _{1,m
	_1} \left(q_{1,m _2} (q_1\cdot \epsilon _1)+t_1 \epsilon _{1,m _2}\right)\\&+\left.v_{m _1} v_{m _2}
\left(\left(q_1\cdot \epsilon _1\right){}^2+t_1\right)\right)]\,,\\
&\\
\left( A_{\mathbf{m}}^{[1,1,1]_9\otimes[1,1,1]_9\rightarrow[2,2,2]_9\rightarrow[2]_4}\right)^2 &=\frac{\left(\alpha '\right)^2}{1920} \left(37 t_1 \left(q_2\cdot \epsilon _1\right){}^2-74 t_1
	(\epsilon _1\cdot \epsilon _2) (q_2\cdot \epsilon _1) (q_2\cdot \epsilon _2)\right.\\&\left.+17 \left(q_1\cdot
	\epsilon _1\right){}^2 \left(\left(q_2\cdot \epsilon _2\right){}^2+t_2\right)+17 t_1
	\left(\left(q_2\cdot \epsilon _2\right){}^2+t_2\right)\right.\\&+\left(q_1\cdot \epsilon
	_2\right){}^2 \left(117 \left(q_2\cdot \epsilon _1\right){}^2+37 t_2\right)\\&-74 (q_1\cdot
	\epsilon _1) (q_1\cdot \epsilon _2) \left((q_2\cdot \epsilon _1) (q_2\cdot \epsilon _2)+t_2
	(\epsilon _1\cdot \epsilon _2)\right)\\&+\left(q_1\cdot q_2\right){}^2 \left(117
	\left(\epsilon _1\cdot \epsilon _2\right){}^2-37\right)\\&\left.+q_1\cdot q_2 \left(2 q_1\cdot
	\epsilon _2 \left(37 q_2\cdot \epsilon _2-117 \epsilon _1\cdot \epsilon _2 q_2\cdot
	\epsilon _1\right)\right.\right.\\&\left.\left.+74 q_1\cdot \epsilon _1 \left(q_2\cdot \epsilon _1-\epsilon _1\cdot
	\epsilon _2 q_2\cdot \epsilon _2\right)\right)-37 t_1 t_2 \left(\epsilon _1\cdot
	\epsilon _2\right){}^2\right)\,,
\eea{eq:list_amplitudes_spin2}

and 8-fold degenerate spin 0 states
\bea
8\, \bullet_4\\
A_{[2]_9 \otimes [2]_9 \rightarrow [4]_9 \rightarrow \bullet_4}&=\frac{1}{48} \sqrt{\frac{35}{286}} \alpha ' \left(4 t_1-15 \left(q_1\cdot \epsilon _1\right){}^2\right)\,,\\
&\\
	A_{[2]_9 \otimes [2]_9 \rightarrow [2,2]_9 \rightarrow \bullet_4}&=\frac{1}{24} \sqrt{\frac{5}{7}} \alpha ' \left(3 \left(q_1\cdot \epsilon _1\right){}^2+t_1\right)\,,\\
	&\\
	A_{[2]_9 \otimes [2]_9 \rightarrow [2]_9 \rightarrow \bullet_4}&=\frac{1}{8} \sqrt{\frac{5}{91}} \alpha ' \left(\left(q_1\cdot \epsilon _1\right){}^2-2 t_1\right)\,,\\
	&\\
	A_{[2]_9 \otimes [2]_9 \rightarrow \bullet_9 \rightarrow\bullet_4}&=\frac{\alpha '}{8 \sqrt{11}} \left(\left(q_1\cdot \epsilon _1\right){}^2-t_1\right)\,,\\
	&\\
	A_{[1,1,1]_9 \otimes [1,1,1]_9 \rightarrow [2,2]_9\rightarrow \bullet_4}&=\frac{1}{8} \sqrt{\frac{3}{7}} \alpha ' \left(-\left(q_1\cdot \epsilon _1\right){}^2-t_1\right)\,,\\
	&\\
\eea{scalarspart1}
\bea	
	A_{[1,1,1]_9 \otimes [1,1,1]_9 \rightarrow [2]_9\rightarrow \bullet_4}&=\frac{1}{8} \sqrt{\frac{5}{21}} \alpha ' \left(\left(q_1\cdot \epsilon _1\right){}^2+t_1\right)\,,\\
	&\\
	A_{[1,1,1]_9 \otimes [1,1,1]_9 \rightarrow \bullet_9\rightarrow \bullet_4}&=\frac{\alpha ' }{8 \sqrt{21}}\left(\left(q_1\cdot \epsilon _1\right){}^2+t_1\right)\,,\\
	&\\
	\left( A_{[1,1,1]_9 \otimes [1,1,1]_9 \rightarrow [2,2,2]_9\rightarrow \bullet_4}\right)^2 &=\frac{\left(\alpha '\right)^2}{4480} \left(-49 t_1 \left(q_2\cdot \epsilon _1\right){}^2+98 t_1
	(\epsilon _1\cdot \epsilon _2) (q_2\cdot \epsilon _1) (q_2\cdot \epsilon _2) \right.\\&-49 \left(q_1\cdot
	\epsilon _2\right){}^2 \left(\left(q_2\cdot \epsilon _1\right){}^2+t_2\right)-29
	\left(q_1\cdot \epsilon _1\right){}^2 \left(\left(q_2\cdot \epsilon
	_2\right){}^2+t_2\right)\\&-29 t_1 \left(\left(q_2\cdot \epsilon
	_2\right){}^2+t_2\right)-49
	\left(q_1\cdot q_2\right){}^2 \left(\left(\epsilon _1\cdot \epsilon
	_2\right){}^2-1\right)\\&+98 (q_1\cdot \epsilon _1) (q_1\cdot \epsilon _2) \left((q_2\cdot
	\epsilon _1) (q_2\cdot \epsilon _2)+t_2 (\epsilon _1\cdot \epsilon _2)\right)\\&\left.-98 q_1\cdot q_2 \left(q_1\cdot \epsilon _2 \left(q_2\cdot
	\epsilon _2-(\epsilon _1\cdot \epsilon _2) (q_2\cdot \epsilon _1)\right)\right.\right.\\&\left.\left.+q_1\cdot \epsilon
	_1 \left(q_2\cdot \epsilon _1-(\epsilon _1\cdot \epsilon _2)( q_2\cdot \epsilon
	_2)\right)\right)+49 t_1 t_2 \left(\epsilon _1\cdot \epsilon _2\right){}^2\right)\,. 
\eea{eq:list_amplitudes_scalar}	
Upon contracting each of these amplitudes (with a Pomeron $P_1$) with the appropriate projector and another three-point amplitude (with a Pomeron $P_2$), we recover, upon summing over all 22 states, the level 1 residue of the four-point amplitude, as described in the main text.

\bibliographystyle{JHEP}
\bibliography{regge_1loop}

\providecommand{\href}[2]{#2}\begingroup\raggedright\begin{thebibliography}{10}

\bibitem{Caron_Huot_2017}
S.~Caron-Huot, \emph{{Analyticity in Spin in Conformal Theories}},
  \href{http://dx.doi.org/10.1007/JHEP09(2017)078}{\emph{JHEP} {\bf 09} (2017)
  078}, [\href{http://arxiv.org/abs/1703.00278}{{\tt 1703.00278}}].

\bibitem{Simmons_Duffin_2018}
D.~Simmons-Duffin, D.~Stanford and E.~Witten, \emph{{A spacetime derivation of
  the Lorentzian OPE inversion formula}},
  \href{http://dx.doi.org/10.1007/JHEP07(2018)085}{\emph{JHEP} {\bf 07} (2018)
  085}, [\href{http://arxiv.org/abs/1711.03816}{{\tt 1711.03816}}].

\bibitem{Kravchuk:2018htv}
P.~Kravchuk and D.~Simmons-Duffin, \emph{{Light-ray operators in conformal
  field theory}}, \href{http://dx.doi.org/10.1007/JHEP11(2018)102}{\emph{JHEP}
  {\bf 11} (2018) 102}, [\href{http://arxiv.org/abs/1805.00098}{{\tt
  1805.00098}}].

\bibitem{Lemos:2017vnx}
M.~Lemos, P.~Liendo, M.~Meineri and S.~Sarkar, \emph{{Universality at large
  transverse spin in defect CFT}},
  \href{http://dx.doi.org/10.1007/JHEP09(2018)091}{\emph{JHEP} {\bf 09} (2018)
  091}, [\href{http://arxiv.org/abs/1712.08185}{{\tt 1712.08185}}].

\bibitem{Liendo:2019jpu}
P.~Liendo, Y.~Linke and V.~Schomerus, \emph{{A Lorentzian inversion formula for
  defect CFT}}, \href{http://dx.doi.org/10.1007/JHEP08(2020)163}{\emph{JHEP}
  {\bf 08} (2020) 163}, [\href{http://arxiv.org/abs/1903.05222}{{\tt
  1903.05222}}].

\bibitem{Carmi:2019cub}
D.~Carmi and S.~Caron-Huot, \emph{{A Conformal Dispersion Relation:
  Correlations from Absorption}},
  \href{http://dx.doi.org/10.1007/JHEP09(2020)009}{\emph{JHEP} {\bf 09} (2020)
  009}, [\href{http://arxiv.org/abs/1910.12123}{{\tt 1910.12123}}].

\bibitem{Caron-Huot:2020adz}
S.~Caron-Huot, D.~Mazac, L.~Rastelli and D.~Simmons-Duffin, \emph{{Dispersive
  CFT Sum Rules}},  \href{http://arxiv.org/abs/2008.04931}{{\tt 2008.04931}}.

\bibitem{Bissi:2019kkx}
A.~Bissi, P.~Dey and T.~Hansen, \emph{{Dispersion Relation for CFT Four-Point
  Functions}}, \href{http://dx.doi.org/10.1007/JHEP04(2020)092}{\emph{JHEP}
  {\bf 04} (2020) 092}, [\href{http://arxiv.org/abs/1910.04661}{{\tt
  1910.04661}}].

\bibitem{Alday:2016njk}
L.~F. Alday, \emph{{Large Spin Perturbation Theory for Conformal Field
  Theories}},
  \href{http://dx.doi.org/10.1103/PhysRevLett.119.111601}{\emph{Phys. Rev.
  Lett.} {\bf 119} (2017) 111601}, [\href{http://arxiv.org/abs/1611.01500}{{\tt
  1611.01500}}].

\bibitem{Alday:2016jfr}
L.~F. Alday, \emph{{Solving CFTs with Weakly Broken Higher Spin Symmetry}},
  \href{http://dx.doi.org/10.1007/JHEP10(2017)161}{\emph{JHEP} {\bf 10} (2017)
  161}, [\href{http://arxiv.org/abs/1612.00696}{{\tt 1612.00696}}].

\bibitem{Aharony:2016dwx}
O.~Aharony, L.~F. Alday, A.~Bissi and E.~Perlmutter, \emph{{Loops in AdS from
  Conformal Field Theory}},
  \href{http://dx.doi.org/10.1007/JHEP07(2017)036}{\emph{JHEP} {\bf 07} (2017)
  036}, [\href{http://arxiv.org/abs/1612.03891}{{\tt 1612.03891}}].

\bibitem{Alday:2017vkk}
L.~F. Alday and S.~Caron-Huot, \emph{{Gravitational S-matrix from CFT
  dispersion relations}},
  \href{http://dx.doi.org/10.1007/JHEP12(2018)017}{\emph{JHEP} {\bf 12} (2018)
  017}, [\href{http://arxiv.org/abs/1711.02031}{{\tt 1711.02031}}].

\bibitem{Alday:2017zzv}
L.~F. Alday, J.~Henriksson and M.~van Loon, \emph{{Taming the
  $\epsilon$-expansion with large spin perturbation theory}},
  \href{http://dx.doi.org/10.1007/JHEP07(2018)131}{\emph{JHEP} {\bf 07} (2018)
  131}, [\href{http://arxiv.org/abs/1712.02314}{{\tt 1712.02314}}].

\bibitem{Maldacena:1997re}
J.~M. Maldacena, \emph{{The Large N limit of superconformal field theories and
  supergravity}}, \href{http://dx.doi.org/10.1023/A:1026654312961}{\emph{Int.
  J. Theor. Phys.} {\bf 38} (1999) 1113--1133},
  [\href{http://arxiv.org/abs/hep-th/9711200}{{\tt hep-th/9711200}}].

\bibitem{Witten:1998qj}
E.~Witten, \emph{{Anti-de Sitter space and holography}},
  \href{http://dx.doi.org/10.4310/ATMP.1998.v2.n2.a2}{\emph{Adv. Theor. Math.
  Phys.} {\bf 2} (1998) 253--291},
  [\href{http://arxiv.org/abs/hep-th/9802150}{{\tt hep-th/9802150}}].

\bibitem{Gubser:1998bc}
S.~Gubser, I.~R. Klebanov and A.~M. Polyakov, \emph{{Gauge theory correlators
  from noncritical string theory}},
  \href{http://dx.doi.org/10.1016/S0370-2693(98)00377-3}{\emph{Phys. Lett. B}
  {\bf 428} (1998) 105--114}, [\href{http://arxiv.org/abs/hep-th/9802109}{{\tt
  hep-th/9802109}}].

\bibitem{Liu:2018jhs}
J.~Liu, E.~Perlmutter, V.~Rosenhaus and D.~Simmons-Duffin,
  \emph{{$d$-dimensional SYK, AdS Loops, and $6j$ Symbols}},
  \href{http://dx.doi.org/10.1007/JHEP03(2019)052}{\emph{JHEP} {\bf 03} (2019)
  052}, [\href{http://arxiv.org/abs/1808.00612}{{\tt 1808.00612}}].

\bibitem{Ponomarev:2019ofr}
D.~Ponomarev, \emph{{From bulk loops to boundary large-N expansion}},
  \href{http://dx.doi.org/10.1007/JHEP01(2020)154}{\emph{JHEP} {\bf 01} (2020)
  154}, [\href{http://arxiv.org/abs/1908.03974}{{\tt 1908.03974}}].

\bibitem{Meltzer:2019nbs}
D.~Meltzer, E.~Perlmutter and A.~Sivaramakrishnan, \emph{{Unitarity Methods in
  AdS/CFT}}, \href{http://dx.doi.org/10.1007/JHEP03(2020)061}{\emph{JHEP} {\bf
  03} (2020) 061}, [\href{http://arxiv.org/abs/1912.09521}{{\tt 1912.09521}}].

\bibitem{Meltzer:2020qbr}
D.~Meltzer and A.~Sivaramakrishnan, \emph{{CFT unitarity and the AdS Cutkosky
  rules}}, \href{http://dx.doi.org/10.1007/JHEP11(2020)073}{\emph{JHEP} {\bf
  11} (2020) 073}, [\href{http://arxiv.org/abs/2008.11730}{{\tt 2008.11730}}].

\bibitem{Fitzpatrick:2011dm}
A.~Fitzpatrick and J.~Kaplan, \emph{{Unitarity and the Holographic S-Matrix}},
  \href{http://dx.doi.org/10.1007/JHEP10(2012)032}{\emph{JHEP} {\bf 10} (2012)
  032}, [\href{http://arxiv.org/abs/1112.4845}{{\tt 1112.4845}}].

\bibitem{Cornalba:2006xk}
L.~Cornalba, M.~S. Costa, J.~Penedones and R.~Schiappa, \emph{{Eikonal
  Approximation in AdS/CFT: From Shock Waves to Four-Point Functions}},
  \href{http://dx.doi.org/10.1088/1126-6708/2007/08/019}{\emph{JHEP} {\bf 08}
  (2007) 019}, [\href{http://arxiv.org/abs/hep-th/0611122}{{\tt
  hep-th/0611122}}].

\bibitem{Cornalba:2006xm}
L.~Cornalba, M.~S. Costa, J.~Penedones and R.~Schiappa, \emph{{Eikonal
  Approximation in AdS/CFT: Conformal Partial Waves and Finite N Four-Point
  Functions}},
  \href{http://dx.doi.org/10.1016/j.nuclphysb.2007.01.007}{\emph{Nucl. Phys. B}
  {\bf 767} (2007) 327--351}, [\href{http://arxiv.org/abs/hep-th/0611123}{{\tt
  hep-th/0611123}}].

\bibitem{Gross:1987kza}
D.~J. Gross and P.~F. Mende, \emph{{The High-Energy Behavior of String
  Scattering Amplitudes}},
  \href{http://dx.doi.org/10.1016/0370-2693(87)90355-8}{\emph{Phys. Lett. B}
  {\bf 197} (1987) 129--134}.

\bibitem{Gross:1987ar}
D.~J. Gross and P.~F. Mende, \emph{{String Theory Beyond the Planck Scale}},
  \href{http://dx.doi.org/10.1016/0550-3213(88)90390-2}{\emph{Nucl. Phys. B}
  {\bf 303} (1988) 407--454}.

\bibitem{Amati:1987wq}
D.~Amati, M.~Ciafaloni and G.~Veneziano, \emph{{Superstring Collisions at
  Planckian Energies}},
  \href{http://dx.doi.org/10.1016/0370-2693(87)90346-7}{\emph{Phys. Lett. B}
  {\bf 197} (1987) 81}.

\bibitem{Amati:1987uf}
D.~Amati, M.~Ciafaloni and G.~Veneziano, \emph{{Classical and Quantum Gravity
  Effects from Planckian Energy Superstring Collisions}},
  \href{http://dx.doi.org/10.1142/S0217751X88000710}{\emph{Int. J. Mod. Phys.}
  {\bf A3} (1988) 1615--1661}.

\bibitem{Amati:1988tn}
D.~Amati, M.~Ciafaloni and G.~Veneziano, \emph{{Can Space-Time Be Probed Below
  the String Size?}},
  \href{http://dx.doi.org/10.1016/0370-2693(89)91366-X}{\emph{Phys. Lett. B}
  {\bf 216} (1989) 41--47}.

\bibitem{Cornalba:2007zb}
L.~Cornalba, M.~S. Costa and J.~Penedones, \emph{{Eikonal approximation in
  AdS/CFT: Resumming the gravitational loop expansion}},
  \href{http://dx.doi.org/10.1088/1126-6708/2007/09/037}{\emph{JHEP} {\bf 09}
  (2007) 037}, [\href{http://arxiv.org/abs/0707.0120}{{\tt 0707.0120}}].

\bibitem{Brower:2007qh}
R.~C. Brower, M.~J. Strassler and C.-I. Tan, \emph{{On the eikonal
  approximation in AdS space}},
  \href{http://dx.doi.org/10.1088/1126-6708/2009/03/050}{\emph{JHEP} {\bf 03}
  (2009) 050}, [\href{http://arxiv.org/abs/0707.2408}{{\tt 0707.2408}}].

\bibitem{Brower:2006ea}
R.~C. Brower, J.~Polchinski, M.~J. Strassler and C.-I. Tan, \emph{{The Pomeron
  and gauge/string duality}},
  \href{http://dx.doi.org/10.1088/1126-6708/2007/12/005}{\emph{JHEP} {\bf 12}
  (2007) 005}, [\href{http://arxiv.org/abs/hep-th/0603115}{{\tt
  hep-th/0603115}}].

\bibitem{Cornalba:2007fs}
L.~Cornalba, \emph{{Eikonal methods in AdS/CFT: Regge theory and multi-reggeon
  exchange}},  \href{http://arxiv.org/abs/0710.5480}{{\tt 0710.5480}}.

\bibitem{Costa:2012cb}
M.~S. Costa, V.~Goncalves and J.~Penedones, \emph{{Conformal Regge theory}},
  \href{http://dx.doi.org/10.1007/JHEP12(2012)091}{\emph{JHEP} {\bf 12} (2012)
  091}, [\href{http://arxiv.org/abs/1209.4355}{{\tt 1209.4355}}].

\bibitem{Cornalba:2008qf}
L.~Cornalba, M.~S. Costa and J.~Penedones, \emph{{Eikonal Methods in AdS/CFT:
  BFKL Pomeron at Weak Coupling}},
  \href{http://dx.doi.org/10.1088/1126-6708/2008/06/048}{\emph{JHEP} {\bf 06}
  (2008) 048}, [\href{http://arxiv.org/abs/0801.3002}{{\tt 0801.3002}}].

\bibitem{Brower:2007xg}
R.~C. Brower, M.~J. Strassler and C.-I. Tan, \emph{{On The Pomeron at Large 't
  Hooft Coupling}},
  \href{http://dx.doi.org/10.1088/1126-6708/2009/03/092}{\emph{JHEP} {\bf 03}
  (2009) 092}, [\href{http://arxiv.org/abs/0710.4378}{{\tt 0710.4378}}].

\bibitem{Cornalba:2009ax}
L.~Cornalba, M.~S. Costa and J.~Penedones, \emph{{Deep Inelastic Scattering in
  Conformal QCD}}, \href{http://dx.doi.org/10.1007/JHEP03(2010)133}{\emph{JHEP}
  {\bf 03} (2010) 133}, [\href{http://arxiv.org/abs/0911.0043}{{\tt
  0911.0043}}].

\bibitem{Meltzer:2019pyl}
D.~Meltzer, \emph{{AdS/CFT Unitarity at Higher Loops: High-Energy String
  Scattering}}, \href{http://dx.doi.org/10.1007/JHEP05(2020)133}{\emph{JHEP}
  {\bf 05} (2020) 133}, [\href{http://arxiv.org/abs/1912.05580}{{\tt
  1912.05580}}].

\bibitem{Karateev:2018oml}
D.~Karateev, P.~Kravchuk and D.~Simmons-Duffin, \emph{{Harmonic Analysis and
  Mean Field Theory}},
  \href{http://dx.doi.org/10.1007/JHEP10(2019)217}{\emph{JHEP} {\bf 10} (2019)
  217}, [\href{http://arxiv.org/abs/1809.05111}{{\tt 1809.05111}}].

\bibitem{Ferrara:1972uq}
S.~Ferrara, A.~Grillo, G.~Parisi and R.~Gatto, \emph{{The shadow operator
  formalism for conformal algebra. Vacuum expectation values and operator
  products}}, \href{http://dx.doi.org/10.1007/BF02907130}{\emph{Lett. Nuovo
  Cim.} {\bf 4S2} (1972) 115--120}.

\bibitem{Simmons_Duffin_2014}
D.~Simmons-Duffin, \emph{{Projectors, Shadows, and Conformal Blocks}},
  \href{http://dx.doi.org/10.1007/JHEP04(2014)146}{\emph{JHEP} {\bf 04} (2014)
  146}, [\href{http://arxiv.org/abs/1204.3894}{{\tt 1204.3894}}].

\bibitem{Dobrev:1977qv}
V.~Dobrev, G.~Mack, V.~Petkova, S.~Petrova and I.~Todorov, \emph{{Harmonic
  Analysis on the n-Dimensional Lorentz Group and Its Application to Conformal
  Quantum Field Theory}}, vol.~63.
\newblock 1977,
  \href{http://dx.doi.org/10.1007/BFb0009678}{10.1007/BFb0009678}.

\bibitem{Hartman:2015lfa}
T.~Hartman, S.~Jain and S.~Kundu, \emph{{Causality Constraints in Conformal
  Field Theory}}, \href{http://dx.doi.org/10.1007/JHEP05(2016)099}{\emph{JHEP}
  {\bf 05} (2016) 099}, [\href{http://arxiv.org/abs/1509.00014}{{\tt
  1509.00014}}].

\bibitem{Camanho:2014apa}
X.~O. Camanho, J.~D. Edelstein, J.~Maldacena and A.~Zhiboedov, \emph{{Causality
  Constraints on Corrections to the Graviton Three-Point Coupling}},
  \href{http://dx.doi.org/10.1007/JHEP02(2016)020}{\emph{JHEP} {\bf 02} (2016)
  020}, [\href{http://arxiv.org/abs/1407.5597}{{\tt 1407.5597}}].

\bibitem{Regge:1959mz}
T.~Regge, \emph{{Introduction to complex orbital momenta}},
  \href{http://dx.doi.org/10.1007/BF02728177}{\emph{Nuovo Cim.} {\bf 14} (1959)
  951}.

\bibitem{Caron-Huot:2020nem}
S.~Caron-Huot and J.~Sandor, \emph{{Conformal Regge Theory at Finite Boost}},
  \href{http://arxiv.org/abs/2008.11759}{{\tt 2008.11759}}.

\bibitem{Ademollo:1989ag}
M.~Ademollo, A.~Bellini and M.~Ciafaloni, \emph{{Superstring Regge Amplitudes
  and Emission Vertices}},
  \href{http://dx.doi.org/10.1016/0370-2693(89)91609-2}{\emph{Phys. Lett. B}
  {\bf 223} (1989) 318--324}.

\bibitem{Ademollo:1990sd}
M.~Ademollo, A.~Bellini and M.~Ciafaloni, \emph{{Superstring Regge Amplitudes
  and Graviton Radiation at Planckian Energies}},
  \href{http://dx.doi.org/10.1016/0550-3213(90)90626-O}{\emph{Nucl. Phys. B}
  {\bf 338} (1990) 114--142}.

\bibitem{DAppollonio:2013mgj}
G.~D'Appollonio, P.~Di~Vecchia, R.~Russo and G.~Veneziano, \emph{{Microscopic
  unitary description of tidal excitations in high-energy string-brane
  collisions}}, \href{http://dx.doi.org/10.1007/JHEP11(2013)126}{\emph{JHEP}
  {\bf 11} (2013) 126}, [\href{http://arxiv.org/abs/1310.1254}{{\tt
  1310.1254}}].

\bibitem{Boels:2014dka}
R.~H. Boels and T.~Hansen, \emph{{String theory in target space}},
  \href{http://dx.doi.org/10.1007/JHEP06(2014)054}{\emph{JHEP} {\bf 06} (2014)
  054}, [\href{http://arxiv.org/abs/1402.6356}{{\tt 1402.6356}}].

\bibitem{Kulaxizi_2018}
M.~Kulaxizi, A.~Parnachev and A.~Zhiboedov, \emph{{Bulk Phase Shift, CFT Regge
  Limit and Einstein Gravity}},
  \href{http://dx.doi.org/10.1007/JHEP06(2018)121}{\emph{JHEP} {\bf 06} (2018)
  121}, [\href{http://arxiv.org/abs/1705.02934}{{\tt 1705.02934}}].

\bibitem{Gillioz:2018mto}
M.~Gillioz, \emph{{Momentum-space conformal blocks on the light cone}},
  \href{http://dx.doi.org/10.1007/JHEP10(2018)125}{\emph{JHEP} {\bf 10} (2018)
  125}, [\href{http://arxiv.org/abs/1807.07003}{{\tt 1807.07003}}].

\bibitem{Costa:2011dw}
M.~S. Costa, J.~Penedones, D.~Poland and S.~Rychkov, \emph{{Spinning Conformal
  Blocks}}, \href{http://dx.doi.org/10.1007/JHEP11(2011)154}{\emph{JHEP} {\bf
  11} (2011) 154}, [\href{http://arxiv.org/abs/1109.6321}{{\tt 1109.6321}}].

\bibitem{Karateev:2017jgd}
D.~Karateev, P.~Kravchuk and D.~Simmons-Duffin, \emph{{Weight Shifting
  Operators and Conformal Blocks}},
  \href{http://dx.doi.org/10.1007/JHEP02(2018)081}{\emph{JHEP} {\bf 02} (2018)
  081}, [\href{http://arxiv.org/abs/1706.07813}{{\tt 1706.07813}}].

\bibitem{Costa:2017twz}
M.~S. Costa, T.~Hansen and J.~Penedones, \emph{{Bounds for OPE coefficients on
  the Regge trajectory}},
  \href{http://dx.doi.org/10.1007/JHEP10(2017)197}{\emph{JHEP} {\bf 10} (2017)
  197}, [\href{http://arxiv.org/abs/1707.07689}{{\tt 1707.07689}}].

\bibitem{Penedones:2010ue}
J.~Penedones, \emph{{Writing CFT correlation functions as AdS scattering
  amplitudes}}, \href{http://dx.doi.org/10.1007/JHEP03(2011)025}{\emph{JHEP}
  {\bf 03} (2011) 025}, [\href{http://arxiv.org/abs/1011.1485}{{\tt
  1011.1485}}].

\bibitem{Li:2017lmh}
D.~Li, D.~Meltzer and D.~Poland, \emph{{Conformal Bootstrap in the Regge
  Limit}}, \href{http://dx.doi.org/10.1007/JHEP12(2017)013}{\emph{JHEP} {\bf
  12} (2017) 013}, [\href{http://arxiv.org/abs/1705.03453}{{\tt 1705.03453}}].

\bibitem{Costa:2014kfa}
M.~S. Costa, V.~Gonçalves and J.~Penedones, \emph{{Spinning AdS Propagators}},
  \href{http://dx.doi.org/10.1007/JHEP09(2014)064}{\emph{JHEP} {\bf 09} (2014)
  064}, [\href{http://arxiv.org/abs/1404.5625}{{\tt 1404.5625}}].

\bibitem{Carmi:2018qzm}
D.~Carmi, L.~Di~Pietro and S.~Komatsu, \emph{{A Study of Quantum Field Theories
  in AdS at Finite Coupling}},
  \href{http://dx.doi.org/10.1007/JHEP01(2019)200}{\emph{JHEP} {\bf 01} (2019)
  200}, [\href{http://arxiv.org/abs/1810.04185}{{\tt 1810.04185}}].

\bibitem{Meltzer:2017rtf}
D.~Meltzer and E.~Perlmutter, \emph{{Beyond $a = c$: gravitational couplings to
  matter and the stress tensor OPE}},
  \href{http://dx.doi.org/10.1007/JHEP07(2018)157}{\emph{JHEP} {\bf 07} (2018)
  157}, [\href{http://arxiv.org/abs/1712.04861}{{\tt 1712.04861}}].

\bibitem{Kotikov:2002ab}
A.~Kotikov and L.~Lipatov, \emph{{DGLAP and BFKL equations in the $N=4$
  supersymmetric gauge theory}},
  \href{http://dx.doi.org/10.1016/S0550-3213(03)00264-5}{\emph{Nucl. Phys. B}
  {\bf 661} (2003) 19--61}, [\href{http://arxiv.org/abs/hep-ph/0208220}{{\tt
  hep-ph/0208220}}].

\bibitem{Kotikov:2007cy}
A.~Kotikov, L.~Lipatov, A.~Rej, M.~Staudacher and V.~Velizhanin,
  \emph{{Dressing and wrapping}},
  \href{http://dx.doi.org/10.1088/1742-5468/2007/10/P10003}{\emph{J. Stat.
  Mech.} {\bf 0710} (2007) P10003}, [\href{http://arxiv.org/abs/0704.3586}{{\tt
  0704.3586}}].

\bibitem{Hanany:2010da}
A.~Hanany, D.~Forcella and J.~Troost, \emph{{The Covariant perturbative string
  spectrum}},
  \href{http://dx.doi.org/10.1016/j.nuclphysb.2011.01.002}{\emph{Nucl. Phys. B}
  {\bf 846} (2011) 212--225}, [\href{http://arxiv.org/abs/1007.2622}{{\tt
  1007.2622}}].

\bibitem{sagemath}
{The Sage Developers}, \emph{{S}ageMath, the {S}age {M}athematics {S}oftware
  {S}ystem ({V}ersion 9.1)}, 2020.

\bibitem{Costa:2016hju}
M.~S. Costa, T.~Hansen, J.~Penedones and E.~Trevisani, \emph{{Projectors and
  seed conformal blocks for traceless mixed-symmetry tensors}},
  \href{http://dx.doi.org/10.1007/JHEP07(2016)018}{\emph{JHEP} {\bf 07} (2016)
  018}, [\href{http://arxiv.org/abs/1603.05551}{{\tt 1603.05551}}].

\bibitem{Costa:2018mcg}
M.~S. Costa and T.~Hansen, \emph{{AdS Weight Shifting Operators}},
  \href{http://dx.doi.org/10.1007/JHEP09(2018)040}{\emph{JHEP} {\bf 09} (2018)
  040}, [\href{http://arxiv.org/abs/1805.01492}{{\tt 1805.01492}}].

\end{thebibliography}\endgroup

\end{document}